\documentclass[%
 reprint,
 amsmath,amssymb,
 aps,
]{revtex4-2}

\usepackage{graphicx}
\usepackage{dcolumn}
\usepackage{bm}
\usepackage{xcolor}
\usepackage{amsmath,amssymb,amsfonts,mathtools}
\usepackage{booktabs}
\usepackage[flushleft]{threeparttable}
\usepackage{indentfirst}
\usepackage{rotating}
\usepackage{makecell}
\usepackage{array}
\usepackage{subfigure}
\usepackage{float}
\usepackage{scalerel}
\usepackage{multirow}
\usepackage{stackengine}
\usepackage{tabularx}

\newcommand\CircArrowLeft[1]{\stackengine{-.3ex}{#1}{\CAL}{O}{c}{F}{F}{L}}
\newcommand\CAL{\scalebox{1.5}{\rotatebox[origin=center]{90}{$\circlearrowleft$}}}
\stackMath
\usepackage{chemformula} 
\usepackage[T1]{fontenc} 
\begin{document}
\preprint{APS/123-QED}
\title{A Comprehensive Multipolar Theory for Periodic Metasurfaces}
\author{Aso Rahimzadegan$^{1,\#}$}
\email{aso.rahimzadegan@kit.edu}
\author{Theodosios D. Karamanos$^{1,\#}$}
\email{theodosios.karamanos@kit.edu}
\author{Rasoul Alaee$^{1,2}$}
\author{Aristeidis G. Lamprianidis$^{1}$}
\author{Dominik Beutel$^{1}$}
\author{Robert W. Boyd$^{2}$}
\author{Carsten Rockstuhl$^{1,3}$}
\affiliation{ 
$^{1}$Institute of Theoretical Solid State Physics, Karlsruhe Institute of Technology, 76131 Karlsruhe, Germany}
\affiliation{ $^{2}$Department of Physics, University of Ottawa, Ottawa, ON K1N 6N5, Canada}
\affiliation{ $^{3}$Institute of Nanotechnology, Karlsruhe Institute of Technology, 76131 Karlsruhe, Germany}
\affiliation{ $^{\#}$These authors contributed equally.}
%
\begin{abstract}
Optical metasurfaces consist of a 2D arrangement of scatterers, and they control the amplitude, phase, and polarization of an incidence field on demand. Optical metasurfaces are the cornerstone for a future generation of flat optical devices in a wide range of applications. The rapidly growing advances in nanofabrication have made the versatile design and analysis of these ultra-thin surfaces an ever-growing necessity. However, despite their importance, a comprehensive theory to describe the optical response of periodic metasurfaces in closed-form and analytical expressions has not been formulated, and prior attempts were frequently approximate. Here, we develop a theory that analytically links the properties of the scatterer, from which a periodic metasurface is made, to its optical response via the lattice coupling matrix. The scatterers are represented by their polarizability or T matrix, and our theory works for normal and oblique incidence. We provide explicit expressions for the optical response up to octupolar order in both spherical and Cartesian coordinates. Several examples demonstrate that our analytical tool constitutes a paradigm shift in designing and understanding optical metasurfaces. Novel fully-diffracting metagratings and particle-independent polarization filters are proposed, and novel insights into the response of Huygens' metasurfaces under oblique incidence are provided. Our analytical expressions are a powerful tool for exploring the physics of metasurfaces and designing novel flat optics devices.
\end{abstract}
\keywords{metasurfaces, optics, T matrix, polarizability, array, scattering}
\maketitle
\section{Introduction}
During the past decade, research in electromagnetic metamaterials has grown into a solid and mature scientific domain. Their two-dimensional counterparts, metasurfaces, have gained particular attention thanks to their easier fabrication combined with exciting application perspectives at microwave, THz, and optical frequencies \cite{koshelev2020dielectric,qiu2021quo,Staude2017,kamali2018review,Kuznetsov2016}. Metasurfaces exhibit a plethora of properties, which make them appealing for a wide range of applications, such as absorbers \cite{landy2008perfect,alaee2017theory}, reciprocal and nonreciprocal polarization rotators \cite{zhao2011conjugated,kodera2011artificial}, holograms \cite{genevet2015holographic,huang2018metasurface,Chong2015,Zhou2013,ni2013metasurface,rahimzadegan2020beyond}, lenses \cite{lalanne2017metalenses,arbabi2015dielectric,khorasaninejad2015achromatic,lin2014dielectric,west2014all}, splitters \cite{khorasaninejad2015efficient}, diffusers \cite{jang2018wavefront,Arslan2021}, light sails for space explorations \cite{siegel2019self,gieseler2021self}, biomedical applications \cite{tittl2018imaging}, as well as computational and quantum applications \cite{stav2018quantum,wang2018quantum}. 

The growing number of metasurface applications and rapid advances in their fabrication and characterization \cite{yoon2021recent} prompt methodologies to accurately analyze and design metasurfaces. While full-wave numerical solutions are always an option, analytical tools can be much more appealing because they facilitate the design and provide valuable insights into the underlying physics of metasurfaces. For periodic metasurfaces that consist of a single scatterer per unit cell, the type of metasurfaces on which we concentrate, herein, several techniques exist for this purpose.
First, comprehensible circuit models of metasurfaces and metamaterials \cite{marques2011metamaterials,epstein2016huygens,chen2018theory} were developed, which are easy to use in industry, especially for microwave applications. A second approach follows the homogenization principle. It aims to replace metasurfaces at stake with surfaces with equivalent surface susceptibilities \cite{holloway2009discussion,holloway2011characterizing,achouri2015general}. Although very helpful for component design, these methods are inadequate to describe the internal physics of the structures under study, such as the interaction of consisting particles. Moreover, circuit modeling and homogenization methodologies involve, sometimes, assumptions that simplify the investigated problem at the expense of accuracy.

More from "first-principles", a third approach aims to construct the response of 2D arrays from the bottom-up by summing the response of its constituting particles. While sharing some characteristics with the two approaches mentioned initially, this bottom-up approach is more general and versatile. It enables the easier handling of a plethora of designs, including mm-wave and optical applications \cite{tretyakov2003analytical,babicheva2018metasurfaces,evlyukhin2020bianisotropy,evlyukhin2010optical,shamkhi2019transparency,alaee2017theory,yazdi2015bianisotropic,ra2013total,ross2016optical}. In this approach, the optical action of the constituting particles is best discussed using a multipolar expansion of fields \cite{Xu1995c,muhlig2011multipole,mishchenko2002scattering,dezert2019complete}. Within the multipolar expansion, the scatterers' optical response is expressed in a series of multipole moments induced by the external illumination and the scattered field from all the other particles forming the metasurface.  Using an ever-increasing number of multipole moments is important to capture the response of meta-atoms and consequently the metasurface more accurately. The involved fields are expanded into an orthonormal basis set to reach an algebraic formulation of the scattering process. The amplitude of each mode used to expand the incident and the scattered field is one element of a dedicated vector. The relation between the expansion coefficient of the incident and the scattered field is, then, merely a matrix multiplication. The connecting matrix can then be considered as the most comprehensive representation of the scatterers' optical properties. Two different formulations for this matrix can be found, namely the polarizability matrix and the T matrix. 

The polarizability matrix expresses the scattering response in Cartesian coordinates. Several methodologies have been developed to acquire the polarizability matrix beyond simple shapes, especially for the lowest, i.e., the dipolar order \cite{asadchy2014determining,liu2016polarizability, karamanos2018full}. On the other hand, the T matrix expresses the scattering response in spherical coordinates. It has attracted a substantial share of interest as it can easily accommodate higher-order multipole moments \cite{fruhnert2017computing, demesy2018scattering, santiago2019decomposition}. Due to the equivalence of Cartesian and spherical coordinates representations \cite{alaee2019exact}, polarizability and T matrices are interchangeable in the sense that they contain the same information. This equivalence has been explicitly documented up to quadrupolar order \cite{bernal2014underpinning}, and, more recently, up to octupolar order \cite{mun2020describing}.

Modeling of metasurfaces via the multipolar analysis initially involved considering particles characterized by only dipole moments \cite{tretyakov2003analytical}, while specific quadrupole moments were added later into the models \cite{alaee2015magnetoelectric,alaee2017theory}. Moreover, the description of the interaction among all the particles forming the periodic metasurfaces is crucial in every modeling attempt. Earlier works involved approximate expression for this purpose and failed to accurately capture the spatial dispersion occurring in many applications where the metasurfaces are not operated in a deep sub-wavelength regime \cite{tretyakov2003analytical}. Following this observation, efforts shifted into expressing this lattice interaction more accurately, particularly via fast converging Green's function summations \cite{belov2005homogenization}. This has resulted in interesting models that could accommodate dipole moments at oblique incidence \cite{dimitriadis2012surface,dimitriadis2015generalized,albooyeh2014resonant}, dipole and quadrupole moments at normal incidence \cite{babicheva2019analytical}, and even up to octupole moments at normal incidence \cite{babicheva2021multipole}. However, these efforts were generally limited in scope, e.g., focusing on specific particles with specific combinations of multipole moments (i.e., isotropic particles most of the time) or were limited to normal incidence. Furthermore, diffracting metasurfaces were not studied because sub-wavelength metasurfaces were considered that sustain only a zeroth-order mode in reflection and transmission. 

To alleviate these problems, D. Beutel {\it et al.} \cite{beutel2021efficient} used spherical coordinates and a T matrix representation to develop a numerical method to calculate the complete response of a metasurface, i.e., propagating and evanescent modes, for any particle and up to a desired multipolar order. Based on previous efforts on isotropic particles \cite{modinos1987scattering,stefanou2000multem}, this approach employs the Ewald summation \cite{ewald1921berechnung,popov2020calculation} for the fast-converging determination of the lattice couplings to achieve a complete description of a 2D array response. Although efficient, this work lacks the interchangeability between spherical and -more popular- Cartesian representations. It also lacks closed-form analytical expressions, which increases the understanding and versatility among users in physics and engineering.
\begin{figure}[t]
    \centering
\includegraphics[scale=0.43]{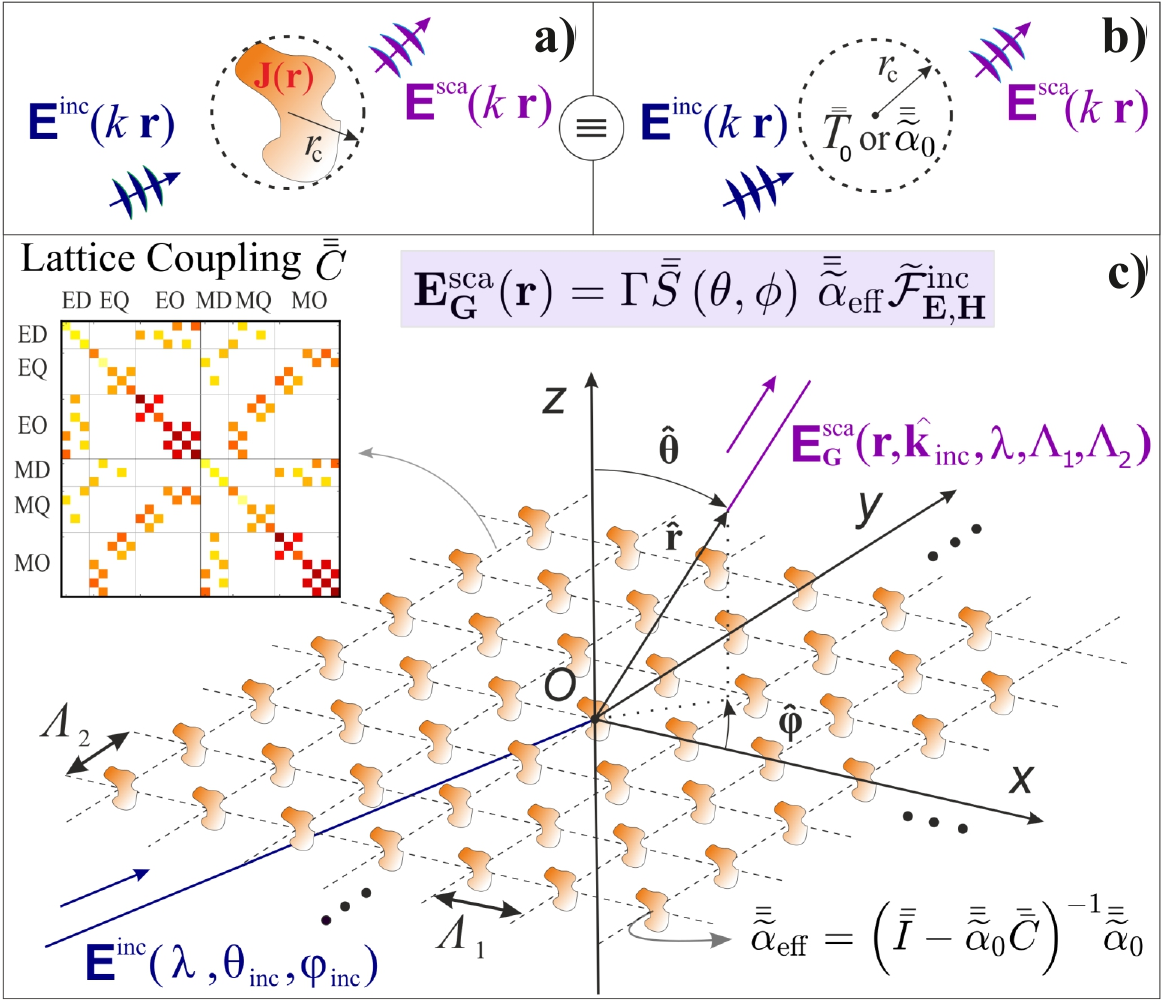}
    \caption{\textbf{The set-up:} a) An arbitrary particle placed into an infinite,  homogeneous space, illuminated by an incident wave, $\mathbf{E}^{\rm inc}$, and its subsequent scattered wave, $\mathbf{E}^{\rm sca}$. The radius of the smallest sphere enclosing the particle is $r_c$, while $\mathbf{J}(\mathbf{r})$ is the induced current volume density. b) Equivalent setup as Fig.~\ref{fig1}a for $r>r_c$, when the respective T or polarizability matrices are used. c) A scattering rectangular 2D lattice of identical scatterers along with the set-up Cartesian and spherical coordinate systems. A simplified equation for the scattered field in Cartesian coordinates is shown in the figure. This response depends on the effective polarizability $\bar{\bar{\widetilde{\alpha}}}_{\rm eff}$, the Cartesian multipole-to-field translation matrix $\bar{\bar{S}}\left(\theta,\phi\right)$, and the incident electromagnetic field $\widetilde{\mathcal{F}}_{\mathbf{E,H}}^{\rm inc}$. The effective polarizability is a function of the lattice coupling matrix $\bar{\bar{C}}$ and the polarizability of the isolated particle $\bar{\bar{\widetilde{\alpha}}}_{\rm 0}$. The inset shows the lattice coupling for a rectangular lattice. The labels denote the electric (E) and magnetic (M) coupling of dipolar (D), quadrupolar (Q), and octupolar (O) orders.}
\label{fig1}
\vspace{-0mm}
\end{figure}

In this work, we derive accessible expressions for the optical response of periodic metasurfaces to provide a unifying and comprehensive framework. The expressions are based on a multipole expansion and accurately express the amplitudes of propagating diffraction orders of periodic metasurfaces upon illumination at normal or oblique incidence. This approach renders our contribution relevant for the study of metasurfaces and diffracting metagratings. While higher-order multipole moments can be accommodated, we express the response from the scattering structure defining the unit cell of the metasurface up to the octupolar order. Unlike previous attempts, the proposed methodology is interchangeable between a Cartesian and spherical basis, meaning that either the polarizability or T matrix of a particle can be considered, making our contribution flexible, general, and convenient to use.
Additionally, we demonstrate reducing our generally valid expressions to handy closed-form analytical formulas for selected specific cases if not all degrees of freedom are accommodated. Such reduction eases physical explorations and simplifies the design. Finally, the robustness of the provided analytical formulas is demonstrated when applying them to selected design challenges for metasurfaces and metagratings.

The paper is structured as follows. The first section defines the multipole moments and fields and provides formulas that transform them between a Cartesian and spherical basis. Additionally, the lattice coupling matrices are defined (Fig.~\ref{FIG:latticeFull} Row I), and the concepts of effective polarizability/T matrices within 2D arrays are elaborated. At the end of this section, closed-form equations to express the optical response from scattering metasurfaces composed of meta-atoms with general symmetries described up to octupolar order and for an arbitrary illumination direction are presented in a Cartesian and vector spherical harmonics basis. Simplified expressions are provided for rectangular and cubic lattices. In the following section, we explore the symmetry of the lattice coupling matrix at the practically most important examples of a square and hexagonal lattice at normal incidence. The isolated and effective polarizability or T matrices of three meta-atoms with distinct symmetries (isotropic, anisotropic, bianisotropic) are explored inside and outside a square/hexagonal lattice.  

Afterward, we demonstrate how to reduce the most comprehensive expressions to some special cases and how to use these expressions in specific design challenges. Subsequently, we derive an analytic expression for the amplitudes of the propagating diffraction orders in transmission and reflection from a square-periodic array decorated with isotropic particles described in dipolar-quadrupolar approximation and illuminated at normal incidence. These analytic equations help in designing a fully diffracting metagrating.

In the last section, we explore obliquely illuminated metasurfaces. We derive an analytic expression for the amplitude of the zeroth-order in transmission and reflection of metasurfaces made from isotropic and dipolar meta-atoms. The analytic equations help to design a metaatom-independent polarization filter. It also helps to analyze Huygens' metasurfaces \cite{Staude2017,decker2015high,chen2018huygensCoreshell,liu2017huygens} under oblique incidence. 

Finally, the \textit{Appendix} includes essential derivations and equations, while the Supplementary Information or \textit{Supp. Info.} includes the step-by-step derivations and complementary information.
\section{Multipolar calculation of the scattering field from a 2D array in Cartesian and vector spherical harmonics basis: General Equations} \label{subsection:basicEq}
\subsection{Isolated particles}
Let us consider an arbitrary particle placed in an infinite, homogeneous surrounding, as shown in Fig.~\ref{fig1}a. In the vector spherical harmonics basis, the scattering response of the particle to an incident electromagnetic wave outside the smallest sphere circumscribing the particle can be described via the T matrix, or $\bar{\bar{T}}_0$, as
\begin{subequations} \label{t-matrix-def}
\begin{gather}
\left[\hspace{-0mm}\begin{array}{c}
\mathbf{b}^{\rm e}_1\\[0.05cm]
\mathbf{b}^{\rm e}_2\\[0.05cm]
\mathbf{b}^{\rm e}_3\\[0.05cm]
\mathbf{b}^{\rm m}_1\\[0.05cm]
\mathbf{b}^{\rm m}_2\\[0.05cm]
\mathbf{b}^{\rm m}_3
\end{array}\hspace{-0mm}\right] \hspace{-0mm} = \hspace{-0mm}\bar{\bar{T}}_0\hspace{-0mm}
\left[\hspace{-0mm}\begin{array}{c}
\mathbf{q}^{\rm e}_1\\[0.05cm]
\mathbf{q}^{\rm e}_2\\[0.05cm]
\mathbf{q}^{\rm e}_3\\[0.05cm]
\mathbf{q}^{\rm m}_1\\[0.05cm]
\mathbf{q}^{\rm m}_2\\[0.05cm]
\mathbf{q}^{\rm m}_3
\end{array}\hspace{-0mm}\right],\\
\intertext{with}
\bar{\bar{T}}_0 = 
    \left[\hspace{-0mm}
\arraycolsep=1mm
\begin{array}{cccccc}
\bar{\bar{T}}_{11}^{\rm ee} & \bar{\bar{T}}_{12}^{\rm ee} & \bar{\bar{T}}_{13}^{\rm ee} & \bar{\bar{T}}_{11}^{\rm em} & \bar{\bar{T}}_{12}^{\rm em} & \bar{\bar{T}}_{13}^{\rm em}  \\[0.05cm]
\bar{\bar{T}}_{21}^{\rm ee} & \bar{\bar{T}}_{22}^{\rm ee} & \bar{\bar{T}}_{23}^{\rm ee} & \bar{\bar{T}}_{21}^{\rm em} & \bar{\bar{T}}_{22}^{\rm em} & \bar{\bar{T}}_{23}^{\rm em}  \\[0.05cm]
\bar{\bar{T}}_{31}^{\rm ee} & \bar{\bar{T}}_{32}^{\rm ee} & \bar{\bar{T}}_{33}^{\rm ee} & \bar{\bar{T}}_{31}^{\rm em} & \bar{\bar{T}}_{32}^{\rm em} & \bar{\bar{T}}_{33}^{\rm em}  \\[0.05cm]
\bar{\bar{T}}_{11}^{\rm me} & \bar{\bar{T}}_{12}^{\rm me} & \bar{\bar{T}}_{13}^{\rm me} & \bar{\bar{T}}_{11}^{\rm mm} & \bar{\bar{T}}_{12}^{\rm mm} & \bar{\bar{T}}_{13}^{\rm mm} \\[0.05cm]
\bar{\bar{T}}_{21}^{\rm me} & \bar{\bar{T}}_{22}^{\rm me} &  \bar{\bar{T}}_{23}^{\rm me} & \bar{\bar{T}}_{21}^{\rm mm} & \bar{\bar{T}}_{22}^{\rm mm} & \bar{\bar{T}}_{23}^{\rm mm} \\[0.05cm]
\bar{\bar{T}}_{31}^{\rm me} & \bar{\bar{T}}_{32}^{\rm me} &  \bar{\bar{T}}_{33}^{\rm me} & \bar{\bar{T}}_{31}^{\rm mm} & \bar{\bar{T}}_{32}^{\rm mm} & \bar{\bar{T}}_{33}^{\rm mm}
\end{array}\hspace{-0mm}\right],
\end{gather}
\end{subequations}
\noindent herein, expressed up to the third (i.e., the octupolar) order. The T matrix represents the electromagnetic response of a scatterer. The scattering coefficient vectors in the vector spherical harmonics (VSH) basis are defined as  $\mathbf{b}^v_j = [b^v_{j,-j}\,\,b^v_{j,-j+1}\,\, ... \,\,b^v_{j,j-1}\,\,b^v_{j,j}]^T$, with $v=\{{\rm e,m}\}$ denoting the electric or magnetic multipoles and $j=\{1,2,3\}$ being the multipolar order corresponding to dipole, quadrupole, and octupole response. The vectors $\mathbf{q}^v_j$ contain the amplitude coefficients expanding the incident field similarly to the scattering coefficient vectors. The subscript "0" for the T matrix refers to the response of an isolated particle. 

Note that VSH functions are defined, herein, as in Ref.~\citenum{mishchenko2002scattering} ($\rightarrow$ \textit{Appendix A}). Alternatively, the scattering coefficients can be calculated from the scattered field of a particle as defined in \textit{Appendix A}.

Alternative to the vector spherical harmonics basis, we can also describe the scattering response in the Cartesian basis by a normalized polarizability matrix, or in short, the polarizability matrix, $\bar{\bar{\widetilde{\alpha}}}_0$, defined up to the octupolar order through 
\begin{subequations} \label{a-matrix-def}
\begin{equation}
\left[\hspace{-0.5mm}\begin{array}{r}
\left(\varepsilon \zeta_1\right)^{-1} \mathbf{p}^{\phantom{a}}\\[0.05cm]
 k \left(\varepsilon \zeta_2\right)^{-1}\mathbf{Q}^{\rm e}\\[0.05cm]
k^2 \left(\varepsilon  \zeta_3\right)^{-1}\mathbf{O}^{\rm e}\\[0.05cm]
\mathrm{i}\eta\left(  \zeta_1\right)^{-1}\mathbf{m}^{\phantom{a}}\\[0.05cm]
\mathrm{i}\eta k \left(  \zeta_2\right)^{-1}\mathbf{Q}^{\rm m}\\[0.05cm]
\mathrm{i}\eta k^2\left(  \zeta_3\right)^{-1} \mathbf{O}^{\rm m}
\end{array}\hspace{-0.5mm}\right] \hspace{-0.8mm}=\hspace{-0.8mm} \frac{\bar{\bar{\widetilde{\alpha}}}_0}{k^3} \widetilde{\mathcal{F}}_{\mathbf{E,H}}^{\rm inc}\hspace{-0.8mm} =\hspace{-0.8mm} \frac{\bar{\bar{\widetilde{\alpha}}}_0}{k^3} \hspace{-0.5mm}
\left[\hspace{-0.5mm}\begin{array}{r}
 \zeta_1\mathbf{E}_1 \\[0.05cm]
k^{-1}\zeta_2\mathbf{E}_2\\[0.05cm]
k^{-2}\zeta_3\mathbf{E}_3\\[0.05cm]
\mathrm{i}\eta\zeta_1\mathbf{H}_1\\[0.05cm]
\mathrm{i}\eta k^{-1}\zeta_2 \mathbf{H}_2\\[0.05cm]
\mathrm{i}\eta  k^{-2} \zeta_3\mathbf{H}_3
\end{array}\hspace{-0.5mm}\right]\hspace{-0.3mm},
\end{equation}
\text{with}
\begin{equation}
\bar{\bar{\widetilde{\alpha}}}_0 \hspace{-0mm} = \hspace{-0mm}
\left[\hspace{-0mm}\arraycolsep=1mm
\begin{array}{cccccc}
\bar{\bar{\widetilde{\alpha}}}_{11}^{\rm ee} & \bar{\bar{\widetilde{\alpha}}}_{12}^{\rm ee} & \bar{\bar{\widetilde{\alpha}}}_{13}^{\rm ee} & \bar{\bar{\widetilde{\alpha}}}_{11}^{\rm em} & \bar{\bar{\widetilde{\alpha}}}_{12}^{\rm em} & \bar{\bar{\widetilde{\alpha}}}_{13}^{\rm em} \\[0.05cm]
\bar{\bar{\widetilde{\alpha}}}_{21}^{\rm ee} & \bar{\bar{\widetilde{\alpha}}}_{22}^{\rm ee} & \bar{\bar{\widetilde{\alpha}}}_{23}^{\rm ee} & \bar{\bar{\widetilde{\alpha}}}_{21}^{\rm em} & \bar{\bar{\widetilde{\alpha}}}_{22}^{\rm em} & \bar{\bar{\widetilde{\alpha}}}_{23}^{\rm em}  \\[0.05cm]
\bar{\bar{\widetilde{\alpha}}}_{31}^{\rm ee} & \bar{\bar{\widetilde{\alpha}}}_{32}^{\rm ee} & \bar{\bar{\widetilde{\alpha}}}_{33}^{\rm ee} & \bar{\bar{\widetilde{\alpha}}}_{31}^{\rm em} & \bar{\bar{\widetilde{\alpha}}}_{32}^{\rm em} & \bar{\bar{\widetilde{\alpha}}}_{33}^{\rm em} \\[0.05cm]
\bar{\bar{\widetilde{\alpha}}}_{11}^{\rm me} & \bar{\bar{\widetilde{\alpha}}}_{12}^{\rm me} & \bar{\bar{\widetilde{\alpha}}}_{13}^{\rm me} & \bar{\bar{\widetilde{\alpha}}}_{11}^{\rm mm} & \bar{\bar{\widetilde{\alpha}}}_{12}^{\rm mm} & \bar{\bar{\widetilde{\alpha}}}_{13}^{\rm mm}  \\[0.05cm]
\bar{\bar{\widetilde{\alpha}}}_{21}^{\rm me} & \bar{\bar{\widetilde{\alpha}}}_{22}^{\rm me} & \bar{\bar{\widetilde{\alpha}}}_{23}^{\rm me} & \bar{\bar{\widetilde{\alpha}}}_{21}^{\rm mm} & \bar{\bar{\widetilde{\alpha}}}_{22}^{\rm mm} & \bar{\bar{\widetilde{\alpha}}}_{23}^{\rm mm} \\[0.05cm]
\bar{\bar{\widetilde{\alpha}}}_{31}^{\rm me} & \bar{\bar{\widetilde{\alpha}}}_{32}^{\rm me} & \bar{\bar{\widetilde{\alpha}}}_{33}^{\rm me} & \bar{\bar{\widetilde{\alpha}}}_{31}^{\rm mm} & \bar{\bar{\widetilde{\alpha}}}_{32}^{\rm mm} & \bar{\bar{\widetilde{\alpha}}}_{33}^{\rm mm} 
\end{array}\hspace{-0mm}\right]\hspace{-0mm},
\end{equation}
\begin{equation}
\zeta_j = \sqrt{(2j+1)!\,\pi},
\end{equation}
\end{subequations}
\noindent where $k=2\pi/\lambda$ is the wavenumber of the scattered field in the embedding medium, $\eta=\sqrt{\mu/\varepsilon}$ is the impedance of the embedding medium, $\varepsilon$ and $\mu$ are the permittivity and permeability of the embedding medium, respectively, and $\widetilde{\mathcal{F}}_{\mathbf{E,H}}^{\rm inc}$ is the normalized electromagnetic incident field. The tilde indicates the normalized polarizability ($\rightarrow$ {\textit{Appendix B}}). This dimensionless and irreducible polarizability matrix $\bar{\bar{\widetilde{\alpha}}}_0$ facilitates analytic calculations and will simplify equations later on in this work. The vectors $\mathbf{E}_n$ and $\mathbf{H}_n$ are the electric and magnetic multipolar amplitudes of the incident field as defined in \textit{Appendix B}, and contain spatial derivatives of the Cartesian incident fields at the origin considered as the center of the particle \cite{feshbach2019methods, bernal2014underpinning, mun2020describing}. The vectors $\mathbf{p}$  ($\mathbf{m}$), $\mathbf{Q}^{\rm e}$ ($\mathbf{Q}^{\rm m}$), and $\mathbf{O}^{\rm e}$ ($\mathbf{O}^{\rm m}$), are the irreducible Cartesian electric (magnetic) dipole, quadrupole, and octupole moments, respectively  ($\rightarrow$ \textit{Appendix C}). The far-fields radiated by a scattering particle as a function of the multipole moments expressed in Cartesian basis are also provided in \textit{Appendix D}. 

If we employ the transformation formulas, we can acquire the elements of the T matrix from the ones of the polarizability matrix and vice versa via
\begin{subequations}\label{T-from-a}
\begin{align}
\bar{\bar{T}}_{jj'}^{vv'} &= \mathrm{i}\,\bar{\bar{F}}_j^{-1}\,{\bar{\bar{\widetilde{\alpha}}}}_{jj'}^{vv'}\,\bar{\bar{F}}_{j'},  \\
\bar{\bar{\widetilde{\alpha}}}_{jj'}^{vv'} &= -\mathrm{i}\,\bar{\bar{F}}_j\,\bar{\bar{T}}_{jj'}^{vv'}\,\bar{\bar{F}}_{j'}^{-1}, 
\end{align}
\end{subequations}
where $\{j,j'\} = \{1,2,3\}$ and $\{v,v'\} = \{\rm e,\rm m\}$. The expressions do not have a subscript '0' as they are valid for the particles independent of whether we consider them isolated in free space or as a part of the lattice. In this work, the use of the normalized polarizability matrix and the choice of transformer tensors such that $\bar{\bar{F}}_j = (\bar{\bar{F}}_j^{-1})^\dagger$ ($\rightarrow$ \textit{Appendix E}) enabled the formulation of simpler formulas than the case of Ref.~\citenum{mun2020describing}. This has become possible by defining, herein, the multipole and the fields in Cartesian coordinates as provided in \textit{Appendix C}.
Therefore, a scattering particle can be described either in the Cartesian or the vector spherical harmonics basis and be replaced by the respective T or polarizability matrix, as depicted in Fig.~\ref{fig1}b, simplifying, subsequently, the analysis extensively. 

\begin{figure*}[t]
    \centering
    \includegraphics[scale=0.883]{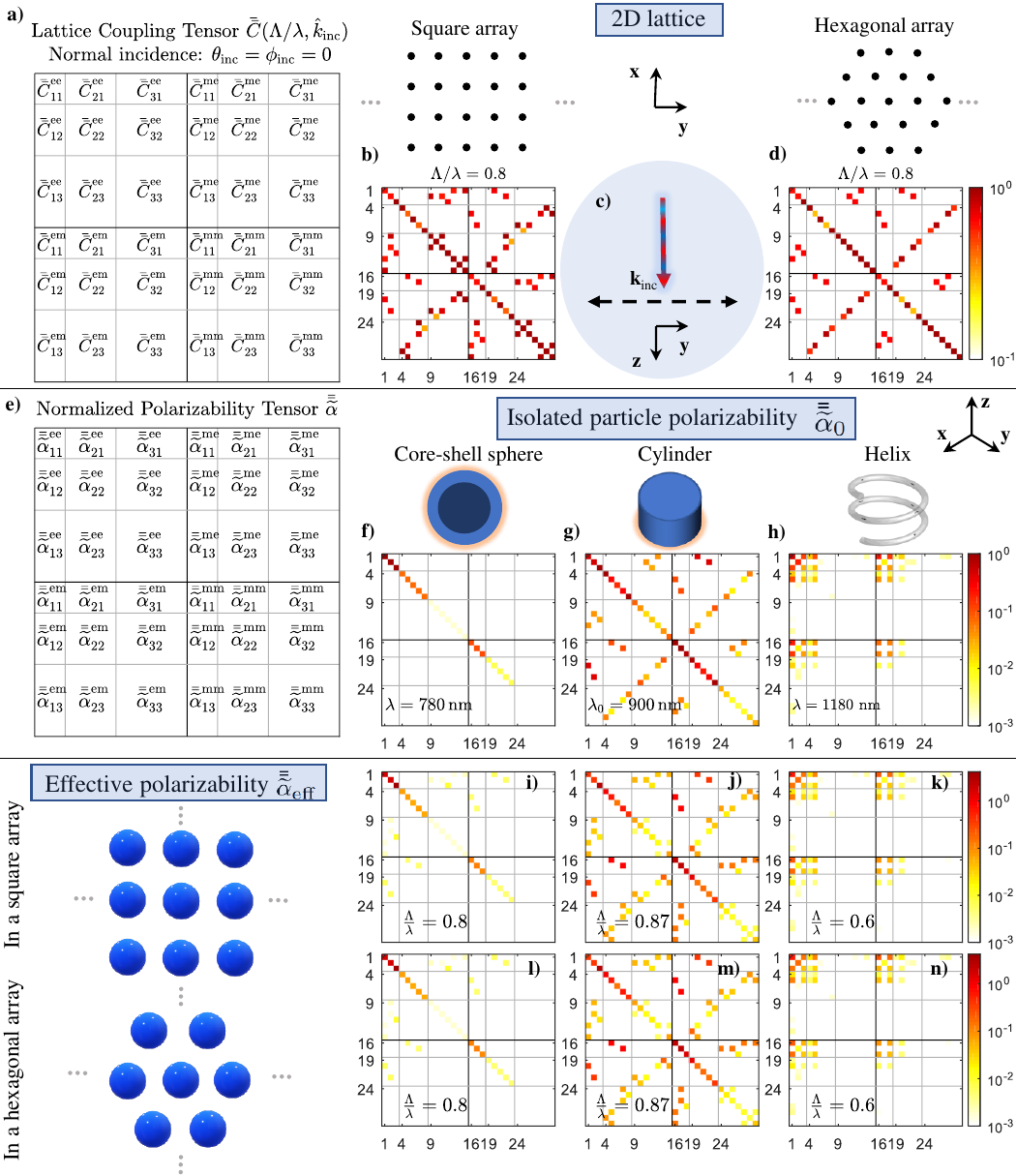}
    \caption{\textbf{Symmetries of scatterers and 2D lattices in Cartesian basis}: \textbf{Row I:} a) The Cartesian lattice coupling matrix amplitude up to octupolar order for b) a square array and d) a hexagonal array under normal incidence (i.e. $\theta_{\rm inc}=0$), as shown in c). The normalized periodicity $\tilde{\Lambda}$ for both arrays is 0.8. \textbf{Row II:}  e) The normalized polarizability matrix amplitude for f) an isolated Ag-core SiO$_2$-shell particle ($r_\mathrm{core}=120\,{\rm nm},~r_\mathrm{shell}=120+30\,{\rm nm}$) in free space at $\lambda=780$~nm, and g) an isolated amorphous silicon ($n=3.959+0.009\mathrm{i}$) cylinder ($r=291\,{\rm nm},h=211\,{\rm nm}$) embedded in silica ($n=1.44$) at $\lambda_0=900$~nm, and h) an isolated silver ($n=0.095+8.675\mathrm{i}$ \cite{mcpeak2015plasmonic}) helix ($R_{\rm axial}=80$~nm, $r_{\rm rod}=20$~nm, $P_{\rm pitch}=105$~nm, and $N_{\rm turn}=2$) in free space at $\lambda=1180$~nm. \textbf{Row III:} The normalized effective polarizability amplitude of i) \& j) the core-shell sphere, j) \& m) the cylinder, and k) \& n) the helix inside an infinitely periodic i)-k) square array or l)-n) hexagonal array.}\vspace{-4.5mm}
\label{FIG:latticeFull}
\end{figure*}

The normalized polarizability matrix of three objects with three different symmetries (spherical, cylindrical, and helical) are shown up to octupolar order in Fig.~\ref{FIG:latticeFull} Row II. For the interested reader and completeness, the vector spherical harmonics counterpart of the figure is plotted in the \textit{Supp. Info.} Fig.~S3. While analytical solutions for the polarizability of isotropic particles are available via the Mie theory, the polarizability of the non-spherical particles has been obtained from full-wave numerical simulations based on the finite-element method as described in \textit{Appendix A}. These numerical simulations are the only full-wave simulations involved in our analysis. They constitute the base, as they provide information on how a single particle scatters light. Nevertheless, once it is calculated and stored, it can be reused in all future calculations that consider the same particle.

An isotropic particle (Fig.~\ref{FIG:latticeFull}f) has a diagonal T and polarizability matrix. The diagonal elements of the T matrix are the Mie coefficients, but with a negative sign, in agreement with the definitions of VSH ($\rightarrow$ \textit{Appendix A}). The diagonal elements of the normalized polarizability are the Mie coefficients with an "$\mathrm{i}$" multiplicand. Unlike isotropic particles, a particle with cylindrical symmetry (Fig.~\ref{FIG:latticeFull}g) only has a diagonal polarizability matrix for the dipolar order. Beyond dipolar approximation, non-diagonal terms appear, which need to be taken into account. For helical structures (Fig.~\ref{FIG:latticeFull}h) that possess a chiral response, non-zero terms exist in the diagonal of the electric-magnetic polarizability matrices. Note that the white elements in the matrices in Figs.~\ref{FIG:latticeFull} Row II are either symmetry-protected strictly zero or express a diminishing response of the small particle for higher multipolar orders. Especially these symmetry-protected zeros are important, as they help construct a theoretical model up to a particular multipolar order and for a specific particle symmetry that leads to simplified analytical equations for the metasurface scattering response. We can ignore them from the very first beginning. The symmetry of the normalized polarizability or the T matrix within the defined bases fully describes the electromagnetic symmetry and response of a particle. This feature makes these two matrices crucial tools in nanophotonics design and analysis.

So far, we have focused on the response of isolated particles. The following subsection explores how these polarizabilities/T matrices are modified inside a 2D lattice and how to derive such effective matrices analytically.
\subsection{Periodic arrangement of identical particles}
Let us assume an infinite number of arbitrary, but identical particles arranged in a 2D lattice described by two unit-cell base vectors, ${\mathbf{u}}_1$ and ${\mathbf{u}}_2$, parallel to the lattice plane \cite{kittel1996introduction}, with $|{\mathbf{u}}_1| = \Lambda_1$ and $|{\mathbf{u}}_2| = \Lambda_2$ being the two periodicities (Fig.~\ref{fig1}c). The arrangement is embedded in a homogeneous material with refractive index $n = \sqrt{\varepsilon_{\rm r}\,\mu_{\rm r}}$, with $\varepsilon_{\rm r}$ and $\mu_{\rm r}$ being the relative permittivity and permeability of the medium, respectively. 

Now, let us assume that the 2D lattice is illuminated by a time-harmonic plane electromagnetic wave with an electric field corresponding to $\mathbf{E}^{\rm inc} = \mathbf{E}_0\,e^{\mathrm{i}\mathbf{k}^{\rm inc}\cdot\mathbf{r}}$, with $\mathbf{k}^{\rm inc} \left(\theta_{\rm inc},\phi_{\rm inc},\lambda\right) = k_0 n (\hat{\mathbf{k}}_{\rm inc}\cdot\hat{\mathbf{r}})={k}^{\rm inc}_x\, \hat{\mathbf{x}} + {k}^{\rm inc}_y\, \hat{\mathbf{y}} + {k}^{\rm inc}_z\, \hat{\mathbf{z}}$ being the incident field wavevector and $E_0=|\mathbf{E}_0|$ being the amplitude of the incident plane wave, which, in this work, is normalized, i.e., $E_0 = 1$ V/m unless explicitly mentioned. Note that $\lambda=\lambda_0/n$ is the wavelength inside the embedding medium.
\subsubsection{Vector spherical harmonics basis} 
For the case of the spherical coordinate, the general equation for the amplitude of each diffraction order propagating in a medium without absorption and supported by the reciprocal lattice $\mathbf{G}$, can be derived as \cite{antonakakis2014gratings} ($\rightarrow$ \textit{Supp. Info. I.A} and \textit{I.B})
\begin{equation}\label{spherical-general-1}
\begin{split}
&\mathbf{E}^{\rm sca}_{\textnormal{s},\mathbf{G}} (\mathbf{r}) = \left[\hspace{-0mm}\begin{array}{c}
E^{\rm sca}_{\mathbf{G},\theta}(\mathbf{r})\\
E^{\rm sca}_{\mathbf{G},\phi}(\mathbf{r})
\end{array} \hspace{-0mm}\right]
= \\ 
&=\frac{\mathrm{i}\sqrt{\pi}}{2A k^2} \frac{e^{\mathrm{i}\mathbf{k}_{\mathbf{G}}^{\pm}\cdot\mathbf{r}}}{|\cos{\theta}|} \sum_{j=1}^{3}\frac{\sqrt{2j+1}}{\mathrm{i}^{\,j}}  \, \bar{\bar{W_j}}\left(\theta,\phi\right) \left[\hspace{-0mm}\begin{array}{c}
\mathbf{b}^{\rm e}_j\\
\mathbf{b}^{\rm m}_j
\end{array} \hspace{-0mm}\right]\hspace{0mm},
\end{split}
\end{equation}
\noindent where $A= ({\mathbf{u}}_1\times{\mathbf{u}}_2)\cdot\hat{\mathbf{z}}$ is the area of the unit cell, $\mathbf{k}^{\pm}_\mathbf{G}$ is the wavevector of the diffraction order of the lattice $\mathbf{G}$, and $(\theta,\phi)$ are the polar and azimuth angles of the wavevector. The "+" and "-" signs refer to forward (i.e., $0\leq
\theta\leq \pi/2$) and backward (i.e., $\pi/2< \theta\leq
\pi$) propagating diffraction orders, respectively. It corresponds to transmission and reflection. The matrix $\bar{\bar{W}}\left(\theta,\phi\right)$ is the spherical multipole-field translation matrix, depending only on the direction of the diffraction order, and contains trigonometric functions. We have calculated its elements as analytic relations up to octupolar order. These elements are provided in \textit{Appendix F}. These analytic formulas facilitate the derivation of closed-form equations beyond the complexity of the semi-analytic, summation approach of \eqref{spherical-general-1}.

The vectors $\mathbf{b}^{\rm e}_j$ and $
\mathbf{b}^{\rm m}_j$ in \eqref{spherical-general-1} are the effective electric and magnetic scattering coefficients of each of the particles, respectively. They include the interaction among all the particles in the array and are identical for all the (identical) particles due to symmetry. These effective parameters are calculated via \eqref{t-matrix-def} by replacing the T matrix of the isolated particle, or $\bar{\bar{T}}_0$, with the effective T matrix calculated via the following equation \cite{xu2013scattering}
\begin{equation}\label{T-eff-def}
\bar{\bar{T}}_{\rm eff} = \left[\bar{\bar{I}} - \bar{\bar{T}}_0\left(\lambda\right)\bar{\bar{C}}_s\left(\hat{\mathbf{k}}_\mathrm{inc},\frac{\Lambda_1}{\lambda},\frac{\Lambda_2}{\lambda}\right)\right]^{-1}\bar{\bar{T}}_0\left(\lambda\right),
\end{equation}
where $\bar{\bar{I}}$ is the identity matrix, and $\bar{\bar{C}}_s$ is the \textit{lattice coupling matrix} expressed in spherical coordinates, which is a function of the normalized periodicities $\tilde{\Lambda}$, i.e., the physical periodicity normalized to the wavelength, and the direction of illumination. The coupling matrix elements are infinite summations over lattice points and can be calculated using various summation methods for the translation matrices \cite{mishchenko2002scattering,moroz2001exponentially} ($\rightarrow$ \textit{Supp. Info. I.C}). To solve these tedious summations efficiently, we divide them into summations in the real and Fourier space using Ewald's method, which results in exponentially convergent summations \cite{beutel2021efficient}. Note that no approximation is used here, up to the considered multipolar order, unlike other references that take approximated Green's functions.

 For the specific case of \textbf{rectangular lattices}, like the one depicted in Fig.~\ref{fig1}c, $\mathbf{k}^{\pm}_\mathbf{G}$ of the respective diffraction orders are calculated as \cite{dimitriadis2012surface, antonakakis2014gratings}
\begin{subequations}\label{modes-wavevectors-1}
\begin{gather}
\mathbf{k}^{\pm}_\mathbf{G} =  k_{\mathbf{G},x}\,\mathbf{\hat{x}} + k_{\mathbf{G},y}\,\mathbf{\hat{y}} + k^{\pm}_{\mathbf{G},z}\,\hat{\mathbf{z}},\\
\intertext{with}
k_{\mathbf{G},x} = k_x^{\rm inc} + \frac{2\pi n_1}{\Lambda_1},\quad k_{\mathbf{G},y} = k_y^{\rm inc} + \frac{2\pi n_2}{\Lambda_2},\\
\begin{split}
 &k_{\mathbf{G},z}^{\pm} =  k\cos{\theta} = k_0n\cos{\theta} = \\
 & =\hspace{-0mm}\pm\hspace{-0mm} \sqrt{k^2\hspace{-0mm} - \hspace{-0mm} \left(\hspace{-0mm}k_x^{\rm inc} \hspace{-0mm} + \hspace{-0mm}\frac{2\pi n_1}{\Lambda_1}\hspace{-0mm}\right)^2 \hspace{-0mm}- \hspace{-0mm} \left(k_y^{\rm inc} + \frac{2\pi n_2}{\Lambda_2}\hspace{-0mm}\right)^2},
 \end{split}\\
\intertext{and}
\cos{\theta} = \frac{k_{\mathbf{G},z}^{\pm}}{|\mathbf{k}^{\pm}_{\mathbf{G}}|}, \qquad \phi = {\rm arctan}\left(\frac{k_{\mathbf{G},y}}{k_{\mathbf{G},x}}\right).  
\end{gather}
\end{subequations}
\noindent where $n_1,n_2\in \mathbb{Z}$ are the diffraction orders. Note that the diffraction orders are propagating only if $k_{\mathbf{G},z}^{\pm}\in \mathbb{R}$. A similar procedure can be used to calculate the wavevector $\mathbf{k}^{\pm}_{\mathbf{G}}$ for other types of lattices, e.g., hexagonal, as elaborated in the \textit{Supp. Info. II.} 
\subsubsection{Cartesian basis} 
The formulations mentioned above in spherical coordinates can be translated into Cartesian coordinates. After employing \eqref{pm-from-ab} and applying the transformations to Cartesian coordinates, expressed in \eqref{spherical-general-1}, and after tedious calculations, we arrive at ($\rightarrow$ \textit{Supp. Info. I.A} and \textit{I.B})
\begin{align}
\label{scat-field-total-2D-octupole-cartesian-matrix-1}
\mathbf{E}^{\rm sca}_{{\rm c},\mathbf{G}}&(\mathbf{r}) =  \frac{\mathrm{i}k\sqrt{\pi}e^{\mathrm{i}\mathbf{k}_{\mathbf{G}}^{\pm}\cdot\mathbf{r}}}{2A\,|\cos{\theta}\,|}\,
\bar{\bar{S}}\left(\theta,\phi\right)\left[\hspace{-0mm}\begin{array}{r}
\left(\varepsilon \zeta_1\right)^{-1} \mathbf{p}^{\phantom{a}}\\[0.05cm]
 k \left(\varepsilon \zeta_2\right)^{-1}\mathbf{Q}^{\rm e}\\[0.05cm]
k^2 \left(\varepsilon  \zeta_3\right)^{-1}\mathbf{O}^{\rm e}\\[0.05cm]
\mathrm{i}\eta\left(  \zeta_1\right)^{-1}\mathbf{m}^{\phantom{a}}\\[0.05cm]
\mathrm{i}\eta k \left(  \zeta_2\right)^{-1}\mathbf{Q}^{\rm m}\\[0.05cm]
\mathrm{i}\eta k^2\left(  \zeta_3\right)^{-1} \mathbf{O}^{\rm m}
\end{array}\hspace{-0mm}\right] = \notag \\ &=\frac{\mathrm{i}\sqrt{\pi}e^{\mathrm{i}\mathbf{k}_{\mathbf{G}}^{\pm}\cdot\mathbf{r}}}{2 A k^2 |\cos{\theta}\,|}\,
\bar{\bar{S}}\left(\theta,\phi\right) \, \bar{\bar{\widetilde{\alpha}}}_{\rm eff}  
\left[\hspace{-0mm}\begin{array}{r}
 \zeta_1\mathbf{E}_1 \\[0.05cm]
k^{-1}\zeta_2\mathbf{E}_2\\[0.05cm]
k^{-2}\zeta_3\mathbf{E}_3\\[0.05cm]
\mathrm{i}\eta\zeta_1\mathbf{H}_1\\[0.05cm]
\mathrm{i}\eta k^{-1}\zeta_2 \mathbf{H}_2\\[0.05cm]
\mathrm{i}\eta  k^{-2} \zeta_3\mathbf{H}_3
\end{array}\hspace{-0mm}\right],
\end{align}
\noindent with  $\mathbf{E}^{\rm sca}_{{\rm c},\mathbf{G}} (\mathbf{r}) = \Big[ 
     {E}^{\rm sca}_{\mathbf{G},x}  \,\,\,
     {E}^{\rm sca}_{\mathbf{G},y} \,\,\,
     {E}^{\rm sca}_{\mathbf{G},z}
\Big]^T$. The matrix $\bar{\bar{S}}\left(\theta,\phi\right)$ is the Cartesian multipole-to-field translation matrix, defined as 
\begin{subequations}\label{General-S}
\begin{gather}
\bar{\bar{S}}\hspace{-0mm}\left(\theta,\phi\right)  =
\left[\hspace{-0mm}
\arraycolsep=1mm
\begin{array}{cccccc}
\mathbf{S}_{1}^{\rm xe} & \mathbf{S}_{2}^{\rm xe} & \mathbf{S}_{3}^{\rm xe} & \mathbf{S}_{1}^{\rm xm} & \mathbf{S}_{2}^{\rm xm} & \mathbf{S}_{3}^{\rm xm} \\[0.05cm]
\mathbf{S}_{1}^{\rm ye} & \mathbf{S}_{2}^{\rm ye} & \mathbf{S}_{3}^{\rm xe} & \mathbf{S}_{1}^{\rm ym} & \mathbf{S}_{2}^{\rm ym} & \mathbf{S}_{3}^{\rm ym} \\[0.05cm]
\mathbf{S}_{1}^{\rm ze} & \mathbf{S}_{2}^{\rm ze} & \mathbf{S}_{3}^{\rm ze} & \mathbf{S}_{1}^{\rm zm} & \mathbf{S}_{2}^{\rm zm} & \mathbf{S}_{3}^{\rm zm}  
\end{array}
\right],\\
\intertext{with the vector elements of the matrix}
\begin{split}
&\mathbf{S}_{j}\left(\theta,\phi\right) = \left[\hspace{-0mm}
\arraycolsep=1mm
\begin{array}{cccccc}
\mathbf{S}_{j}^{\rm xe} & \mathbf{S}_{j}^{\rm xm} \\[0.05cm]
\mathbf{S}_{j}^{\rm ye} & \mathbf{S}_{j}^{\rm ym}  \\[0.05cm]
\mathbf{S}_{j}^{\rm ze} & \mathbf{S}_{j}^{\rm zm}  
\end{array}\hspace{-0mm}\right] = \\
&= \frac{\sqrt{2j+1}}{\mathrm{i}^{1-j}}\,\, \left[\bar{\bar{R}}\hspace{-0mm}\left(\theta,\phi\right)^T \bar{\bar{W}}\hspace{-0mm}\left(\theta,\phi\right)\right]\hspace{-0mm} \bar{\bar{F_j}}^{-1}.
\end{split}
\end{gather}
\end{subequations}
The $\bar{\bar{R}}\left(\theta,\phi\right)$ matrix is the transformation operator from spherical to Cartesian coordinates ($\rightarrow$ \textit{Appendix F}). 
The matrix $\bar{\bar{\widetilde{\alpha}}}_{\rm eff}$ in \eqref{scat-field-total-2D-octupole-cartesian-matrix-1} is the normalized effective polarizability matrix. The matrix includes the coupling between particles on the lattice, in the same way as $\bar{\bar{T}}_{\rm eff}$, as explained above. The effective polarizability can either be calculated from the $\bar{\bar{T}}_{\rm eff}$ via applying the transformations of \eqref{T-from-a}, or directly via ($\rightarrow$ \textit{Supp. Info. VIII.})
\begin{equation}\label{a-eff-def}
\bar{\bar{\widetilde{\alpha}}}_{\rm eff} = \left[\bar{\bar{I}} - \bar{\bar{\widetilde{\alpha}}}_0\left(\lambda\right)\bar{\bar{C}}\left(\hat{\mathbf{k}}_\mathrm{inc},\frac{\Lambda_1}{\lambda},\frac{\Lambda_2}{\lambda}\right)\right]^{-1} \bar{\bar{\widetilde{\alpha}}}_0\left(\lambda\right), 
\end{equation}
where $\bar{\bar{\widetilde{\alpha}}}_0$ is the normalized polarizability matrix of the isolated particle and $\bar{\bar{C}}=\mathrm{i}F \bar{\bar{C}}_{\mathrm{s}}F^{-1}$ is the lattice coupling matrix expressed in Cartesian coordinates. The matrix $\bar{\bar{C}}$ can be, alternatively, calculated using various summation methods for dyadic Green's functions, and, hence, some elements of $\bar{\bar{C}}$ have been analytically obtained  \cite{tretyakov2003analytical,babicheva2018metasurfaces,belov2005homogenization}. However, only certain simplified metasurface cases are investigated in these publications, or an approximated Green's function is considered. 

Therefore, with the closed-form formulas \eqref{spherical-general-1} and \eqref{scat-field-total-2D-octupole-cartesian-matrix-1}, introduced in this work, one can effectively calculate the response of a 2D lattice of particles up to octupolar order when illuminated by a plane wave. In particular, \eqref{scat-field-total-2D-octupole-cartesian-matrix-1} that employs Cartesian coordinates, which enjoy popularity in the metasurface community, is a notable contribution \cite{dimitriadis2012surface, tretyakov2015metasurfaces, Alaee2015reciprocal,babicheva2019analytical,mobini2018theory}. However, we want to point out that both representations, spherical or Cartesian, are physically equivalent \cite{alaee2019exact}, and the choice depends on the geometry of the problem or the user comfort.   

The lattice coupling matrix expressed in Cartesian coordinates, $\bar{\bar{C}}$, has the exact dimensions as the polarizability matrix and is defined as
\vspace{5mm}
\begin{equation} 
\bar{\bar{C}}= \left[
\arraycolsep=1mm
\begin{array}{cccccc}
\bar{\bar{C}}_{11}^{\rm ee} & \bar{\bar{C}}_{12}^{\rm ee} & \bar{\bar{C}}_{13}^{\rm ee} & \bar{\bar{C}}_{11}^{\rm em} & \bar{\bar{C}}_{12}^{\rm em} & \bar{\bar{C}}_{13}^{\rm em}  \\[0.05cm]
\bar{\bar{C}}_{21}^{\rm ee} & \bar{\bar{C}}_{22}^{\rm ee} & \bar{\bar{C}}_{23}^{\rm ee} & \bar{\bar{C}}_{21}^{\rm em} & \bar{\bar{C}}_{22}^{\rm em} & \bar{\bar{C}}_{23}^{\rm em}  \\[0.05cm]
\bar{\bar{C}}_{31}^{\rm ee} & \bar{\bar{C}}_{32}^{\rm ee} & \bar{\bar{C}}_{33}^{\rm ee} & \bar{\bar{C}}_{31}^{\rm em} & \bar{\bar{C}}_{32}^{\rm em} & \bar{\bar{C}}_{33}^{\rm em}  \\[0.05cm]
\bar{\bar{C}}_{11}^{\rm me} & \bar{\bar{C}}_{12}^{\rm me} & \bar{\bar{C}}_{13}^{\rm me} & \bar{\bar{C}}_{11}^{\rm mm} & \bar{\bar{C}}_{12}^{\rm mm} & \bar{\bar{C}}_{13}^{\rm mm} \\[0.05cm]
\bar{\bar{C}}_{21}^{\rm me} & \bar{\bar{C}}_{22}^{\rm me} &  \bar{\bar{C}}_{23}^{\rm me} & \bar{\bar{C}}_{21}^{\rm mm} & \bar{\bar{C}}_{22}^{\rm mm} & \bar{\bar{C}}_{23}^{\rm mm} \\[0.05cm]
\bar{\bar{C}}_{31}^{\rm me} & \bar{\bar{C}}_{32}^{\rm me} &  \bar{\bar{C}}_{33}^{\rm me} & \bar{\bar{C}}_{31}^{\rm mm} & \bar{\bar{C}}_{32}^{\rm mm} & \bar{\bar{C}}_{33}^{\rm mm}
\end{array}\right],
\end{equation}
\noindent where due to electromagnetic duality symmetry \cite{Jackson1999,tung1985group,FernandezCorbaton2014a,rahimzadegan2016optical} $\bar{\bar{C}}_{jj'}^\mathrm{ee}=\bar{\bar{C}}_{jj'}^\mathrm{mm}$ and $\bar{\bar{C}}_{jj'}^\mathrm{me}=\bar{\bar{C}}_{jj'}^\mathrm{em}$. 
Generally, based on the 2D lattice symmetry and incidence angle, the coupling matrices take different arrangements. Specifically, at normal incidence, i.e., $\theta_{\rm inc}=0$, the coupling matrix takes a much simpler form. Hence, in Fig.~\ref{FIG:latticeFull}, we show the Cartesian coupling matrix $\bar{\bar{C}}$ for square and hexagonal lattices. The spherical coupling matrix counterpart is shown in Fig.~S3 in the \textit{Supp. Info. IX.A.}. Note that the coupling matrix is a function of the $\hat{\mathbf{k}}_{\rm inc}$. Therefore, for normal incidence, the choice of $\phi_{\rm inc}$ does not make any difference. Throughout the manuscript, normal incidence refer to $\theta_{\rm inc}=0$, unless, explicitly, a constraint on $\phi_{\rm inc}$ is mentioned.

Following the calculation of $\bar{\bar{C}}$ for a specific lattice, the normalized effective polarizability can, then, be obtained via \eqref{a-eff-def}. The Row III of the Fig.~\ref{FIG:latticeFull} shows $\bar{\bar{\widetilde{\alpha}}}_{\rm eff}$ of the spherical, cylindrical, and helical particles inside the square and hexagonal lattices, i.e., including the coupling influence of all particles on the 2D array.
%
%
%
%
\begin{table*}[t]
  \caption{Main equations for the scattering of 2D lattices in spherical and Cartesian coordinates.}
  \centering
    \begin{tabular}{c|c|c}
      \addlinespace[-0ex]
    \midrule
     Spherical$^*$ & Conversion & Cartesian$^*$ \\
         \midrule\midrule 
$\begin{aligned}     
 \left[\begin{array}{c}
E^{\rm sca}_{\mathbf{G},\theta}\\[0.1cm]
E^{\rm sca}_{\mathbf{G},{\phi}}
\end{array} \right]
= \Gamma \sum_{j=1}^{3} \frac{\sqrt{2j+1}}{\mathrm{i}^{j}} \, \bar{\bar{W_j}}\left[\begin{array}{c}
\mathbf{b}^{\rm e}_j\\[0.1cm]
\mathbf{b}^{\rm m}_j
\end{array} \right]
\end{aligned}$ & 
$\begin{aligned}  \mathbf{S}_{j}= \frac{\sqrt{2j+1}}{\mathrm{i}^{1-j}} \left(\bar{\bar{R}}^T \bar{\bar{W}}_j\right) \bar{\bar{F_j}}^{-1} \end{aligned}$    &  
$\begin{aligned}     
 \left[\begin{array}{c}
     {E}^{\rm sca}_{\mathbf{G},x}  \\[0.1cm]
     {E}^{\rm sca}_{\mathbf{G},y} \\[0.1cm]
     {E}^{\rm sca}_{\mathbf{G},z}
\end{array}\right] = \Gamma k^3
\bar{\bar{S}}\left[\begin{array}{r}
\frac{1}{\zeta_1\varepsilon}\mathbf{p}^{\phantom{a}}\\[0.1cm]
\frac{k}{\zeta_2\varepsilon}\mathbf{Q}^{\rm e}\\[0.1cm]
\frac{k^2}{\zeta_3\varepsilon}\mathbf{O}^{\rm e}\\[0.1cm]
\frac{\mathrm{i}\eta}{\zeta_1}\mathbf{m}^{\phantom{a}}\\[0.1cm]
\frac{\mathrm{i}k\eta}{\zeta_2}\mathbf{Q}^{\rm m}\\[0.1cm]
\frac{\mathrm{i} k^2 \eta}{\zeta_3}\mathbf{O}^{\rm m}
\end{array}\right] 
\end{aligned}$ \\
         \midrule
     $\begin{aligned} \left[\begin{array}{c}
\mathbf{b}^{\rm e}_1\\[0.05cm]
\mathbf{b}^{\rm e}_2\\[0.05cm]
\mathbf{b}^{\rm e}_3\\[0.05cm]
\mathbf{b}^{\rm m}_1\\[0.05cm]
\mathbf{b}^{\rm m}_2\\[0.05cm]
\mathbf{b}^{\rm m}_3
\end{array}\right]=
\bar{\bar{T}}
\left[\begin{array}{c}
\mathbf{q}^{\rm e}_1\\[0.05cm]
\mathbf{q}^{\rm e}_2\\[0.05cm]
\mathbf{q}^{\rm e}_3\\[0.05cm]
\mathbf{q}^{\rm m}_1\\[0.05cm]
\mathbf{q}^{\rm m}_2\\[0.05cm]
\mathbf{q}^{\rm m}_3
\end{array}\right]\end{aligned}$   
&  
$\begin{aligned} \left[\begin{array}{r}
\frac{\mathrm{i}{k^3}}{\zeta_1\varepsilon}\mathbf{p}^{\phantom{a}}\\[0.1cm]
\frac{\mathrm{i}{k^4}}{\zeta_2\varepsilon}\mathbf{Q}^{\rm e}\\[0.1cm]
\frac{\mathrm{i}{k^5}}{\zeta_3\varepsilon}\mathbf{O}^{\rm e}\\[0.1cm]
\frac{-k^2\eta}{\zeta_1}\mathbf{m}^{\phantom{a}}\\[0.1cm]
\frac{-k^4\eta}{\zeta_2}\mathbf{Q}^{\rm m}\\[0.1cm]
\frac{-k^5 \eta}{\zeta_3}\mathbf{O}^{\rm m}
\end{array}\right]  = 
\left[\begin{array}{c}
\bar{\bar{F}}_1\mathbf{b}^{\rm e}_1\\[0.1cm]
\bar{\bar{F}}_2\mathbf{b}^{\rm e}_2\\[0.1cm]
\bar{\bar{F}}_3\mathbf{b}^{\rm e}_3\\[0.1cm]
\bar{\bar{F}}_1\mathbf{b}^{\rm m}_1\\[0.1cm]
\bar{\bar{F}}_2\mathbf{b}^{\rm m}_2\\[0.1cm]
\bar{\bar{F}}_3\mathbf{b}^{\rm m}_3
\end{array}\right]\end{aligned}$                           &            
$\begin{aligned} \left[\begin{array}{r}
\frac{1}{\zeta_1\varepsilon}\mathbf{p}^{\phantom{a}}\\[0.1cm]
\frac{k}{\zeta_2\varepsilon}\mathbf{Q}^{\rm e}\\[0.1cm]
\frac{k^2}{\zeta_3\varepsilon}\mathbf{O}^{\rm e}\\[0.1cm]
\frac{\mathrm{i}\eta}{\zeta_1}\mathbf{m}^{\phantom{a}}\\[0.1cm]
\frac{\mathrm{i}k\eta}{\zeta_2}\mathbf{Q}^{\rm m}\\[0.1cm]
\frac{\mathrm{i} k^2 \eta}{\zeta_3}\mathbf{O}^{\rm m}
\end{array}\right] = \frac{\bar{\bar{\widetilde{\alpha}}}}{k^3}\,
\left[\begin{array}{r}
\zeta_1\mathbf{E}_1 \\[0.1cm]
\frac{\zeta_2}{k}\mathbf{E}_2\\[0.1cm]
\frac{\zeta_3}{k^2}\mathbf{E}_3\\[0.1cm]
\mathrm{i}\eta\zeta_1\mathbf{H}_1\\[0.1cm]
\frac{\mathrm{i}\eta \zeta_2}{k} \mathbf{H}_2\\[0.1cm]
\frac{\mathrm{i}\eta \zeta_3}{k^2} \mathbf{H}_3
\end{array}\right]\end{aligned}$ \\
         \midrule
        $\begin{aligned} \bar{\bar{T}}_0=\left[\begin{array}{cc}
\bar{\bar{T}}^{\rm ee} & \bar{\bar{T}}^{\rm em}  \\[0.05cm]
\bar{\bar{T}}^{\rm me} & \bar{\bar{T}}^{\rm mm}  
\end{array}\right]  \end{aligned}$ &  
$\begin{aligned} \bar{\bar{\widetilde{\alpha}}}_{jj'}^{vv'} = -\mathrm{i}\,\bar{\bar{F}}_j\,\bar{\bar{T}}_{jj'}^{vv'}\,\bar{\bar{F}}_{j'}^{-1} \end{aligned}$ &  
$\begin{aligned} \bar{\bar{\widetilde{\alpha}}}_0=\left[\begin{array}{cc}
\bar{\bar{\widetilde{\alpha}}}^{\rm ee} & \bar{\bar{\widetilde{\alpha}}}^{\rm em}  \\[0.05cm]
\bar{\bar{\widetilde{\alpha}}}^{\rm me} & \bar{\bar{\widetilde{\alpha}}}^{\rm mm} 
\end{array}\right] \end{aligned}$ \\
         \midrule
        $\begin{aligned}  \bar{\bar{T}}_\mathrm{eff}=	(\bar{\bar{I}}-\bar{\bar{T}}_0\bar{\bar{C}}_{\mathrm{s}})^{-1}\bar{\bar{T}}_0 \end{aligned}$ & 
        $\begin{aligned} \bar{\bar{C}}=\mathrm{i}F \bar{\bar{C}}_{\mathrm{s}}F^{-1} \end{aligned}$ & 
        $\begin{aligned} \bar{\bar{\widetilde{\alpha}}}_{\rm eff} = \left(\bar{\bar{I}} - \bar{\bar{\widetilde{\alpha}}}_0\bar{\bar{C}}\right)^{-1} \bar{\bar{\widetilde{\alpha}}}_0 \end{aligned}$ \\
          \midrule\midrule
    \end{tabular}
    \begin{center}
  \begin{tablenotes}
  \centering
\item[*] 
$^*\quad\Gamma=\frac{\mathrm{i}\sqrt{\pi}\,e^{\mathrm{i}\mathbf{k}_{\mathbf{G}}^{\pm}\cdot\mathbf{r}}}{2A k^2 |\cos{\theta}|}$,\, $A=\Lambda_1\Lambda_2$,\, $\mathbf{k}^{\pm}_{\mathbf{G}} =  \mathbf{k}_{\parallel} + \mathbf{G} \pm \hat{\mathbf{z}} \sqrt{k^2 - |\mathbf{k}_{\parallel} + \mathbf{G}|^2 }$, \, $\mathbf{G} = \frac{2\pi n_1}{\Lambda_1}\hat{\mathbf{x}} + \frac{2\pi n_2}{\Lambda_1}\hat{\mathbf{y}},\,  \zeta_j = \sqrt{(2j+1)!\,\pi}.$
\end{tablenotes}
\end{center}
  \end{table*}
%

Let us now explore a commonly considered case for metasurfaces, the \textbf{square lattice}, with $\Lambda_1=\Lambda_2=\Lambda$ as the periodicity. If we calculate the coupling matrix for this case and, afterward, the effective polarizability via \eqref{a-eff-def}, the scattered field in \eqref{scat-field-total-2D-octupole-cartesian-matrix-1} is further simplified to
\begin{widetext}
\begin{equation}\label{scat-field-total-2D-octupole-cartesian-matrix-argument}
   \mathbf{E}^{\rm sca}_{\mathbf{G},{\rm c}} \left(\mathbf{r},\hat{\mathbf{k}}_\mathrm{inc},\Lambda,\lambda\right) = \frac{\mathrm{i}e^{\mathrm{i}\mathbf{k}_{\mathbf{G}}^{\pm}\cdot\mathbf{r}}}{8 \pi^{3/2}  |\cos{\theta}| \widetilde{\Lambda}^2}\,\bar{\bar{S}}\left(\theta,\phi\right)
 \frac{\bar{\bar{\widetilde{\alpha}}}_0\left(\lambda\right)}{\bar{\bar{I}} - \bar{\bar{\widetilde{\alpha}}}_0\left(\lambda\right)\bar{\bar{C}}\left(\hat{\mathbf{k}}_\mathrm{inc},\widetilde{\Lambda}\right)} 
\left[\hspace{-0mm}\begin{array}{r}
 \zeta_1\mathbf{E}_1\left(\hat{\mathbf{k}}_\mathrm{inc},\lambda\right) \\[0.05cm]
k^{-1}\zeta_2\mathbf{E}_2\left(\hat{\mathbf{k}}_\mathrm{inc},\lambda\right)\\[0.05cm]
k^{-2}\zeta_3\mathbf{E}_3\left(\hat{\mathbf{k}}_\mathrm{inc},\lambda\right)\\[0.05cm]
\mathrm{i}\eta\zeta_1\mathbf{H}_1\left(\hat{\mathbf{k}}_\mathrm{inc},\lambda\right)\\[0.05cm]
\mathrm{i} \eta k^{-1} \zeta_2 \mathbf{H}_2\left(\hat{\mathbf{k}}_\mathrm{inc},\lambda\right)\\[0.05cm]
\mathrm{i}\eta k^{-2} \zeta_3\mathbf{H}_3\left(\hat{\mathbf{k}}_\mathrm{inc},\lambda\right)
\end{array}\hspace{-0mm}\right].
\end{equation}
\end{widetext}
\noindent Here, we have included all the arguments in the equation for clarity. The factors that control the response of the particle square array are evident from \eqref{scat-field-total-2D-octupole-cartesian-matrix-argument}; specifically, the surrounding material, represented by $k=2\pi n/\lambda_0$, the incident field direction, represented by the field vector and $\hat{\mathbf{k}}_\mathrm{inc}$, the properties of the individual particle, represented by $\bar{\bar{\widetilde{\alpha}}}_0$, and the dimensions of the lattice, represented by $\widetilde{\Lambda}$. Hence, modifying each of these parameters could change the response of the metasurface according to one's goals. As inserting the arguments in \eqref{scat-field-total-2D-octupole-cartesian-matrix-argument},  can make the equations lengthy and might distract the reader, we ignore them from now on unless needed. However, implicitly they are always assumed. 

The summations \eqref{spherical-general-1} and \eqref{scat-field-total-2D-octupole-cartesian-matrix-1} can describe the diffracted field from any 2D array in spherical and Cartesian coordinates, respectively, in terms of the incident fields, the lattice, and the consisting particles attributes, provided that the problem can be sufficiently described with octupoles as the maximum multipolar order. Moreover, unlike previous efforts \cite{dimitriadis2012surface, babicheva2018metasurfaces}, these formulas include all the propagating, diffraction orders and not only a zeroth-order. It allows the study of metagratings operating at wavelengths shorter than the array periodicity and further facilitates the design of related optical structures, as demonstrated later.
For convenience, we provide a summary of these equations in Table I. The verification of the proposed equations is performed in the \textit{Supp. Info. I.D.} using COMSOL Multiphysics\textsuperscript{\textregistered} \cite{comsol} for an isotropic particle, as well as a particle with broken symmetries, namely a metallic helix, and under oblique incidence. The T matrix of the metallic helix was obtained via an extraction algorithm \cite{demesy2018scattering} using JCMsuite \cite{burger2008jcmsuite}.

It should be noted that the form of the matrices and their subsequent symmetries are essentially dependent on the combinations we choose as an irreducible representation of the Cartesian multipole moments and, subsequently, the selection of the transformation matrices $\bar{\bar{F}}_j, \,\, j=\{1,2,3\}$ ($\rightarrow$ \textit{Appendix E}). The utilization of the specific transformation matrices, in this work, originates from real spherical harmonic corresponding to atomic orbitals, $p, d$, and $f$ \cite{chisholm1976group}, and enable the conservation of certain symmetries between Cartesian and spherical bases, such as diagonality for T matrices for isotropic particles. Different setups of multipole moments in Cartesian basis or a different choice of transformation matrices will lead to different matrices in Cartesian basis from those demonstrated in Fig.~\ref{FIG:latticeFull}, as shown in Ref.~\citenum{mun2020describing}. Additionally, the polarizability matrix via a different Cartesian basis can be retrieved from the current one through simple algebraic transformations ($\Rightarrow\textit{Supp. Info. IV.}$). 

Let us, now, assume specific symmetries about the geometry of the particles in use or a particular wave incidence onto the metasurface. In that case, the closed-form equations provided in this section can be further simplified to accessible analytical formulas that could greatly assist the metasurface design process. This approach will be demonstrated in the following sections. The next section is dedicated to normal incidence, while the next but one section is dedicated to oblique incidence.
\begin{figure*}[t]
    \centering
    \includegraphics[width=130mm,trim={0 0.25cm 0 0.1cm},clip]{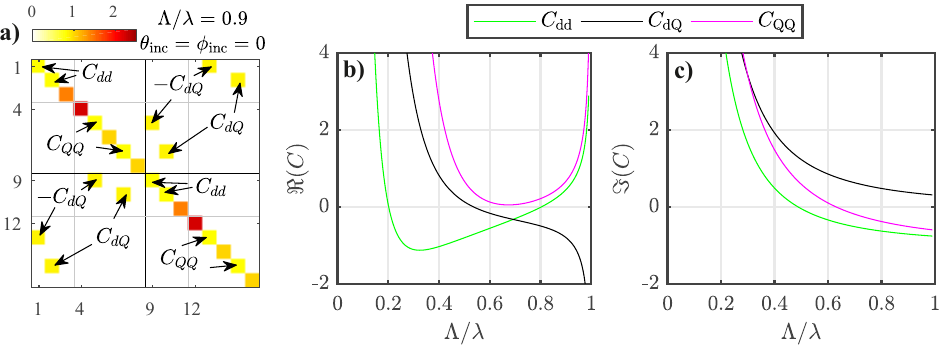}
    \caption{\textbf{The Cartesian coupling matrix:} a) The Cartesian coupling matrix amplitude for a square array with a normalized periodicity of $\Lambda/\lambda=0.9$ under normal incidence ($\theta_{\rm inc}=\phi_{\rm inc}=0$) up to quadrupolar order. The relevant matrix elements for a metasurface made from isotropic particles are dipole-dipole $C_{\rm dd}$, quadrupole-quadrupole $C_{\rm QQ}$, and dipole-quadrupole $C_{\rm dQ}$ couplings. They are marked with an arrow. b) The real and c) imaginary part of the relevant lattice couplings as a function of the normalized periodicity $\Lambda/\lambda$. For a square array, the electric and magnetic parts are equal.}
    \label{FIG:lattice}
    \vspace{-2mm}
\end{figure*}
\section{Analytic equations: Normal incidence}
\subsection{Propagating diffraction orders of dipole-quadrupole metasurfaces made from isotropic meta-atoms}
In this subsection, originating from \eqref{scat-field-total-2D-octupole-cartesian-matrix-argument}, we derive a general, simplified, closed-form, analytic expression for the amplitudes of the propagating diffraction orders from dipole-quadrupole metasurfaces and metagratings made from isotropic meta-atoms and illuminated at normal incidence. An isotropic particle has spherical symmetry and includes homogeneous/core-multishell spheres and isotropic colloidal particles \cite{dezert2017isotropic}. Unlike previous modeling efforts, our analytic procedure captures the amplitudes in reflection and transmission of the zeroth diffraction orders, as well as higher propagating diffraction ones. 

For isolated isotropic particles, as shown in Fig.~\ref{FIG:latticeFull}f, the polarizability matrix is diagonal. Thus, the elements of the matrix $\bar{\bar{\widetilde{\alpha}}}_0$  can be written via \eqref{a-matrix-def} as
\begin{subequations}\label{a-matrix-defined_diag} 
\begin{align}
\bar{\bar{\widetilde{\alpha}}}_{11}^{\rm ee} = \widetilde{\alpha}_{\rm p}\, \bar{\bar{I}} &,&
\bar{\bar{\widetilde{\alpha}}}_{11}^{\rm mm} = \widetilde{\alpha}_{\rm m}\, \bar{\bar{I}}, \\
\bar{\bar{\widetilde{\alpha}}}_{22}^{\rm ee} = \widetilde{\alpha}_{\rm Q^{\rm e}}\, \bar{\bar{I}} &,&
\bar{\bar{\widetilde{\alpha}}}_{22}^{\rm mm} = \widetilde{\alpha}_{\rm Q^{\rm m}}\, \bar{\bar{I}},
\end{align}
\end{subequations}
\noindent where $\widetilde{\alpha}_{\rm p}$ ($\widetilde{\alpha}_{\rm m}$), and $\widetilde{\alpha}_{\rm Q^e}$ ($\widetilde{\alpha}_{\rm Q^m}$) are the normalized electric (magnetic) dipole and quadrupole polarizabilities, respectively. The dimensions of the unitary matrix, $\bar{\bar{I}}$, changes according to the multipolar order, i.e. $3\times3$ for dipole and $5\times5$ for quadrupole.
These polarizabilities can be linked to the Mie coefficients via \eqref{T-from-a}, as \cite{mun2020describing,rahimzadegan2020beyond}
\begin{subequations}\label{a-vs-Mie} 
\begin{align} 
    \widetilde{\alpha}_{\rm p}= \mathrm{i} a_1,&\quad \widetilde{\alpha}_{\rm m}= \mathrm{i} b_1,\\
    \widetilde{\alpha}_{\rm Q^e}= \mathrm{i} a_2,&\quad\widetilde{\alpha}_{\rm Q^m}= \mathrm{i}b_2.
\end{align}
\end{subequations}
\noindent where $a_j$ ($b_j$) is the electric (magnetic) Mie coefficient of $j$'s order. When inside the lattice, these polarizabilities will be modulated and coupled to each other according to  \eqref{a-eff-def}. 
\settowidth{\rotheadsize}{\bfseries Long long long \quad}
\begin{table*}
  \caption{Reflection and transmission coefficients for metasurfaces under \textbf{A)} normal and \textbf{B)} oblique incidence. \textbf{A.1 \& B.1)} Effective Mie coefficients (coupled and modulated). \textbf{A.2 \& B.2)} Modulated Mie coefficients. \textbf{A.3)} General equations. }
  \centering 
  \begin{threeparttable}
    \begin{tabular}{m{1em} c}
\addlinespace[2ex]
\multicolumn{2}{c}{\textbf{A) Normal Incidence$^*$}: $\theta_{\rm inc}=\phi_{\rm inc}=0$}\\
\midrule
\multicolumn{2}{c}{$\begin{aligned}
 \left[\begin{array}{cc}
\{t,r\}_{\textnormal{TE}_{0,0} \rightarrow  \textnormal{TE}_{n_1,n_2}} & \{t,r\}_{\textnormal{TE}_{0,0} \rightarrow  \textnormal{TM}_{n_1,n_2}} \\[0.5cm]
\{t,r\}_{\textnormal{TM}_{0,0} \rightarrow  \textnormal{TE}_{n_1,n_2}} & \{t,r\}_{\textnormal{TM}_{0,0} \rightarrow  \textnormal{TM}_{n_1,n_2}}
\end{array}\right] = \frac{1\pm1}{2} \left[\begin{array}{cc}
\delta_{n_1n_20} & 0 \\[0.5cm]
0 &\delta_{n_1n_20}
\end{array}\right] - \frac{\pm\cos{\phi}}{4\pi\widetilde{\Lambda}^2}\times  \quad  \quad    \quad   \quad  \quad  \\\\ \left[\begin{array}{cc}
\left(\frac{ 3a_{\mathrm{1,eff}}}{\cos{\theta}} + 3b_{\mathrm{1,eff}} + 5a_{\mathrm{2,eff}}+\frac{ 5\cos{2\theta}b_{\mathrm{2,eff}}}{\cos{\theta}}\right) &  \left(3a_{\mathrm{1,eff}}+\frac{3b_{\mathrm{1,eff}}}{\cos{\theta}}+\frac{5 \cos{2\theta}a_{\mathrm{2,eff}}}{\cos{\theta}}+5b_{\mathrm{2,eff}}\right)\tan{\phi} \\[0.5cm]
-\left(\frac{3a_{\mathrm{1,eff}}}{\cos{\theta}}+3b_{\mathrm{1,eff}}+5a_{\mathrm{2,eff}}+\frac{5\cos{2\theta} b_{\mathrm{2,eff}}}{\cos{\theta}}\right)\tan{\phi} & \left(3a_{\mathrm{1,eff}}+\frac{3b_{\mathrm{1,eff}}}{\cos{\theta}}+\frac{5 \cos{2\theta}a_{\mathrm{2,eff}}}{\cos{\theta}}+5b_{\mathrm{2,eff}}\right)
\end{array}\right],   
\end{aligned}$} \\
 \cmidrule(l  r ){1-2}
   \multirow{1}{*}[75pt]{\rothead{\textbf{A.1}}}
  & $\begin{aligned}
 \frac{1}{a_{\mathrm{1,eff}}}=\frac{1+C_{\rm dQ}^{2}\text{\ensuremath{b_{\mathrm{2,mod}}}}a_{\mathrm{1,mod}}}{a_{\mathrm{1,mod}}\left(1+\mathrm{i}\sqrt{5/3}C_{\rm dQ}\text{\ensuremath{b_{\mathrm{2,mod}}}}\right)},\quad \frac{1}{b_{\mathrm{1,eff}}}=\frac{1+C_{\rm dQ}^{2}\text{\ensuremath{a_{\mathrm{2,mod}}}}b_{\mathrm{1,mod}}}{b_{\mathrm{1,mod}}\left(1+\mathrm{i}\sqrt{5/3}C_{\rm dQ}\text{\ensuremath{a_{\mathrm{2,mod}}}}\right)}, &
\\\frac{1}{a_{\mathrm{2,eff}}}=\frac{1+C_{\rm dQ}^{2}\text{\ensuremath{b_{\mathrm{1,mod}}}}a_{\mathrm{2,mod}}}{a_{\mathrm{2,mod}}\left(1+\mathrm{i}\sqrt{3/5}C_{\rm dQ}\text{\ensuremath{b_{\mathrm{1,mod}}}}\right)},\quad\frac{1}{b_{\mathrm{2,eff}}}=\frac{1+C_{\rm dQ}^{2}\text{\ensuremath{a_{\mathrm{1,mod}}}}b_{\mathrm{2,mod}}}{b_{\mathrm{2,mod}}\left(1+\mathrm{i}\sqrt{3/5}C_{\rm dQ}\text{\ensuremath{a_{\mathrm{1,mod}}}}\right)},
\end{aligned}
$   \\
  \cmidrule(l  r ){1-2}
  \multirow[c]{1}{*}[80pt]{\rothead{\textbf{A.2}}} 
  & $\begin{aligned}
 \frac{1}{a_{\mathrm{1,mod}}}=\frac{1}{a_{1}}-\mathrm{i}C_{\rm dd},\,\frac{1}{b_{\mathrm{1,mod}}}=\frac{1}{b_{1}}-\mathrm{i}C_{\rm dd}\, \frac{1}{a_{\mathrm{2,mod}}}=\frac{1}{a_{2}}-\mathrm{i}C_{\rm QQ},\,\frac{1}{b_{\mathrm{2,mod}}}=\frac{1}{b_{2}}-\mathrm{i}C_{\rm QQ},
\end{aligned}
$              \\
  \cmidrule(l r ){1-2}
    \multirow[c]{1}{*}[80pt]{\rothead{\textbf{A.3}}} 
  &  $\begin{aligned}
\theta &=&\arccos{[\pm\sqrt{1-\frac{1}{\widetilde{\Lambda}^2}(n_1^2+n_2^2)}]},\qquad 
\phi &=& \arctan{(\frac{n_2}{n_1})} ,\qquad 
\widetilde{\Lambda} &=& \frac{\Lambda}{\lambda},
\end{aligned}
$   \\   
\midrule
\midrule
\\
\multicolumn{2}{c}{\textbf{B) Oblique Incidence}: $\phi_{\rm inc}=0$}\\
\midrule
\multicolumn{2}{c}{$\begin{aligned}
 \left[\begin{array}{c}
t_{\textnormal{TE}_{0,0} \rightarrow  \textnormal{TE}_{0,0}}\\[0.25cm]
t_{\textnormal{TM}_{0,0} \rightarrow  \textnormal{TM}_{0,0}}
\end{array}\right] =  \left[\begin{array}{c}
1  \\[0.25cm]
1
\end{array}\right] - \frac{3}{4\pi\widetilde{\Lambda}^2|\cos{\theta_\mathrm{inc}}|} \left[\begin{array}{c}
b_{\mathrm{1,mod.xx}}\cos\theta_\mathrm{inc}^2+a_{\mathrm{1,eff.yy}}+b_{\mathrm{1,eff.zz}}\sin\theta_\mathrm{inc}^2 \\[0.5cm]
a_{\mathrm{1,mod.xx}}\cos\theta_\mathrm{inc}^2+b_{\mathrm{1,eff.yy}}+a_{\mathrm{1,eff.zz}}\sin\theta_\mathrm{inc}^2
\end{array}\right],   
\end{aligned}
$}             \\ \\ 
\multicolumn{2}{c}{ $\begin{aligned}
 \left[\begin{array}{c}
r_{\textnormal{TE}_{0,0} \rightarrow  \textnormal{TE}_{0,0}}\\[0.25cm]
r_{\textnormal{TM}_{0,0} \rightarrow  \textnormal{TM}_{0,0}}
\end{array}\right] =  \frac{3}{4\pi\widetilde{\Lambda}^2|\cos{\theta_\mathrm{inc}}|}\left[\begin{array}{c}
b_{\mathrm{1,mod.xx}}\cos\theta_\mathrm{inc}^2-a_{\mathrm{1,eff.yy}}-b_{\mathrm{1,eff.zz}}\sin\theta_\mathrm{inc}^2 \\[0.5cm]
a_{\mathrm{1,mod.xx}}\cos\theta_\mathrm{inc}^2-b_{\mathrm{1,eff.yy}}-a_{\mathrm{1,eff.zz}}\sin\theta_\mathrm{inc}^2
\end{array}\right],
\end{aligned} 
$}    \\
    \cmidrule(l r ){1-2}
  \multirow{1}{*}[85pt]{\rothead{\textbf{B.1}}} 
&  $\begin{aligned}
\frac{1}{a_{\mathrm{1,eff.yy}}}=\frac{1-C_{\rm yz}^{2}\text{\ensuremath{b_{\mathrm{1,mod.zz}}}}a_{\mathrm{1,mod.yy}}}{a_{\mathrm{1,mod.yy}}\left(1-C_{\rm yz}\text{\ensuremath{b_{\mathrm{1,mod.zz}}}}\sin\theta_\mathrm{inc}\right)}
,&\,
\frac{1}{b_{\mathrm{1,eff.yy}}}=\frac{1-C_{\rm yz}^{2}\text{\ensuremath{b_{\mathrm{1,mod.yy}}}}a_{\mathrm{1,effzz}}}{b_{\mathrm{1,mod.yy}}\left(1-C_{\rm yz}\text{\ensuremath{a_{\mathrm{1,mod.zz}}}}\sin\theta_\mathrm{inc}\right)},
\\
\frac{1}{a_{\mathrm{1,eff.zz}}}=\frac{1-C_{\rm yz}^{2}\text{\ensuremath{b_{\mathrm{1,mod.yy}}}}a_{\mathrm{1,mod.zz}}}{a_{\mathrm{1,mod.zz}}\left(1-C_{\rm yz}\text{\ensuremath{b_{\mathrm{1,mod.yy}}}}\csc\theta_\mathrm{inc}\right)}
,& \,
\frac{1}{b_{\mathrm{1,eff.zz}}}=\frac{1-C_{\rm yz}^{2}\text{\ensuremath{b_{\mathrm{1,mod.zz}}}}a_{\mathrm{1,mod.yy}}}{b_{\mathrm{1,mod.zz}}\left(1-C_{\rm yz}\text{\ensuremath{a_{\mathrm{1,mod.yy}}}}\csc\theta_\mathrm{inc}\right)},
\end{aligned} 
$               \\
  \cmidrule(l r ){1-2}
  \multirow[c]{2}{*}[80pt]{\rothead{\textbf{B.2}}} 
&  $\begin{aligned}
\frac{1}{b_{\mathrm{1,mod.xx}}}=\frac{1}{b_{1}}-\mathrm{i}C_{\rm xx},\quad
\frac{1}{b_{\mathrm{1,mod.zz}}}=\frac{1}{b_{1}}-\mathrm{i}C_{\rm zz},\quad
\frac{1}{b_{\mathrm{1,mod.yy}}}=\frac{1}{b_{1}}-\mathrm{i}C_{\rm yy},\\
\frac{1}{a_{\mathrm{1,mod.xx}}}=\frac{1}{a_{1}}-\mathrm{i}C_{\rm xx},\quad
\frac{1}{a_{\mathrm{1,mod.yy}}}=\frac{1}{a_{1}}-\mathrm{i}C_{\rm yy},\quad
\frac{1}{a_{\mathrm{1,mod.zz}}}=\frac{1}{a_{1}}-\mathrm{i}C_{\rm zz}.
\end{aligned} 
$               \\
    \midrule\midrule
    \end{tabular}
 \begin{tablenotes}
\item[*] Note that $\theta$ and $\phi$ are the angles of the scattering field  (i.e., for the reflection $\cos{\theta}<0$). In \textbf{A)} for '$\pm$' use '$+$' for transmission and '$-$' for reflection. In \textbf{B)} $k^{\mathrm{inc}}_y=0$.
\end{tablenotes}
\end{threeparttable}
\vspace{0mm}
  \end{table*}

For normal incidence, or $\theta_{\rm inc}=0$, the incident TE and TM polarizations can be defined as
\begin{subequations}
\begin{align}
\mathbf{E}^{\rm inc}_{\rm TM}&=\left[
\begin{array}{c}E_x\\ E_y\\ E_z\end{array}\right] = E_0\left[\begin{array}{c}1\\ 0\\ 0\end{array}\right]\mathrm{e}^{\mathrm{ik}z},\\
\mathbf{E}^{\rm inc}_{\rm TE}&=\left[
\begin{array}{c}E_x\\ E_y\\ E_z\end{array}\right]= E_0\left[\begin{array}{c}0\\ 1\\ 0\end{array}\right]\mathrm{e}^{\mathrm{ik}z}.
\end{align}
\end{subequations}
In this work, TE/TM polarizations are equivalent to s-/p- polarizations, where the E-field vector direction is the same as the unit-vectors of the spherical coordinates' system, $\hat{\bm{\phi}}$ and $\hat{\bm{\theta}}$, respectively ($\rightarrow$ \textit{Supp. Info. I.A} and \textit{I.B}).

Let us now consider the calculation of the lattice coupling matrix. For a square array and under normal incidence, $\bar{\bar{C}}$ has a simpler symmetry with many symmetry-protected zeros. 
Additionally, for isotropic particles with elements described in \eqref{a-matrix-defined_diag}, not all coupling coefficients enter \eqref{scat-field-total-2D-octupole-cartesian-matrix-argument}, further simplifying calculations. In Fig.~\ref{FIG:lattice}a, we show the Cartesian coupling matrix under normal incidence for an exemplary normalized periodicity of $\widetilde{\Lambda}=0.9$ up to quadrupolar order and the relevant elements of $\bar{\bar{C}}$ for isotropic constituents are marked. In Figs.~\ref{FIG:lattice}b and \ref{FIG:lattice}c, the real and imaginary parts of the relevant coupling coefficients are shown versus the normalized lattice periodicity. The relevant lattice coefficients for the specific normal incidence and square lattice calculations are dipole-dipole $C_{\rm dd}$, quadrupole-quadrupole $C_{\rm QQ}$, and dipole-quadrupole $C_{\rm dQ}$ couplings. As their name suggests, they are coefficients for the coupling of multipoles of a different order. The imaginary parts of these coefficients can be analytically calculated using energy conservation relations. We have calculated, herein, the imaginary part of these coefficients for sub-wavelength metasurfaces using the analytic equations for transmission and reflection ($\rightarrow$ \textit{Supp. Info. VII.}) as
\begin{subequations}\label{imagCdd}
\begin{align}
\Im{(C_{\rm dd})} &= \frac{3}{4\pi\widetilde{\Lambda}^2}-1,\\
\Im{(C_{\rm QQ})} &= \frac{5}{4\pi\widetilde{\Lambda}^2}-1,\\
\Im{(C_{\rm dQ})} &= \frac{\sqrt{15}}{4\pi\widetilde{\Lambda}^2}.
\end{align}
\end{subequations}

The imaginary parts above can be used to determine fundamental limits. The imaginary part of the dipole-dipole lattice coupling has already been identified in the literature \cite{tretyakov2003analytical}. The real part of the coupling coefficients (Fig.~\ref{FIG:lattice}b) cannot be analytically derived, and infinite summations, as discussed before, are required for accurate calculations. The real part of the coupling coefficients can be linked to the detuning of the response of the isolated meta-atoms \cite{bettles2016enhanced,shahmoon2017cooperative}. For the dipole-dipole coupling, $\Re{(C_{\rm dd})}$ crosses zero for two normalized periodicities $\widetilde{\Lambda}\approx0.2,0.8$ ($\widetilde{\Lambda}\approx0.21,0.88$ for a Hexagonal lattice). As we will show later, this "magic lattice spacing" \cite{bettles2016enhanced} is where the "cooperative resonance" \cite{shahmoon2017cooperative} of the meta-atoms reflects all the incident light. At this point of operation, the resonance of the isolated meta-atom experiences no detuning, and hence the effective resonance also occurs at the same spacing. Needless to say, that changing the meta-atom changes the point of the collective resonance. For the quadrupole-quadrupole coupling, $\Re{(C_{\rm QQ})}$ does not cross the zero point. It is worth mentioning that for a normalized periodicity of 1, i.e., at the onset of a new diffraction order, the real part of the coupling constants diverges to infinity. This finding is important to consider when performing numerical calculations.

In the next step, we calculate the effective multipole moments induced in each particle in the lattice as a function of the effective Mie coefficients up to quadrupolar order for both TM and TE polarized excitation. For TM incidence, they are analytically expressed as
\begin{subequations}\label{inducedMoments}
\vspace{-0.25mm}
\begin{gather} 
\left[\begin{array}{r}
 p_x/\varepsilon\\[0.05cm]
 Q_{\rm xz}^{\rm e}/\varepsilon\\[0.05cm]
\eta m_y\\[0.05cm]
\eta Q_{\rm yz}^{\rm m}
\end{array}\right] =
\left[\begin{array}{r}
\frac{6\pi\mathrm{i}}{k^3} a_{\mathrm{1,mod.}}^{\CircArrowLeft{{\scaleto{b_2}{4.5pt}}}}\\[0.05cm]
\frac{-60\pi}{k^4} a_{\mathrm{2,mod.}}^{\CircArrowLeft{{\scaleto{b_1}{4.5pt}}}}\\[0.05cm]
\frac{6\pi\mathrm{i}}{k^3} b_{\mathrm{1,mod.}}^{\CircArrowLeft{{\scaleto{a_2}{4.5pt}}}}\\[0.05cm]
\frac{-60\pi}{k^4} b_{\mathrm{2,mod.}}^{\CircArrowLeft{{\scaleto{a_1}{4.5pt}}}}
\end{array}\right] E_0,\label{inducedMomentsTM_N} \\
\intertext{and for a TE polarized incidence,} 
\left[\begin{array}{r}
 p_y/\varepsilon\\[0.05cm]
 Q_{\rm yz}^{\rm e}/\varepsilon\\[0.05cm]
\eta m_x\\[0.05cm]
\eta Q_{\rm xz}^{\rm m}
\end{array}\right] =
\left[\begin{array}{r}
\frac{6\pi\mathrm{i}}{k^3} a_{\mathrm{1,mod.}}^{\CircArrowLeft{{\scaleto{b_2}{4.5pt}}}}\\[0.05cm]
\frac{-60\pi}{k^4} a_{\mathrm{2,mod.}}^{\CircArrowLeft{{\scaleto{b_1}{4.5pt}}}}\\[0.05cm]
\frac{-6\pi\mathrm{i}}{k^3} b_{\mathrm{1,mod.}}^{\CircArrowLeft{{\scaleto{a_2}{4.5pt}}}}\\[0.05cm]
\frac{60\pi}{k^4} b_{\mathrm{2,mod.}}^{\CircArrowLeft{{\scaleto{a_1}{4.5pt}}}}
\end{array}\right] E_0.\label{inducedMomentsTE_N}
\end{gather}
\end{subequations}
The effective Mie coefficients/polarizabilities of the particles in these expressions above depend on (a) the modulation of the elements of the same multipolar order through $C_{\rm dd}$ and $C_{\rm QQ}$  and (b) the coupling with other multipole moments in the lattice through $C_{\rm dQ}$. The $\CircArrowLeft{{\scaleto{b_2}{4.5pt}}}$ superscript shows the coupled parameters, and the coupling term is written in the circle. Electric (magnetic) dipole moments are coupled to magnetic (electric) quadrupole moments and vice versa. The modulated and coupled Mie coefficients, that are commonly called the \textit{effective Mie coefficients}, can be written as \vspace{5mm}
\begin{subequations}\label{couplingMie}
\begin{align}
\frac{1}{a_{\mathrm{1,eff}}}\hspace{-0.65mm}&=\hspace{-0.65mm}\frac{1}{a_{\mathrm{1,mod.}}^{\CircArrowLeft{{\scaleto{b_2}{4.5pt}}}}}\hspace{-0.65mm} 
=\hspace{-0.65mm}\frac{1+C_{\rm dQ}^{2}\text{\ensuremath{b_{\mathrm{2,mod.}}}}a_{\mathrm{1,mod}}}{a_{\mathrm{1,mod.}}\hspace{-0.8mm}\left(1+\mathrm{i}\sqrt{5/3}C_{\rm dQ}\text{\ensuremath{b_{\mathrm{2,mod.}}}}\right)},\\
\frac{1}{b_{\mathrm{1,eff}}}\hspace{-0.65mm}&=\hspace{-0.65mm}\frac{1}{b_{\mathrm{1,mod.}}^{\CircArrowLeft{{\scaleto{a_2}{4.5pt}}}}}\hspace{-0.65mm}
= \hspace{-0.65mm} \frac{1+C_{\rm dQ}^{2}\text{\ensuremath{a_{\mathrm{2,mod.}}}}b_{\mathrm{1,mod.}}}{b_{\mathrm{1,mod.}}\hspace{-0.8mm}\left(1+\mathrm{i}\sqrt{5/3}C_{\rm dQ}\text{\ensuremath{a_{\mathrm{2,mod}}}}\right)},\\
\frac{1}{a_{\mathrm{2,eff}}}\hspace{-0.65mm}&=\hspace{-0.65mm}\frac{1}{a_{\mathrm{2,mod.}}^{\CircArrowLeft{{\scaleto{b_1}{4.5pt}}}}} 
\hspace{-0.65mm} = \hspace{-0.65mm}\frac{1+C_{\rm dQ}^{2}\text{\ensuremath{b_{\mathrm{1,mod.}}}}a_{\mathrm{2,mod.}}}{a_{\mathrm{2,mod.}}\hspace{-0.8mm}\left(1+\mathrm{i}\sqrt{3/5}C_{\rm dQ}\text{\ensuremath{b_{\mathrm{1,mod.}}}}\right)},\\
\frac{1}{b_{\mathrm{2,eff}}}\hspace{-0.65mm}&=\hspace{-0.65mm}\frac{1}{b_{\mathrm{2,mod.}}^{\CircArrowLeft{{\scaleto{a_1}{4.5pt}}}}}
\hspace{-0.65mm} = \hspace{-0.65mm}\frac{1+C_{\rm dQ}^{2}\text{\ensuremath{a_{\mathrm{1,mod.}}}}b_{\mathrm{2,mod.}}}{b_{\mathrm{2,mod.}}\hspace{-0.8mm}\left(1+\mathrm{i}\sqrt{3/5}C_{\rm dQ}\text{\ensuremath{a_{\mathrm{1,mod.}}}}\right)}.
\end{align}
\end{subequations}
The modulated Mie coefficients in \eqref{couplingMie} are explicitly written as
\begin{subequations}\label{effectiveMie-1}
\begin{align}
\frac{1}{a_{\mathrm{1,mod.}}}=\frac{1}{a_{1}}-\mathrm{i}C_{\rm dd} &,&
\frac{1}{b_{\mathrm{1,mod.}}}=\frac{1}{b_{1}}-\mathrm{i}C_{\rm dd},\\
\frac{1}{a_{\mathrm{2,mod.}}}=\frac{1}{a_{2}}-\mathrm{i}C_{\rm QQ} &,&
\frac{1}{b_{\mathrm{2,mod.}}}=\frac{1}{b_{2}}-\mathrm{i}C_{\rm QQ}.
\end{align}
\end{subequations}
Writing the above expression in such a manner has the immediate advantage of distinguishing the effect of the lattice and the single meta-atom response on the effective response. It is worth mentioning that despite a different approach, the above equations are very similar to the equations in Ref.~\cite{babicheva2019analytical}. In Ref.~\cite{rahimzadegan2021colossal}, we have exploited the effective moments in Eqs.~\ref{inducedMomentsTM_N}-\ref{inducedMomentsTE_N} to find operation regimes in which the magnetic dipole moment can be colossally enhanced.

By calculating the relevant lattice coupling matrix elements and polarizabilities of the isolated, isotropic particle and by deriving the effective Mie coefficients with \eqref{couplingMie}, the general equation for square lattices  \eqref{scat-field-total-2D-octupole-cartesian-matrix-argument}, 
can be simplified for both polarizations to
\begin{widetext}
\begin{flalign}
 \label{Sca_iso}
\begin{split}
 &\left[\begin{array}{cc}
E^{\rm sca}_{\textnormal{TE}_{0,0} \rightarrow  \textnormal{TE}_{n_1,n_2}} & E^{\rm sca}_{\textnormal{TE}_{0,0} \rightarrow  \textnormal{TM}_{n_1,n_2}} \\[0.5cm]
E^{\rm sca}_{\textnormal{TM}_{0,0} \rightarrow  \textnormal{TE}_{n_1,n_2}} & E^{\rm sca}_{\textnormal{TM}_{0,0} \rightarrow  \textnormal{TM}_{n_1,n_2}}
\end{array}\right] = - \frac{\cos{\phi}\,\,{\rm sgn}(\cos{\theta})}{4\pi\widetilde{\Lambda}^2}\,\times \\\\
& \qquad\qquad\quad \left[\arraycolsep=1mm
\begin{array}{cc}
\left(\frac{ 3a_{\mathrm{1,eff}}}{\cos{\theta}} + 3b_{\mathrm{1,eff}} + 5a_{\mathrm{2,eff}}+\frac{ 5\cos{2\theta}b_{\mathrm{2,eff}}}{\cos{\theta}}\right) &  \left(3a_{\mathrm{1,eff}}+\frac{3b_{\mathrm{1,eff}}}{\cos{\theta}}+\frac{5 \cos{2\theta}a_{\mathrm{2,eff}}}{\cos{\theta}}+5b_{\mathrm{2,eff}}\right)\tan{\phi} \\[0.5cm]
-\left(\frac{3a_{\mathrm{1,eff}}}{\cos{\theta}}+3b_{\mathrm{1,eff}}+5a_{\mathrm{2,eff}}+\frac{5\cos{2\theta} b_{\mathrm{2,eff}}}{\cos{\theta}}\right)\tan{\phi} & \left(3a_{\mathrm{1,eff}}+\frac{3b_{\mathrm{1,eff}}}{\cos{\theta}}+\frac{5 \cos{2\theta}a_{\mathrm{2,eff}}}{\cos{\theta}}+5b_{\mathrm{2,eff}}\right)
\end{array}\right],  
\end{split}
\end{flalign}
\end{widetext}
\noindent with
\begin{subequations}
\begin{align}
    \theta = \arccos&{\left[\pm\sqrt{1-\frac{1}{\widetilde{\Lambda}^2}(n_1^2+n_2^2)}\,\right]},\\
    \phi =& \arctan{\left(\frac{n_2}{n_1}\right)},
\end{align}
\end{subequations}
\noindent where ($n_1,n_2$) label the different diffraction orders. Note that the modes are propagating if $\theta$ is real. For $\theta>\pi/2$, i.e., scattering in the backward half-sphere, $\cos{\theta}$ is negative.
Hence, using \eqref{Sca_iso}, the transmission and reflection through a dipolar-quadrupolar metasurface with isotropic constituents can be calculated. The derived formulas from this subsection are conveniently summarized in Table II(A). 
\subsection{Zeroth-order transmission and reflection}
\begin{figure*}
    \centering
    \includegraphics[width=130mm,trim={0 0 0 0.5cm},clip]{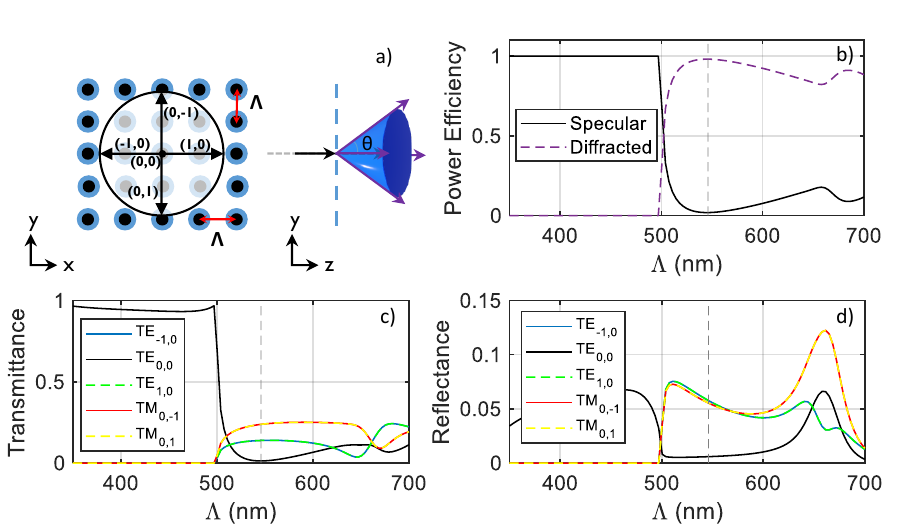}
    \caption{{\bf Fully diffracting metagrating}. a) A square array of core-shell spheres ($r_{\rm core}=0.34\lambda,r_{\rm shell}=(0.34+0.06)\lambda, n_{\rm core}= 1.86, n_{\rm shell}=1.43$) illuminated with a TE polarized plane wave at normal incidence shall diffract all the light to the first diffraction orders at an operational wavelength of $\lambda=500$~nm.  b) The lattice's normalized specular and diffracted power as a function of the periodicity $\Lambda$. c) The transmittance and d) the reflectance of the zeroth and the higher diffraction orders of the lattice. The dashed line shows the periodicity where diffraction to $\theta = 64^\circ$ occurs.}
\label{FIG:dif}
\end{figure*}

Under normal incidence ($\theta_{\rm inc}=\phi_{\rm inc}=0$), if only the zeroth-order mode is considered, or, simply, a non-diffracting square array with $\Lambda < \lambda$, then the calculation formulas for the reflection ($\theta=\pi, \phi=0$) and transmission ($\theta=\phi=0$) coefficients for TE/TM polarization can be simplified to
\begin{subequations}\label{dQEq}
\begin{gather}
   t_{\rm TM}\hspace{-0.4mm}=\hspace{-0.4mm}1\hspace{-0.4mm}-\hspace{-0.4mm}\frac{3a_{\mathrm{1,mod.}}^{\CircArrowLeft{{\scaleto{b_2}{6pt}}}}\hspace{-0.5mm}+\hspace{-0.5mm}3b_{\mathrm{1,mod.}}^{\CircArrowLeft{{\scaleto{a_2}{4.5pt}}}}\hspace{-0.5mm}+\hspace{-0.5mm}5a_{\mathrm{2,mod.}}^{\CircArrowLeft{{\scaleto{b_1}{6pt}}}}\hspace{-0.5mm}+\hspace{-0.5mm}5b_{\mathrm{2,mod.}}^{\CircArrowLeft{{\scaleto{a_1}{4.5pt}}}}}{4\pi\widetilde{\Lambda}^2},\\ 
   r_{\rm TM}\hspace{-0mm}=\hspace{-0mm}\frac{3a_{\mathrm{1,mod.}}^{\CircArrowLeft{{\scaleto{b_2}{6pt}}}}\hspace{-0mm}-\hspace{-0mm}3b_{\mathrm{1,mod.}}^{\CircArrowLeft{{\scaleto{a_2}{4.5pt}}}}\hspace{-0mm}-\hspace{-0mm}5a_{\mathrm{2,mod.}}^{\CircArrowLeft{{\scaleto{b_1}{6pt}}}}\hspace{-0mm}+\hspace{-0mm}5b_{\mathrm{2,mod.}}^{\CircArrowLeft{{\scaleto{a_1}{4.5pt}}}}}{4\pi\widetilde{\Lambda}^2} \hspace{-0mm},\\
   t_{\rm TE}=t_{\rm TM},\\
 r_{\rm TE}=-r_{\rm TM}.
\end{gather}
\end{subequations}
Note that the reflection phase for the TE and TM polarizations are the same, and the sign difference is only due to the specific definition of the TE and TM vectors. The final equations above, for this specific case of a square lattice, normal incidence, and isotropic consisting particles are very similar to the equations derived by Ref.~\citenum{evlyukhin2012collective}.
\subsection{Single Resonance}
If we assume a single magnetic dipole on a square lattice and at the resonance (i.e., $b_1=1$), it is straightforward from \eqref{dQEq} to derive the resonance condition for the 2D lattice, or
\begin{equation}
    t_{\rm TE}= 1 + r_{\rm TE} =0 \Leftrightarrow \frac{3}{4\pi\widetilde{\Lambda}^2} \left(\frac{1}{1/b_1-\mathrm{i}C_{\rm dd}}\right)=1.
\end{equation}
At a normalized periodicity $\widetilde{\Lambda}\approx 0.2$ or $\widetilde{\Lambda}\approx 0.8$, the real part of the dipole-dipole coupling coefficient, or $\Re{(C_{\rm dd})}$, vanishes, as shown in Fig.~\ref{FIG:lattice}b. Hence, inserting the imaginary part of the coefficient from \eqref{imagCdd}, the transmission vanishes completely and the metasurface is perfectly reflecting.  

It might be interesting to find out the scattering cross-section of such a meta-atom inside and outside the lattice for the two "magic lattice spacing." In \textit{Appendix G}, we have calculated the scattering, extinction, and absorption cross-sections in spherical and Cartesian coordinates. For a single magnetic resonance, the scattering cross-section is calculated as 
\begin{equation}\label{scaMag}
\sigma^{\rm c}_{\rm sca} =\frac{k^4\eta^2}{|E_0|^2\,6\pi}\,|\mathbf{m}|^2.
\end{equation}
For a single magnetic dipole at resonance and outside the lattice, for a plane wave excitation, and exploiting  \eqref{a-matrix-def} and \eqref{scaMag}, the Cartesian scattering cross-section turns to \cite{Ruan2011a,rahimzadegan2017fundamental}
\begin{equation}
\sigma^{\rm c}_{\rm sca,0}\left(\widetilde{\Lambda}=\infty\right) =\frac{6 \pi }{k^2}=\frac{3 \lambda^2}{2\pi},
\end{equation}
and, now, for the meta-atom inside the lattice \cite{tretyakov2014maximizing}
\begin{equation}
\frac{\sigma^{\rm c}_{\rm sca}\left(\widetilde{\Lambda}\approx0.2,0.8\right)}{\sigma^{\rm c}_{\rm sca,0}\left(\widetilde{\Lambda}=\infty\right)} =\left(\frac{4\pi\widetilde{\Lambda}^2}{3}\right)^2\approx 0.02,7.18.
\end{equation}
It is clear from the results that at the two resonances, one can achieve much weaker (at $\widetilde{\Lambda}\approx0.2$) or more substantial (at $\widetilde{\Lambda}\approx0.8$) scattering cross-section in comparison to the response of the isolated object. %
\subsection{Fully diffracting metagratings} 
This subsection demonstrates the applicability of the analytical expressions from Table II(A) to a design challenge. We seek a non-absorbing metagrating that diffracts all the light to a polar angle of $\theta=64^{\rm o}$ into the first diffraction orders, or for modes for which it holds $|n_1|+|n_2|=1$, and with no other propagating modes. The goal of the design process is presented in Fig.~\ref{FIG:dif}a. 
Note that for the first diffraction orders, there are four distinct branches with azimuth angles of $\phi=0^{\rm o},90^{\rm o},180^{\rm o},270^{\rm o}$, or, alternatively, the diffraction order pairs,  $(n_1,n_2)=\left\{(1,0),(-1,0),(0,-1),(0,1)\right\}$. For each of the four branches, two different polarizations are possible, TE and TM. However, due to the isotropy of the constituents and lattice symmetry, some modes are zero. Herein, without losing generality, we have assumed a TE polarized plane wave excitation at normal incidence.  

As presented in the equation in Table II(A.3), for normal incidence, or, $\theta_{\rm inc}=0$, the diffraction angles ($\theta,\phi$) are uniquely determined by the diffraction order pair $(n_1,n_2)$. For a set diffraction angle of $\theta=64^\circ$, the required normalized periodicity is, then, calculated as $\widetilde{\Lambda}=\Lambda/\lambda=1.12$. Note, that the 2D array lattice dimension is not subwavelength.
In the next step, using the analytics expressions in Table II(A) for the amplitudes of the diffraction modes, we seek to find the required Mie coefficients that suppress the amplitude into the zeroth diffraction order, or $n_1=n_2 = 0$, transmission. For this purpose, we rely on representing the possible Mie coefficients using the Mie angles \cite{rahimzadegan2020minimalist}. Representing the possible values of the Mie coefficients using the Mie angles allows to systematically search through all possible electric and magnetic dipole and quadrupole Mie coefficients for regimes where the zeroth-order transmission is zero at the fixed periodicity. It should be mentioned that due to the analytic equations at hand, the numerical calculations are computationally cheap, and this fosters multidimensional investigation.
After the identification of a set of Mie coefficients that cancel the zeroth-order transmission, we use a particle swarm optimization method \cite{lee2008modern} to find the dimensions and material parameters of a core-shell sphere that provide the required Mie coefficients. At the operational wavelength of $\lambda=500$~nm, we, finally, find that a core-shell sphere with $r_{\rm core}=170~{\rm nm},r_{\rm shell}=170+30~{\rm nm}, n_{\rm core}= 1.86$, and $n_{\rm shell}=1.43$ satisfies the requirement.

 In Figs.~\ref{FIG:dif}b-d, we show the simulated response of a metagrating made from the designed core-shell spheres arranged in a square array at $\lambda = 500$ nm. Figure~\ref{FIG:dif}b shows the power efficiency of the metagrating in diffracting the power at the chosen periodicity $\Lambda=1.12\lambda=556$ nm (marked with vertical dashed line). Thus, most of the power is diffracted. Figures~\ref{FIG:dif}c-d show the transmittance and reflectance of the individual diffraction orders.
The reflectance and transmittance can be expressed as a function of reflection and transmission coefficients, respectively, as
\begin{equation}
    T = |t|^2 \frac{|\cos{\theta}|}{|\cos{\theta_{\rm inc}}|}, \quad R= |r|^2  \frac{|\cos{\theta}|}{|\cos{\theta_{\rm inc}}|},
\end{equation}
\noindent where for our case $\theta=64^{\rm o}$ and $\theta_{\rm inc}=0^{\rm o}$.
 It can be seen that the zeroth-order transmission and the zeroth-order reflection are successfully suppressed at the assigned periodicity. This novel fully diffracting metagrating is a prime example of how our analytic equations are a powerful design tool for on-demand applications.
\section{Oblique incidence: Analytic Equations}
\subsection{Zeroth-order modes of metasurfaces with isotropic dipole meta-atoms}

\begin{figure}[t]
    \centering
    \includegraphics[scale=0.45]{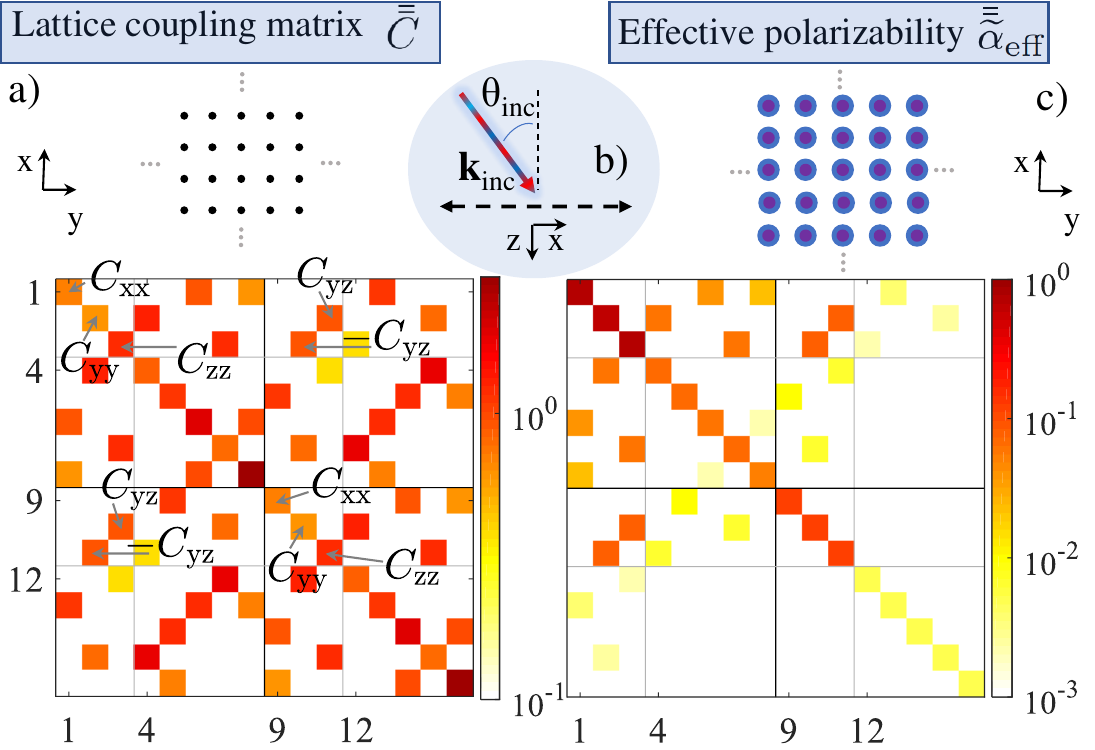}
    \caption{\textbf{Cartesian coupling matrix under oblique angles} a) The Cartesian coupling matrix amplitude for a square array with a normalized periodicity of $\Lambda/\lambda=0.5$ b) under oblique incidence ($\theta_{\rm inc}=\pi/6, \phi_{\rm inc}=0$) up to quadrupolar. The relevant matrix elements for a metasurface made out of dipolar isotropic particles are the parallel $C_{\rm xx}$ and $C_{\rm yy}$, the perpendicular   $C_{\rm zz}$ and the oblique  $C_{\rm yz}$ couplings.  They are marked in the figure. c) The normalized effective polarizability matrix amplitude of the Ag-core SiO$_2$-shell particle considered in Fig.~\ref{FIG:latticeFull}f inside the square lattice when illuminated at the oblique angle.}
\label{FIG:obliqueC}
\end{figure}

Even though metasurfaces have been extensively studied under normal incidence, the oblique incidence is less explored. However, understanding the complete angular response is essential for designing metasurface-based photonic devices \cite{albooyeh2014resonant}. Therefore, this section provides analytical equations to express the optical response of dipolar metasurfaces illuminated at oblique incidence.
For the case of an oblique incidence, the symmetry of the lattice coupling matrix changes with respect to normal incidence, and there are more non-zero elements on the matrix $\bar{\bar{C}}$. Nevertheless, the symmetry of the coupling matrix at oblique incidence is only affected by the lattice type. Therefore, changing the lattice constant or the incidence angles for a specific 2D lattice does not alter the symmetry of the matrix, but only the values of the coefficients.  Figure~\ref{FIG:obliqueC}a shows the dipole-quadrupolar Cartesian coupling matrix for a square array under oblique incidence angles of $\theta_\mathrm{\,inc}=\pi/6,\phi_\mathrm{\,inc}=0$ and at a normalized periodicity of $\Lambda/\lambda=0.5$. These coupling coefficients are a function of the incident angles, $\theta_\mathrm{\,inc}$ and $\phi_\mathrm{\,inc}$ (setup at Fig.~\ref{FIG:obliqueC}b) and the normalized periodicity of the square array, $\widetilde{\Lambda}=\Lambda/\lambda$. In Fig.~\ref{FIG:obliqueC}c, the effective polarizability matrix for an Ag-core SiO$_2$-shell spherical particle (with geometry also depicted in Fig.~\ref{FIG:latticeFull}f) inside the lattice above is shown up to quadrupolar order. The effective polarizability is derived via \eqref{a-eff-def}. It can be observed that, unlike the polarizability of the isolated spherical object, the effective polarizability is not diagonal but takes the symmetry of the lattice.

Let us assume that a metasurface consists of isotropic dipoles, and we try to derive the zeroth-order response. In this case, not all the coupling coefficients are relevant; the elements of $\bar{\bar{C}}$ that determine the amplitudes of the propagating diffraction orders of the metasurface made from isotropic dipolar particles are $C_{xx}$, $C_{yy}$, $C_{zz}$, and $C_{yz}$. These coupling coefficients are marked in the matrix in Fig.~\ref{FIG:obliqueC}a. Note that for simplicity, we have ignored the "${\rm dd}$" subscript, compared to the previous section, as we anyhow only consider dipolar particles. Prior knowledge of the non-zero and relevant elements of the coupling coefficients for this case of study enables us to construct an analytic model that we can use for the upcoming expressions.
Next, we follow a similar procedure to the normal incidence case of the previous section to derive an analytic equation for the zeroth diffraction order, characterized by $(n_1, n_2) = (0,0)$, from a metasurface made from dipolar isotropic particles. Additionally, in all the following equations, without losing generality, we assume the azimuthal angle to be zero, i.e., $\phi_\mathrm{inc}=0$.
Assuming, then, a plane wave oblique incidence, the TE and TM polarizations can be defined as
\begin{subequations}
\begin{align}
\mathbf{E}^{\rm \, inc}_{\rm \, TM}=&\left[\begin{array}{c}E_x\\ E_y\\  E_z\end{array}\right]
= E_0\left[\begin{array}{c} \cos{\theta_\mathrm{inc}} \\ 0 \\ -\sin{\theta_\mathrm{inc}}\end{array}\right]\mathrm{e}^{\mathrm{i}(k^{\mathrm{inc}}_x x+k^{\mathrm{inc}}_z z)},\\
\mathbf{E}^{\rm \, inc}_{\rm \, TE}
=&\left[\begin{array}{c} E_x \\ E_y \\ E_z\end{array}\right] = E_0\left[\begin{array}{c} 0 \\ 1 \\ 0\end{array}\right]\mathrm{e}^{\mathrm{i}(k^{\mathrm{inc}}_x x+k^{\mathrm{inc}}_z z)},
\end{align}
\end{subequations}
\noindent where $k^{\mathrm{inc}}_x=k\sin{\theta_\mathrm{inc}}$ and $k^{\mathrm{inc}}_z=k \cos{\theta_\mathrm{inc}}$ are the $x$ and $z$ component of the impinging wavevector.

Following \eqref{a-matrix-def}, the induced multipole moments in each particle for TM polarization as a function of the effective Mie coefficients are derived as
\begin{subequations}\label{Induced_ob}
\begin{align} 
\left[\begin{array}{c}
 p_x/\varepsilon\\[0.05cm]
 p_z/\varepsilon\\[0.05cm]
\eta m_y\\[0.05cm]
\end{array}\right] = \frac{6\pi\mathrm{i}}{k^3}
\left[\begin{array}{c}
 a_{\mathrm{1,mod.xx}}\cos{\theta_\mathrm{inc}}\\[0.05cm]
-a_{\mathrm{1,mod.zz}}^{\CircArrowLeft{{\scaleto{b_{yy}}{5.5pt}}}}\sin{\theta_\mathrm{inc}}\\[0.05cm]
b_{\mathrm{1,mod.yy}}^{\CircArrowLeft{{\scaleto{a_{zz}}{4.5pt}}}}
\end{array}\right] E_0.
\intertext{Similarly, for TE polarization, the following relation can be derived for the effective dipole moments,}
\left[\begin{array}{r}
 p_y/\varepsilon\\[0.05cm]
\eta m_x\\[0.05cm]
\eta m_z\\[0.05cm]
\end{array}\right] =  \frac{6\pi\mathrm{i}}{k^3}
\left[\begin{array}{c}
 a_{\mathrm{1,mod.yy}}^{\CircArrowLeft{{\scaleto{b_{zz}}{5.5pt}}}}\\[0.05cm]
 -b_{\mathrm{1,mod.xx}}\cos{\theta_\mathrm{inc}}\\[0.05cm]
b_{\mathrm{1,mod.zz}}^{\CircArrowLeft{{\scaleto{a_{yy}}{4.5pt}}}}\sin{\theta_\mathrm{inc}}
\end{array}\right] E_0.
\end{align}
\end{subequations}
For a normal incidence excitation, i.e., $\theta_\mathrm{inc}=0$, the induced moments are the same as what we derived in \eqref{inducedMoments}. However, at oblique incidence, the main difference is the induced moments normal to the metasurface plane. These exited moments bring about novel interference patterns not seen at $\theta_\mathrm{inc}=0$ incidence. We will explore, later on, how these normal moments modify the optical response of a metasurface.
The effective dipolar polarizabilities of the particles in \eqref{Induced_ob} depend on (a) the modulation of the elements with the coupling coefficients $C_{\rm xx}$, $C_{\rm yy}$, and $C_{\rm zz}$ and (b) the coupling with other multipole moments through the $C_{\rm yz}$ coefficient. The $\CircArrowLeft{{\scaleto{b_{yy}}{4.5pt}}}$ shows the coupled parameters, and the coupling term is written inside the circle. Electric (magnetic) dipole moments are coupled to magnetic (electric) dipole moments. The \textit{effective Mie coefficients} are, therefore, expressed as,
\begin{subequations}\label{couplingMie_Ob}
\begin{align}
\begin{split}
\frac{1}{a_{\mathrm{1,eff.yy}}}&=\frac{1}{a_{\mathrm{1,mod.yy}}^{\CircArrowLeft{{\scaleto{b_{zz}}{5.5pt}}}}} =\\
&=\frac{1-C_{\rm yz}^{2}\text{\ensuremath{b_{\mathrm{1,mod.zz}}}}a_{\mathrm{1,mod.yy}}}{a_{\mathrm{1,mod.yy}}\left(1-C_{\rm yz}\text{\ensuremath{b_{\mathrm{1,mod.zz}}}}\sin\theta_\mathrm{inc}\right)},
\end{split}\\
\begin{split}
\frac{1}{b_{\mathrm{1,eff.yy}}}&=\frac{1}{b_{\mathrm{1,mod.yy}}^{\CircArrowLeft{{\scaleto{a_{zz}}{4.5pt}}}}} =\\
&=\frac{1-C_{\rm yz}^{2}\text{\ensuremath{a_{\mathrm{1,mod.zz}}}}b_{\mathrm{1,mod.yy}}}{b_{\mathrm{1,mod.yy}}\left(1-C_{\rm yz}\text{\ensuremath{a_{\mathrm{1,mod.zz}}}}\sin\theta_\mathrm{inc}\right)},
\end{split}\\
\begin{split}
\frac{1}{a_{\mathrm{1,eff.zz}}}&=\frac{1}{a_{\mathrm{1,mod.zz}}^{\CircArrowLeft{{\scaleto{b_{yy}}{5.5pt}}}}} =\\
&=\frac{1-C_{\rm yz}^{2}\text{\ensuremath{b_{\mathrm{1,mod.yy}}}}a_{\mathrm{1,mod.zz}}}{a_{\mathrm{1,mod.zz}}\left(1-C_{\rm yz}\text{\ensuremath{b_{\mathrm{1,mod.yy}}}}\csc\theta_\mathrm{inc}\right)},
\end{split}\\
\begin{split}
\frac{1}{b_{\mathrm{1,eff.zz}}}&=\frac{1}{b_{\mathrm{1,mod.zz}}^{\CircArrowLeft{{\scaleto{a_{yy}}{4.5pt}}}}} =\\
&=\frac{1-C_{\rm yz}^{2}\text{\ensuremath{a_{\mathrm{1,mod.yy}}}}b_{\mathrm{1,mod.zz}}}{b_{\mathrm{1,mod.zz}}\left(1-C_{\rm yz}\text{\ensuremath{a_{\mathrm{1,mod.yy}}}}\csc\theta_\mathrm{inc}\right)},
\end{split}
\end{align}
\end{subequations}
while the modulated Mie coefficients are calculated as,
%
\begin{subequations}\label{effectiveMie-2}
\begin{align}
\frac{1}{a_{\mathrm{1,mod}.vv}}=\frac{1}{a_{1}}-\mathrm{i}\,C_{ vv} &,&
\frac{1}{b_{\mathrm{1,mod}.vv}}=\frac{1}{b_{1}}-\mathrm{i}\,C_{ vv},
\end{align}
\end{subequations}
\noindent where $a_1$ and $b_1$ are the Mie coefficient of the isolated scatterer and ${ v} = \{{\rm x,y,z}\}$.
If we follow a procedure similar to the one performed for normal incidence, the zeroth-order reflection and transmission coefficients of a dipolar metasurface under normal incidence can be calculated. The resulting formulas are provided in Table II(B). 

The analytical equations for the transmission and reflection of an obliquely incident metasurface enable efficient and more accessible exploration of the physics involving metasurface structure. In the following subsections, we derive a particular Brewster angle for a single metasurface and further explore the transmission through a Huygens' metasurface under oblique incidence.

Note that higher-order analytic equations, i.e., involving quadrupoles, can also be derived from equations in Table I. However, in this contribution, we focus on the simpler dipolar expressions for the oblique case. If a particular application for a specific scenario is demanded, other analytic equations can be further derived. 
\subsection{Brewster angle: Particle-independent polarization filter}
As a direct application of the analytic equations for metasurfaces under oblique incidence presented in Table II(B), we can search for specific metasurfaces that offer a desired and predefined optical response. Polarization, for example, is a crucial property of light, and its control is a fundamental necessity for wave modulation. A question at hand concerns the Brewster angle for metasurfaces. The Brewster angle is the angle at which the reflection for TM or TE polarization vanishes \cite{paniagua2016generalized,yin2019metagrating,wang2019all}. Therefore, it provides an important tool to separate different polarizations of light. 
In this subsection, we provide an example to demonstrate the strength of the analytical equations in finding important regimes for on-demand applications. Here, the set goal is to give the recipe for a metasurface that can separate the two polarizations in reflection. First, we assume a metasurface consisting of only isotropic magnetic dipolar scatterers, or equivalently $a_{n>0}=b_{n>1}=0$. Magnetic dipoles constitute the lowest-energy resonance in homogeneous high-refractive-index spheres and are easier to achieve. From the equations in Table II(B), the reflection coefficient for such a metasurface can be written as,
%
\begin{equation}\label{BrewsterEq}
\begin{split}
\left[\begin{array}{c}
r_{\textnormal{TE}_{0,0} \rightarrow  \textnormal{TE}_{0,0}}\\[0.25cm]
r_{\textnormal{TM}_{0,0} \rightarrow  \textnormal{TM}_{0,0}}
\end{array}\right]= \frac{3}{4\pi\widetilde{\Lambda}^2\cos{\theta_\mathrm{inc}}}\times \qquad \qquad \qquad \\ \qquad \left[\begin{array}{c}
b_{\mathrm{1,mod.xx}}\cos\theta_\mathrm{inc}^2-b_{\mathrm{1,mod.zz}}\sin\theta_\mathrm{inc}^2 \\[0.5cm]
-b_{\mathrm{1,mod.yy}}
\end{array}\right].
\end{split}
\end{equation}
%

The equation above shows that the moment induced normal to the metasurface interferes destructively with the moment induced in-plane. To derive the Brewster angle, one needs to find the metasurface parameters where reflection vanishes for the TE polarization. Therefore, the Brewster angle condition further simplifies to
%
\begin{align}\label{Brew}
r_{\textnormal{TE}_{0,0} \rightarrow  \textnormal{TE}_{0,0}} &= 0 \Rightarrow  \notag \\
b_{\mathrm{1,mod.xx}}\cos\theta_\mathrm{inc}^2&=b_{\mathrm{1,mod.zz}}\sin\theta_\mathrm{inc}^2 \Rightarrow \notag\\ 
\frac{\cos\theta_\mathrm{inc}^2}{1/b_1-\mathrm{i}C_{\rm zz}} &= \frac{\sin\theta_\mathrm{inc}^2} {1/b_1-\mathrm{i}C_{\rm xx}}.
\end{align}
%
At an incidence angle of $\theta_{\rm inc}= \pi/4$, the  Brewster condition becomes independent of the scatterer's Mie coefficient, and thus, simplifies to $C_{\rm xx}=C_{\rm zz}$. In other words, at an oblique incidence angle of $45$ degrees, a metasurface can separate the two polarizations irrespective of its constituents, as far as the dipolar approximation holds and the condition mentioned above is satisfied. Sweeping through the lattice constants, we find out that for a Brewster angle of $45$ degrees, the required lattice constant is 
\begin{equation}\label{Brewsterd}
\frac{{\Lambda}_{\theta_\mathrm{B}=\pi/4}}{\lambda} = 0.5352, 
\end{equation}
\noindent which is smaller than the dimension where the first diffraction order appears for this incidence angle (i.e., $\widetilde{\Lambda}=0.58$) and, hence, no diffraction occurs.

Therefore, a metasurface at a normalized periodicity of $0.5352$ suppresses the reflection for TE polarization as far as the spheres are small enough compared to the operational wavelength to be described with a magnetic dipole response. The reflected amplitude for the other polarization can be determined via \eqref{Brew} by the strength of the effective magnetic Mie coefficient.
A similar Brewster angle can be derived to suppress reflection in TM polarization, i.e., the s-polarization, with a metasurface made from electric dipolar particles at the same normalized periodicity point of operation.

Using the analytical formulas of \eqref{Brew}, we can, generally, derive the condition for the Brewster angle for different metasurfaces, depending on combinations of the Mie coefficients and the lattice dimension. Here, we demonstrated a simple but powerful case.
\subsection{Huygens' metasurfaces under oblique angle}
\begin{figure}[t]
    \centering
    \includegraphics[scale=0.6]{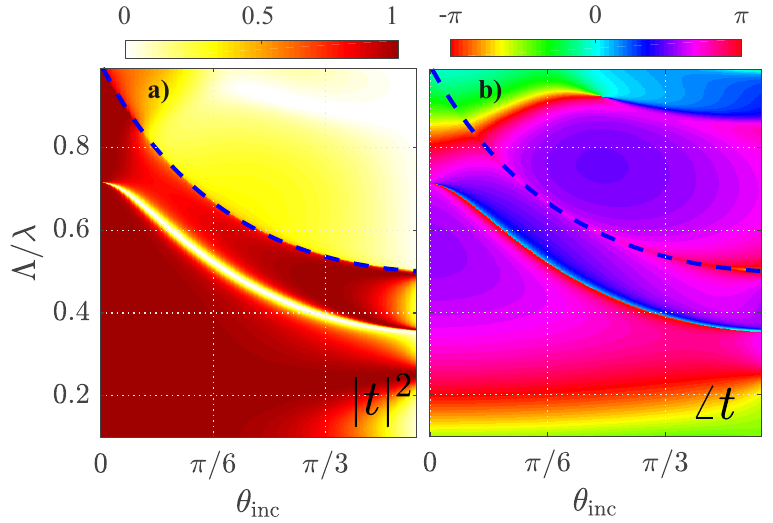}
    \caption{ \textbf{Obliquely illuminated Huygens' metasurface:} a) The amplitude and b) phase of the transmission coefficient of a Huygens' metasurface as a function of the incident angle $\theta_{\rm inc}$ and the normalized periodicity $\Lambda/\lambda$. The metasurface is made from a particle with an isolated electric and magnetic Mie coefficient of 1, i.e., at resonance ($a_1=b_1=1$). The dashed blue line indicates the onset of diffraction orders. Note that for a Huygens' metasurface, the TE and TM excitations are equivalent.}
\label{FIG:Huygens}
\end{figure}
Huygens' metasurfaces have attracted significant attention in the optics community due to their ability to provide unity transmittance, and broad phase coverage \cite{epstein2016huygens,Staude2017,decker2015high,chen2018huygensCoreshell,liu2017huygens,love1901integration}. Although Huygens' metasurfaces are extensively studied at normal incidence, the oblique incidence case is not well studied yet \cite{arslan2017angle,gigli2020fundamental}, the main reason potentially being the lack of analytical tools. Questions of retaining the unity transmittance or additionally providing a phase coverage are still under-explored.

This subsection, shortly, studies the transmission of Huygens' metasurfaces in the subwavelength regime, depending on the lattice constant and the incidence angle. The Huygens' metasurface we study is made from dipolar isotropic particles at resonance when considered isolated (i.e., $a_1=b_1=1$). Figure~\ref{FIG:Huygens} shows the amplitude and phase of transmission versus the normalized periodicity and the angle of incidence via the formulas of Table II(B). The dashed line shows the onset of the first diffraction order. We ignore the results above this dashed line to assure a single-mode operation.

For a Huygens' metasurface, due to the electromagnetic duality symmetry of its constituents, both TE and TM excitations result in the same reflection and transmission. 

When illuminated at normal incidence, or $\theta_{\rm inc}=0$, regardless of the polarization, transmission amplitude is 1, as seen from Fig.~\ref{FIG:Huygens}a. Furthermore, a broad phase shift coverage of almost $3\pi/4$ is achievable when changing the normalized periodicity in the range [$0.7,1$] \cite{rahimzadegan2020beyond}. If different but equal electric and magnetic Mie coefficients are considered, the periodicity range in which a broad phase coverage is accessible changes. Therefore, depending on the particle size and the permissible densities, a proper point of operation can be chosen for the metasurface.

For those currently chosen Mie coefficient values ($a_1=b_1=1$), it is apparent from the results in Fig.~\ref{FIG:Huygens}, that for small oblique incidence (i.e.  $\theta_{\rm inc}<\pi/6$), the broad phase coverage range (i.e. $\Delta(\angle t) > \pi$) falls, primarily, into the regime for which the metasurface is diffracting, and hence the transmission is suppressed. Therefore, we find out that for a better phase coverage with high transmittance under oblique incidence, one must carefully choose a different Mie coefficient to avoid the diffracting region and the sharp zero-transmittance resonance. 

If the incidence is only slightly tilted at a normalized periodicity of $0.71$, a sharp resonance appears in the transmittance of the metasurface. This resonance can be traced back to the destructive interference of the in-plane and out-plane induced moments, as described previously. Although the sudden drop of the transmittance may spoil the functionality of a Huygens' metasurface for its phase coverage, the prior information of the regime in which it occurs can help avoid this point of operation. Moreover, this interesting region for a metasurface can be exploited in other applications in which normal transparency and oblique opacity are sought after.

In short, our analytic tool can help in designing a Huygens metasurface for a specific application, avoiding the certain undesired point of operations.
%
%
\section{Conclusion}
This paper provides exact, robust, and accessible equations to calculate the amplitudes of all propagating diffraction orders from a 2D lattice decorated with identical but otherwise arbitrarily shaped particles. We provided explicit expressions in both Cartesian and spherical bases up to the octupolar order. 
By utilizing the polarizability/T matrix of the individual particles and the lattice coefficients, we calculated the effective polarizability/T matrix of the particles. Besides the primary one, the proposed formulas enable the explicit calculation of the amplitudes of all propagating diffraction orders.
In addition, tools for the convenient transformation of the two equivalent bases are also provided. Although the main manuscript is focused on the Cartesian basis, the supplementary information provides complimentary graphs on the spherical basis.

We investigated the impact of the lattice and that of the decorating particle on the optical response of the metasurfaces. Our analytical framework constitutes an extraordinary tool to disentangle the individual impact. For this purpose, we have introduced the coupling matrix of a 2D lattice and explored it explicitly for square and hexagonal lattices. Moreover, based on the defined bases, symmetries of polarizability and T matrix of isotropic, anisotropic, and helical objects were investigated. Stemming from the symmetry-protected zeros of the particle's polarizability and the lattice' coupling matrix in the Cartesian basis, we introduced simplified, efficient, and closed-form analytical formulas, which we used to conveniently design and explore three contemporary metasurface applications, namely a fully-diffracting metagrating, a polarization filter, and a Huygens' metasurface.

The authors hope that the techniques proposed herein will allow physicists and engineers to conduct investigations related to metasurface phenomena and propose novel photonic designs. Our comprehensive multipolar theory not only paves the way for further exploration of the rich physics of metasurfaces but also enables a paradigm shift in designing next-generation optical devices.
As far as further endeavors are concerned, the presented work could be expanded to incorporate evanescent modes from the 2D arrays and extract more analytical expressions from the existing general equations for on-demand, specific metasurface applications.
%
\appendix
\section{Field expansion via spherical wave functions}
Assume a particle positioned inside an infinite, non-isotropic, linear, homogeneous, and isotropic medium. An electromagnetic field illuminates the particle. The total electric field in the spatial domain outside and around the particle at an angular frequency $\omega$ consists of the incident and scattered fields. Each of these fields can be expanded using vector spherical harmonics (VSH) as \cite{mishchenko2002scattering} as,
\begin{subequations}\label{fields-vsh}
\begin{equation}
\mathbf{E}_{\rm inc}(\mathbf{r}) = 
\sum_{j=1}^{\infty} \sum_{m=-j}^{j} q^{\rm e}_{jm}\mathbf{N}^{(1)}_{jm}(k\mathbf{r}) + q^{\rm m}_{jm}\mathbf{M}^{(1)}_{jm}(k\mathbf{r}),
\end{equation}
\begin{equation}
\mathbf{E}_{\rm sca}(\mathbf{r}) = 
\sum_{j=1}^{\infty} \sum_{m=-j}^{j} b^{\rm e}_{jm}\mathbf{N}^{(3)}_{jm}(k\mathbf{r}) + b^{\rm m}_{jm}\mathbf{M}^{(3)}_{jm}(k\mathbf{r}),
\end{equation}
\end{subequations}
with $q^v_{jm}$ and $b^v_{jm}$, $j=\{1,2,3\}$, $m=\{-j,...,j\}$, $v=\{{\rm e,m}\}$, the electric/magnetic incident and scattered field expansion coefficients, respectively. The wavenumber $k$ corresponds to the medium that surrounds the particle, and $r > r_c$, with $r_c$ being the radius of the sphere that circumscribes the particle.  Moreover, the VSH functions are defined as, 
\begin{subequations}\label{VSW-Hankel}
\begin{gather}
\mathbf{M}^{(l)}_{jm} = \gamma_{jm}\,\nabla\hspace{-0.5mm}\times\hspace{-0.5mm}\left[\hat{\mathbf{r}}z_j^{(l)}\hspace{-0.1mm}(kr)P_j^m( \mathrm {cos} \theta)e^{{\rm i}m\phi}\right],\\
\mathbf{N}^{(l)}_{jm} = \frac{1}{k}\nabla \hspace{-0.5mm}\times\hspace{-0.5mm} \mathbf{M}^{(l)}_{jm},\\
\intertext{with}
\gamma_{jm} =  \sqrt{\frac{(2j+1)}{4\pi j(j+1)}} \sqrt{\frac{(j-m)!}{(j+m)!}},
\end{gather}
\end{subequations}
with $l=1$ for the incident field and $l=3$ for the scattered one. $z^{(1)}_j(x) = j_j(x)$ is the spherical Bessel function of the first kind, while $z^{(3)}_j(x) = h_j(x)$ is the spherical Hankel function of the first kind. Finally, $P^{m}_j(x)$ is the associated Legendre polynomial. 

The scattering coefficients can be calculated, using the orthogonality relations, as a function of the incident field via the following formulas,
\begin{align}\label{scat-coeff-calc}
b^{\rm e}_{jm} &= \frac{\int_S \mathbf{N}^{(3)}_{jm}(\hat{\mathbf{r}})\cdot\mathbf{E}_{\rm sca}(\hat{\mathbf{r}})\,{\rm d}S}{\int_S |\mathbf{N}^{(3)}_{jm}(\hat{\mathbf{r}})|^2\,{\rm d}S},\\
b^{\rm m}_{jm} &= \frac{\int_S \mathbf{M}^{(3)}_{jm}(\hat{\mathbf{r}})\cdot\mathbf{E}_{\rm sca}(\hat{\mathbf{r}})\,{\rm d}S}{\int_S |\mathbf{M}^{(3)}_{jm}(\hat{\mathbf{r}})|^2\,{\rm d}S},
\end{align}
where the integrating surface $S$ is any sphere enclosing the particle, as shown in Fig.~\ref{fig1}. Generally, \eqref{scat-coeff-calc} cannot be calculated analytically, except for simple geometries, like spheres \cite{Bohren2008}. Therefore, the elements of the T matrix of the particle under study can be numerically obtained after a series of simulations for a sufficient number of either plane-wave incidences \cite{fruhnert2017computing} or normalized VSH functions of \eqref{VSW-Hankel} as incidence \cite{demesy2018scattering,santiago2019decomposition}.   
\section{Normalized polarizability and denormalized polarizability in SI units}
In this work, the normalized polarizabilities are used in the definition of \eqref{a-matrix-def}. They are all dimensionless and, hence, can be directly compared to each other. Nevertheless, if the polarizabilities in SI units are required, they can be directly obtained from the normalized ones as,
\begin{subequations}\label{denormalize_a}
\begin{align}
\bar{\bar{\alpha}}_{jj'}^{vv'} &= \frac{\zeta_{j}\zeta_{j'}}{k^{j+j'+1}}\,\bar{\bar{\widetilde{\alpha}}}_{jj'}^{vv'},\\
\zeta_j &= \sqrt{(2j+1)!\,\pi},
\end{align}
\end{subequations}
with $\{j,j'\} = \{1,2,3\}$ and $\{v,v'\} = \{\mathrm{e},\mathrm{m}\}$. More information about calculating  the prefactors $\zeta_j$ can be found in \textit{Supp. Info. III.}.
\section{Field and multipole vector definitions}
The irreducible multipole moment vectors in \eqref{a-matrix-def} are defined as 
\begin{subequations}\label{fields-def-1}
\begin{gather}
 \mathbf{p} = \left[\begin{array}{c}
p_x\\[0.05cm]
p_y\\[0.05cm]
p_z
\end{array}\right],\\
 k\varepsilon^{-1}\mathbf{Q}^{\rm e} = \frac{k}{\varepsilon\sqrt{3}} \,\left[\begin{array}{c}Q^{\rm e}_{xy}\\ Q^{\rm e}_{yz} \\ \frac{\sqrt{3}}{2}\,Q^{\rm e}_{zz}\\ Q^{\rm e}_{xz} \\\frac{1}{2}\left(Q^{\rm e}_{xx} - Q^{\rm e}_{yy}\right)\end{array}\right],\\
 k^2\varepsilon^{-1}\mathbf{O}^{\rm e} = \frac{k^2\sqrt{6}}{4\varepsilon}\,
\left[\begin{array}{c}
\frac{1}{2}\left(3O^{\rm e}_{xxy} - O^{\rm e}_{yyy}\right)\\[1.2mm]
\frac{\sqrt{6}}{4}O^{\rm e}_{xyz}\\[1.2mm]
\frac{\sqrt{15}}{2}O^{\rm e}_{yzz}\\[1.2mm]
\sqrt{\frac{5}{2}}O^{\rm e}_{zzz}\\[1.2mm]
\frac{\sqrt{15}}{2}O^{\rm e}_{xzz}\\[1.2mm]
\frac{\sqrt{6}}{2}\left(O^{\rm e}_{zxx} - O^{\rm e}_{zyy}\right)\\[1.2mm]
\frac{1}{2}\left(O^{\rm e}_{xxx} - 3O^{\rm e}_{xyy}\right) \end{array} \right],
\end{gather}
\end{subequations}
while the fields are defined as in \eqref{fields-def-2},
\begin{subequations}\label{fields-def-2}
\begin{gather}
\mathbf{E}_1 =  
\left[\begin{array}{c}
 E_x \\ 
E_y \\
E_z 
\end{array}\right]\hspace{-0.5mm},\quad\mathbf{E}_2 =  \frac{1}{2\sqrt{3}}
\left[\begin{array}{c}
\partial_y E_x + \partial_x E_y \\ 
\partial_y E_z + \partial_z E_y \\
\sqrt{3}\, \partial_z E_z  \\
\partial_x E_z + \partial_z E_x\\
\partial_x E_x - \partial_y E_y\end{array}\right]\hspace{-0.5mm},\\
\mathbf{E}_3\hspace{-0.8mm} 
=\hspace{-0.8mm}\frac{\sqrt{6}}{24}\hspace{-0.8mm}
\left[\hspace{-1mm}\begin{array}{c}
\frac{1}{2}\hspace{-0.3mm}\{ 2\partial_{xy} E_x\hspace{-0.3mm}+\hspace{-0.3mm} \left(\partial^2_{x} \hspace{-0.3mm}-\hspace{-0.3mm} \partial^2_{y}\right)\hspace{-0.3mm} E_y \} \\ [1mm]
\frac{2}{\sqrt{6}}\hspace{-0.3mm}\left(\partial_{yz} E_x \hspace{-0.3mm} +\hspace{-0.3mm} \partial_{xz} E_y \hspace{-0.3mm}+\hspace{-0.3mm} \partial_{xy} E_z \right) \\[1mm]
\frac{1}{\sqrt{10}}\hspace{-0.3mm}\{\hspace{-0.3mm} \frac{1}{4}\hspace{-0.3mm}(2\hspace{-0.1mm} \partial_{xy}\hspace{-0.6mm}-\hspace{-0.6mm}\partial^2_{y})\hspace{-0.3mm} E_x\hspace{-0.6mm}+\hspace{-0.6mm} (\partial^2_{z}\hspace{-0.6mm}-\hspace{-0.6mm} \frac{3}{4} \hspace{-0.3mm}\partial^2_{y})\hspace{-0.3mm} E_y \hspace{-0.6mm} +\hspace{-0.6mm} 2\hspace{-0.1mm}\partial_{yz} E_z  \}  \\[1mm]
\frac{2}{\sqrt{15}}\hspace{-0.3mm}\{-  2\partial_{xz} E_x \hspace{-0.5mm}-\hspace{-0.5mm} 2 \partial_{yz} E_y\hspace{-0.5mm}+\hspace{-0.5mm}(2\partial^2_{z} \hspace{-0.5mm}-\hspace{-0.5mm} \partial^2_{x} \hspace{-0.5mm}  -\hspace{-0.5mm} \partial^2_{y})E_z   \}
\\[1mm]
\frac{1}{\sqrt{10}}\hspace{-0.3mm}\{(\partial^2_{z}\hspace{-0.5mm}-\hspace{-0.5mm} \frac{1}{4}\hspace{-0.3mm}\partial^2_{y}\hspace{-0.5mm} - \hspace{-0.5mm} \frac{3}{4}\hspace{-0.3mm}\partial^2_{x}) E_x \hspace{-0.5mm}-\hspace{-0.5mm} \frac{1}{2}\hspace{-0.3mm}\partial_{xy} E_y \hspace{-0.5mm}+\hspace{-0.5mm} 2\partial_{xz} E_z   \}
\\[1mm]
\frac{2}{\sqrt{15}}\hspace{-0.3mm}\{2\partial_{xz} E_x \hspace{-0.5mm} -\hspace{-0.5mm} 2\partial_{yz} E_y \hspace{-0.5mm}+\hspace{-0.5mm}(\partial^2_{x} \hspace{-0.5mm} -\hspace{-0.5mm} \partial^2_{y}) E_z  \} \\[1mm]
\frac{1}{2}\hspace{-0.3mm}\{(\partial^2_{x}\hspace{-0.5mm} - \hspace{-0.5mm} \partial^2_{y})E_x  \hspace{-0.5mm}- \hspace{-0.5mm} 2\,\partial_{xy} E_y \}
\end{array}\hspace{-1.2mm}\right]\hspace{-0.5mm},
\end{gather}
\end{subequations}
%
with the magnetic multipoles $\mathbf{m}, \mathbf{Q}^{\rm m}$ and $\mathbf{O}^{\rm m}$ and fields $\mathbf{H}_j,\,j=\{1,2,3\}$, defined in similar fashion as the electric ones above, but with a multiplication with the prefactor $\mathrm{i}\eta$, as in the case of \eqref{a-matrix-def}.

The components of multipole moment vectors used, herein, and defined in \eqref{fields-def-1}, form an irreducible set of Cartesian multipole moments that are sufficient for the representation of a scatterer's response up to the respective expansion order. We have used the real spherical harmonics corresponding to atomic orbitals, $p, d$, and $f$ \cite{chisholm1976group} for the combination and order of the vectors. Other irreducible sets of multipole moments can be formed from the quadrupole, and octupolar matrices \cite{Jackson1999}, depending on the problem under study or for convenience ($\rightarrow$ \textit{Supp. Info. IV.}).
%
\section{Radiation field definition in Cartesian coordinates}
The scattering far-field from a particle described up to octupolar order in Cartesian coordinates is defined as
\begin{subequations}\label{radiation-fields-cart-def}
\begin{align}
\mathbf{E}&^{\rm c}(k\mathbf{r}) = \frac{k^2}{4\pi}\frac{e^{\mathrm{i}kr}}{r}\,\Bigg[\hat{\mathbf{r}}\times\left(\frac{1}{{\varepsilon}}\mathbf{p}\times\hat{\mathbf{r}}\right) + \left(\eta\,\mathbf{m}\times\hat{\mathbf{r}}\right) \nonumber\\
&- \frac{\mathrm{i}k}{6} \hat{\mathbf{r}}\times\left(\frac{1}{\varepsilon}\boldsymbol{\mathcal{Q}}^{\rm e}\times\hat{\mathbf{r}}\right) - \frac{\mathrm{i}k}{6} \left(\eta\,\boldsymbol{\mathcal{Q}}^{\rm m}\times\hat{\mathbf{r}}\right) \nonumber\\
&- \frac{k^2}{16} \hat{\mathbf{r}}\times\left(\frac{1}{\varepsilon}\boldsymbol{\mathcal{O}}^{\rm e}\times\hat{\mathbf{r}}\right) - \frac{k^2}{16} \left(\eta\,\boldsymbol{\mathcal{O}}^{\rm m}\times\hat{\mathbf{r}}\right)\Bigg], 
\end{align}
\begin{align}
\hat{\mathbf{r}} &=  \hat{r}_{x}\,\hat{\mathbf{x}} + \hat{r}_{y}\,\hat{\mathbf{y}} + \hat{r}_{z}\,\hat{\mathbf{z}} = \notag\\
&= {\rm sin}\theta{\rm cos}\phi\,\hat{\mathbf{x}} + {\rm sin}\theta{\rm sin}\phi\,\hat{\mathbf{y}} + {\rm cos}\theta\,\hat{\mathbf{z}},
\end{align}
\end{subequations}
where $\hat{\mathbf{r}}$ is the position unit-vector.

The components of the vectors $\boldsymbol{\mathcal{Q}}$ and $\boldsymbol{\mathcal{O}}$ are defined as,
\begin{subequations}\label{vectors-Q-O}
\begin{align}
\mathcal{Q}^v_{\alpha} &= \sum_{\beta} Q^{v}_{\alpha\beta}\,\hat{r}_{\beta},\\
\mathcal{O}^v_{\alpha} &= \sum_{\beta\gamma} {O}^v_{\alpha\beta\gamma}\,\hat{r}_{\beta}\,\hat{r}_{\gamma},
\end{align}
\end{subequations}
with $\{\alpha,\beta,\gamma\}=\{x,y,z\}$ and $v =\{ {\rm e, m}\}$. The Cartesian multipole moments $Q_{\alpha\beta}^v$ and $O_{\alpha\beta\gamma}^v$ are the elements of the two-dimensional quadrupolar and three-dimensional octupolar matrices, respectively, as defined in Ref.~\citenum{Jackson1999}. The vectors of \eqref{vectors-Q-O} are used only for the calculation of the far-field in \eqref{radiation-fields-cart-def} and are different from the irreducible Cartesian multipole moment vectors $\mathbf{Q}^v$ and $\mathbf{O}$ used in this work and defined in \textit{Appendix C}.
\section{Transformations between spherical and Cartesian coordinates for multipoles and fields}
Following the procedure of Ref.~\citenum{mun2020describing}, the induced Cartesian multipole moments are related to the scattering coefficients of \eqref{t-matrix-def} as
\begin{subequations}\label{pm-from-ab}
\begin{align}
\varepsilon^{-1} \, \mathbf{p}  &= \frac{\zeta_1}{\mathrm{i}k^3}\,\bar{\bar{F}}_1\,\mathbf{b}^{\rm e}_1 , \\
k\varepsilon^{-1} \, \mathbf{Q}^{\rm e}  &= \frac{\zeta_2}{\mathrm{i}k^3}\,\bar{\bar{F}}_2\,\mathbf{b}^{\rm e}_2 , \\
k^2\varepsilon^{-1} \, \mathbf{O}^{\rm e}  &= \frac{\zeta_3}{\mathrm{i}k^3}\,\bar{\bar{F}}_3\,\mathbf{b}^{\rm e}_3 , 
\end{align}
\end{subequations}
with
\begin{subequations}\label{transformation-matrices}
\begin{gather}
\bar{\bar{F}}_1 = \frac{1}{\sqrt{2}} \left[\begin{array}{ccc}
1 & 0 & -1 \\
-\mathrm{i} & 0 & -\mathrm{i} \\
0 & \sqrt{2} & 0 
\end{array}\right],\\
\bar{\bar{F}}_2 =\frac{1}{\sqrt{2}} \left[\begin{array}{ccccc}
-\mathrm{i} & 0 & 0 & 0 & \mathrm{i} \\
0 & -\mathrm{i} & 0 & -\mathrm{i} & 0 \\
0 & 0 & \sqrt{2} & 0 & 0  \\
0 & 1 & 0 & -1 & 0 \\
1 & 0 & 0 & 0 & 1 
\end{array}\right],\\
\bar{\bar{F}}_3=\frac{1}{\sqrt{2}} \left[
\arraycolsep=1mm
\begin{array}{ccccccc}
-\mathrm{i} & 0 & 0 & 0 & 0 & 0 & -\mathrm{i}  \\
0 & -\mathrm{i} & 0 & 0 & 0 & \mathrm{i} & 0 \\
0 & 0 & -\mathrm{i} & 0 & -\mathrm{i} & 0 & 0 \\
0 & 0 & 0 & \sqrt{2} & 0 & 0 & 0  \\
0 & 0 & 1 & 0 & -1 & 0 & 0 \\
0 & 1 & 0 & 0 & 0 & 1 & 0 \\
1 & 0 & 0 & 0 & 0 & 0 & -1 
\end{array}\right].
\end{gather}
\end{subequations}
Note that $\bar{\bar{F}}_j = (\bar{\bar{F}}_j^{-1})^\dagger$. 

The transformations between Cartesian and spherical magnetic multipole moments are performed similarly, according to the definitions of \eqref{t-matrix-def} and \eqref{a-matrix-def}. Moreover, the local or incident fields on the scatterer are related to the incidence coefficients of \eqref{t-matrix-def} as
\begin{subequations}\label{EH-from-pq}
\begin{align}
k^{-j+1}\mathbf{E}_j =& \frac{1}{\zeta_j}\,\bar{\bar{F}}_j\,\mathbf{q}^{\rm e}_j,
\\
k^{-j+1}\mathrm{i}\eta\mathbf{H}_j =& \frac{1}{\zeta_j}\,\bar{\bar{F}}_j\,\mathbf{q}^{\rm m}_j.
\end{align}
\end{subequations}
Herein, for the sake of simplicity, normalized incident field amplitudes are employed.

The induced multipole moments of a particle in the Cartesian coordinates
when illuminated by an incident wave can be calculated as a function of the induced currents, as depicted in Fig.~\ref{fig1}a. Specifically, the multipoles in Cartesian coordinate up to the quadrupolar order can be calculated via \cite{alaee2018electromagnetic}, 
\begin{subequations}\label{multipole-cart-calc}
\begin{flalign} 
&p_{\alpha} = -\frac{1}{{\rm i}\,\omega} \nonumber \Bigg\{\frac{k^2}{2}\int_V\left[3(\mathbf{r}\cdot\mathbf{J})r_\alpha - r^2J_\alpha\right]\frac{j_2(kr)}{(kr)^2}\,{\rm d}V &\\& \qquad \quad  + \int_{V}J_\alpha\,j_0(kr)\,{\rm d}V \Bigg\}, & \\ & m_{\alpha} = \frac{3}{2}\int_V\left(\mathbf{r}\times\mathbf{J}\right)_{\alpha}\frac{j_1(kr)}{kr}\,{\rm d}V, &
\end{flalign}
\begin{flalign} 
&Q^{\rm e}_{\alpha\beta} =
-\frac{3}{{\rm i}\,\omega} \times  &\\ & \Bigg\{\int_V{\rm d}V  \left[3\left(r_{\beta}J_{\alpha}+r_{\alpha}J_{\beta}\right) - 2\left(\mathbf{r}\cdot\mathbf{J}\right)\delta_{\alpha\beta}\right]\frac{j_1(kr)}{kr}+ &\nonumber \\&
 \frac{2j_3(kr)}{kr}\left[\frac{5 r_{\alpha}r_{\beta}\left(\mathbf{r}\cdot\mathbf{J}\right)}{r^2} - {r_{\alpha}J_{\beta} - r_{\beta}J_{\alpha}} -{ \mathbf{r}\cdot\mathbf{J}\delta_{\alpha\beta}}\right]\Bigg\}, \nonumber &
\end{flalign} 
\begin{equation}
Q^{\rm m}_{\alpha\beta}\hspace{-0.4mm} = \hspace{-0.4mm}15\hspace{-0.4mm}\int_V\hspace{-0.4mm}\left[r_{\alpha}\left(\mathbf{r}\hspace{-0.4mm}\times\hspace{-0.4mm}\mathbf{J}\right)_{\beta}  \hspace{-0.4mm}+\hspace{-0.4mm} r_{\beta}\left(\mathbf{r}\hspace{-0.4mm}\times\hspace{-0.4mm}\mathbf{J}\right)_{\alpha}\right]\hspace{-0.4mm}\frac{j_2(kr)}{(kr)^2}{\rm d}V,
\end{equation}
\end{subequations}
\noindent where $\{\alpha,\beta\} = \{x,y,z\}$. The connection between the multipole moments defined above in Cartesian coordinates and the multipole moments vectors defined in \eqref{fields-def-1} is further elaborated in the Supplementary Material.
%
\section{Multipole-to-field translation tensor $\bar{\bar{W}}$ and coordinates transformation tensor $\bar{\bar{R}}$}\label{subsection:W}
In \eqref{spherical-general-1}, the $\bar{\bar{W}}_j$ is defined as
\begin{equation}\label{W-matrices-a}
\bar{\bar{W_j}}\left(\theta,\phi\right) = \left[\begin{array}{c}
 \mathbf{W}_j \quad  \mathbf{W}^{\prime}_j\\
\mathrm{i}\, \mathbf{W}^{\prime}_j \quad \mathrm{i}\, \mathbf{W}_j
\end{array} \right].
\end{equation}
The elements of the tensor $\bar{\bar{W}}$ for each multipolar order, required for the scattered field calculation in \eqref{spherical-general-1} and \eqref{scat-field-total-2D-octupole-cartesian-matrix-1}, are given as in 
\begin{subequations}\label{W-matrices-b}
\begin{gather}
\mathbf{W}_{1} = \left[\begin{array}{c}e^{-\mathrm{i}\phi}{\rm cos}\theta \\ -\sqrt{2}{\rm sin}\theta \\ -e^{\mathrm{i}\phi}{\rm cos}\theta \end{array} 
\right]^T\hspace{-1mm},
\qquad \mathbf{W}^{\prime}_{1} = \left[\begin{array}{c}-e^{-\mathrm{i}\phi}\\ 0 \\ -e^{\mathrm{i}\phi}  \end{array}
\right]^T\hspace{-1mm},
\\
\mathbf{W}_{2} = \left[\begin{array}{c}\frac{1}{2}
e^{-\mathrm{i}2\phi}{\rm sin}2\theta \\ 
e^{-\mathrm{i}\phi}{\rm cos}2\theta \\ - \sqrt{\frac{3}{2}}{\rm sin}2\theta  \\
-e^{\mathrm{i}\phi}{\rm cos}2\theta   \\ \frac{1}{2}
e^{\mathrm{i}2\phi}{\rm sin}2\theta
\end{array}
\right]^T\hspace{-2.8mm},\,\,\,
\mathbf{W}^{\prime}_{2} = \left[\begin{array}{c}
-
e^{-\mathrm{i}2\phi}{\rm sin}\theta \\ -
e^{-\mathrm{i}\phi}{\rm cos}\theta \\ 0 \\ -
e^{\mathrm{i}\phi}{\rm cos}\theta  \\
e^{\mathrm{i}2\phi}{\rm sin}\theta 
\end{array}
\right]^T\hspace{-2.8mm},
\\
\mathbf{W}_{3} = 
\left[\begin{array}{c}
      \frac{\sqrt{15}}{4} e^{-i3\phi}{\rm sin}^2\theta\,{\rm cos}\theta \\
      \frac{1}{4}\sqrt{\frac{5}{2}} e^{-i2\phi}{\rm sin}\theta\,\left(3{\rm cos}2\theta + 1\right) \\
      \frac{1}{16} e^{-\mathrm{i}\phi}\left({\rm cos}\theta + 15{\rm cos}3\theta\right) \\
      -\frac{\sqrt{3}}{8} \left({\rm sin}\theta + 5{\rm sin}3\theta\right)  \\
      -\frac{1}{16} e^{-\mathrm{i}\phi}\left({\rm cos}\theta + 15{\rm cos}3\theta\right) \\
      \frac{1}{4}\sqrt{\frac{5}{2}} e^{-\mathrm{i}2\phi}{\rm sin}\theta\,\left(3{\rm cos}2\theta + 1\right)   \\
      -\frac{\sqrt{15}}{4} e^{-\mathrm{i}3\phi}{\rm sin}^2\theta\,{\rm cos}\theta \end{array}\right]^{T},
\\
\mathbf{W}^{\prime}_{3} = 
\left[\begin{array}{c}
      -\frac{\sqrt{15}}{4} e^{-\mathrm{i}3\phi}{\rm sin}^2\theta \\
      -\frac{1}{2}\sqrt{\frac{5}{2}} e^{-\mathrm{i}2\phi}{\rm sin}2\theta \\
      -\frac{1}{4} e^{-\mathrm{i}\phi}\left(5{\rm cos}^2\theta - 1\right) \\
      0  \\
      -\frac{1}{4} e^{\mathrm{i}\phi}\left(5{\rm cos}^2\theta - 1\right) \\
      \frac{1}{2}\sqrt{\frac{5}{2}} e^{\mathrm{i}2\phi}{\rm sin}2\theta   \\
      -\frac{\sqrt{15}}{4} e^{\mathrm{i}3\phi}{\rm sin}^2\theta 
\end{array}\right]^{T},
\end{gather}
\end{subequations}
where $\theta$ is the polar angle of the wavevector of the respective diffraction order.
The vectors $\hat{\bm{\theta}}$ and $\hat{\bm{\phi}}$ can be expressed as a function of the vectors $\hat{\mathbf{x}}$, $\hat{\mathbf{y}}$, and $\hat{\mathbf{z}}$ using the tensor $\bar{\bar{R}}$ as,
\begin{equation}
\left[\hspace{-0.5mm}\begin{array}{c}
\hat{\bm{\theta}} \\
\hat{\bm{\phi}} 
\end{array}\hspace{-0.5mm}
\right]\hspace{-0.5mm} = \hspace{-0.5mm}
\bar{\bar{R}} \hspace{-0.5mm}\left[\hspace{-0.5mm}\begin{array}{c}
\hat{\mathbf{x}} \\
\hat{\mathbf{y}} \\ 
\hat{\mathbf{z}}
\end{array}\hspace{-0.5mm}
\right] \hspace{-0.5mm}
= \hspace{-0.5mm}
\left[\hspace{-0.5mm}\arraycolsep=1mm
\begin{array}{ccc}
{\rm cos}\theta{\rm cos}\phi & {\rm cos}\theta{\rm sin}\phi & {\rm sin}\theta \\
{\rm sin}\phi & -{\rm cos}\phi & 0
\end{array}\hspace{-0.5mm}
\right]\hspace{-0.5mm}\left[\hspace{-0.5mm}\begin{array}{c}
\hat{\mathbf{x}} \\
\hat{\mathbf{y}} \\ 
\hat{\mathbf{z}}
\end{array}\hspace{-0.5mm}
\right].
\end{equation}
%
%
\section{Transformations between spherical and Cartesian coordinates for scattering and extinction cross-sections}
The scattering cross-section, expressed in spherical coordinates, is defined as \cite{mishchenko2002scattering}
\begin{equation}\label{scattering-cross-section-def-oct}
\sigma^{\rm s}_{\rm sca} = \frac{1}{|E_0|^2 k^2}\sum_{j=1}^{3}\left(\,|\mathbf{b}^{\rm e}_j|^2 +|\mathbf{b}^{\rm m}_j|^2\,\right),
\end{equation}
where $|\cdot|^2$ is the 2-norm of a vector or a matrix. Herein, the multipole order is limited to the octupole or $j=3$, as it is the scope of this work.

To express $\sigma_{\rm sca}$ as a function of the multipole moments in Cartesian coordinates, transformations  \eqref{pm-from-ab} and  \eqref{transformation-matrices} are employed. Therefore,
\eqref{scattering-cross-section-def-oct} is rewritten with the multipole moments expressed in Cartesian coordinates, defined in Appendix A. However, for the specific transformation matrices $|[F_j]^{-1}|^2 = 1$, \eqref{scattering-cross-section-def-oct} can be manipulated further if $\zeta_j$ are, also, calculated. Therefore,
%
\begin{align}\label{scattering-cross-section-def-cart-2}
\sigma&^{\rm c}_{\rm sca} = \nonumber \,\frac{1}{|E_0|^2}\left(\frac{k^4\varepsilon^{-2}}{3!\pi}\,\left|\mathbf{p}\right|^2 + \frac{k^4\eta^2}{3!\pi}\,\left|\mathbf{m}\right|^2
 + \frac{k^6\varepsilon^{-2}}{5!\pi}\,\left|\mathbf{Q}^{\rm e}\right|^2  \right.\\ &\left.+ \frac{k^6\eta^2}{5!\pi}\,\left|\mathbf{Q}^{\rm m}\right|^2 
 +  \frac{k^{8}\varepsilon^{-2}}{7!\pi}\,\left|\mathbf{O}^{\rm e}\right|^2 + \frac{k^{8}\eta^2}{7!\pi}\,\left|\mathbf{O}^{\rm m}\right|^2\right).
\end{align}

Similarly, the extinction cross-section can be calculated as a function of the multipole moments and the fields represented in Cartesian coordinates. The extinction cross-section, herein up to $j=3$ order, is defined as \cite{mishchenko2002scattering},
\begin{align}\label{extinction-cross-section-def-2}
\sigma&_{\rm ext}^{\rm s} = \frac{-1}{|E_0|^2k^2}{\Re} \Bigg\{\hspace{-0mm}\sum_{j=1}^{3}\left[\mathbf{q}^{\rm e}_j\cdot\mathbf{b}^{\rm e,*}_j + \mathbf{q}^{\rm m}_j\cdot\mathbf{b}^{\rm m,*}_j \right]
\hspace{-0mm}\Bigg\}\nonumber \\
&= \frac{-1}{|E_0|^2k^2}{\Re} \Bigg\{\hspace{-0mm}\sum_{j=1}^{3}\left[\,\mathbf{q}^{\rm e,T}_j\,\mathbf{b}^{\rm e,*}_j + \mathbf{q}^{\rm m,T}_j\,\mathbf{b}^{\rm m,*}_j \right]\hspace{-0mm}
\Bigg\},
\end{align}
\noindent where the superscript $^*$ denotes the conjugate operation and the superscript $^{\rm T}$ denotes the transpose operation. If the transformations  \eqref{pm-from-ab},  \eqref{transformation-matrices}, and \eqref{EH-from-pq} are employed, and after the identity $\bar{\bar{F}}_j^{-1,{\rm T}}\bar{\bar{F}}_j^{-1,{*}} = \bar{\bar{I}}$ is utilized, \eqref{extinction-cross-section-def-2} finally arrives to,
\begin{eqnarray}\label{extinction-cross-section-cart-2}
\sigma_{\rm ext}^{\rm c} &=& -\frac{k}{|E_0|^2}\,{\Im}\left(
\frac{1}{\varepsilon}\mathbf{E}_1\cdot\mathbf{p}^* + \eta^2\mathbf{H}_1\cdot\mathbf{m}^* 
+ \frac{1}{\varepsilon}\mathbf{E}_2\cdot\mathbf{Q}^{\mathrm{e},*}
+\right.\nonumber\\ &{\eta^2}&\left.\mathbf{H}_2\cdot\mathbf{Q}^{\mathrm{m},*} 
+ \frac{1}{\varepsilon}\mathbf{E}_3\cdot\mathbf{O}^{\mathrm{e},*}
+ {\eta^2}\mathbf{H}_3\cdot\mathbf{O}^{\mathrm{m},*}\right).
\end{eqnarray}
The absorption cross-section can be, afterwards, calculated from \eqref{scattering-cross-section-def-cart-2} and \eqref{extinction-cross-section-cart-2}, as $\sigma_{\rm abs} = \sigma_{\rm ext} - \sigma_{\rm sca}$.

More information about deriving the formulas of this Appendix can be found in \textit{Supp. Info. V.}.

\begin{acknowledgements}
 We acknowledge support by the German Research Foundation through the priority program SPP 1839 Tailored Disorder (CR 3640/7-2 under project number 278747906) and Germany’s Excellence Strategy via the Excellence Cluster 3D Matter Made to Order (EXC-2082/1 - 390761711) and by the Carl Zeiss Foundation through the “Carl‐Zeiss‐Focus@HEiKA". A. R.\ and A. G. L. acknowledge support from the Karlsruhe School of Optics and Photonics (KSOP). R. A. acknowledges the support of the Alexander von Humboldt Foundation through the Feodor Lynen (Return) Research Fellowship. T. K. acknowledges the support of the Alexander von Humboldt Foundation through the Humboldt Research Fellowship for postdoctoral researchers. A. G. L. acknowledges support from the Max Planck School of Photonics, which is supported by BMBF, Max Planck Society, and Fraunhofer Society. The authors would like to thank Xavier Garcia-Santiago, Ivan Fernandez-Corbaton, Mun Jungho, and Rho Junsuko for the fruitful discussions. We are grateful to the company JCMwave for their free provision of the FEM Maxwell solver JCMsuite.
\end{acknowledgements}


%
%
%
%

\pagebreak
\clearpage
\widetext
\begin{center}
\textbf{\Large Supplementary Information}
\end{center}
\setcounter{section}{0}
\setcounter{equation}{0}
\setcounter{figure}{0}
\setcounter{table}{0}
\setcounter{page}{1}
\makeatletter
\renewcommand{\theequation}{S\arabic{equation}}
\renewcommand{\thefigure}{S\arabic{figure}}
\renewcommand{\thepage}{S\arabic{page}}
\renewcommand{\bibnumfmt}[1]{[S#1]}

\section{\quad Calculation of the scattered field from a 2D lattice of particles for oblique incidence}
\label{subsection:A}
Assume a particle placed inside an infinite, homogeneous, and isotropic medium. An electromagnetic field illuminates the particle. The total electric field in the spatial domain outside and around the particle at an angular frequency $\omega$ consists of the incident and scattered field, or,
\begin{equation}\label{total-e-field}
\mathbf{E}^{\rm tot}_{\rm ind}(\mathbf{r}) = \mathbf{E}^{\rm inc}_{\rm ind}(\mathbf{r}) + \mathbf{E}^{\rm sca}_{\rm ind}(\mathbf{r}).
\end{equation}
The fields above can be expanded using vector spherical harmonics (VSH) \cite{mishchenko2002scattering,fruhnert2017computing}. Thus, the scattered field in \eqref{total-e-field} can be expressed as,
\begin{equation}\label{scat-field-gen}
\mathbf{E}^{\rm sca}_{\rm ind}(\mathbf{r}) = \sum_{j=1}^{\infty} \sum_{m=-j}^{j} b^{\rm e}_{jm}\mathbf{N}^{(3)}_{jm}(k\mathbf{r}) + b^{\rm m}_{jm}\mathbf{M}^{(3)}_{jm}(k\mathbf{r}),
\end{equation}
with $b^{\rm e/m}_{jm}$, being the electric/magnetic scattered field expansion coefficients. Moreover, 

\begin{equation}\label{VSW-Hankel-sup}
\mathbf{M}^{(3)}_{jm} = h_j(kr)\mathbf{X}_{jm}(kr,\theta,\phi), \qquad \mathbf{N}^{(3)}_{jm} = \frac{1}{k}\nabla \times \mathbf{M}^{(3)}_{jm},
\end{equation}

are the outgoing VSH, with $h_j(x)$ being the spherical Hankel function, and $\mathbf{X}_{jm}(kr,\theta,\phi)$, the vector spherical harmonic (VSH) calculated as
\begin{subequations}\label{vsh-X}
\begin{equation}
\mathbf{X}_{jm}(kr,\theta,\phi) = \gamma_{jm}  [\mathrm{i}\pi_{jm}({\rm cos}\theta)\bm{\hat{\theta}} -\tau_{jm}({\rm cos}\theta)\bm{\hat{\phi}}]e^{\mathrm{i}m\phi},
\end{equation}
\text{with}
\begin{equation}
\pi_{jm}({\rm cos}\theta)= \frac{m}{{\rm sin}\theta} P^{m}_j({\rm cos}\theta), \qquad \tau_{jm}({\rm cos}\theta)= \frac{\partial}{\partial\theta} P^{m}_j({\rm cos}\theta),\qquad \gamma_{jm}=  \sqrt{\frac{(2j+1)}{4\pi j(j+1)}} \sqrt{\frac{(j-m)!}{(j+m)!}},
\end{equation}
\end{subequations}
where $P^{m}_j(x)$ is the associated Legendre polynomial. 

Let us now consider an infinite 2D periodic distribution of identical particles in the $z=0$ plane. The 2D lattice can be described by two base vectors $\mathbf{u}_1$ and $\mathbf{u}_2$ parallel to a $xy$ plane, hence, the position of every particle in the lattice is derived as $\mathbf{R}=n_1\mathbf{u}_1+n_2\mathbf{u}_2$, with $n_1, n_2 \in \mathbb{Z}$. Additionally, the area of the unit cell is $A = (\mathbf{u}_1 \times \mathbf{u}_2)\cdot \hat{\mathbf{z}}$. Therefore, the total scattering contribution of the 2D lattice by the summation of the contribution of each particle is calculated via \eqref{VSW-Hankel-sup} as \cite{antonakakis2014gratings}
\begin{equation}\label{scat-field-total-2D}
\mathbf{E}^{\rm sca}_{\rm s}(\mathbf{r}) = \sum_{\mathbf{R}}\sum_{j=1}^{\infty} \sum_{m=-j}^{j}\left\{ b^{\rm e}_{jm}\mathbf{N}^{(3)}_{jm}\big(k(\mathbf{r} - \mathbf{R})\big) 
+ b^{\rm m}_{jm}\mathbf{M}^{(3)}_{jm}\big(k(\mathbf{r} - \mathbf{R}) \big)\right\}e^{\mathrm{i}\mathbf{k}_{\parallel}\cdot\mathbf{R}},
\end{equation}
where $\mathbf{k}_{\parallel}$ is the wavevector component parallel to the plane of the lattice. Subsequently, the calculation is reduced to the summations of the VSH over the particle positions, which are in turn calculated by applying the Poisson summation formula as \cite{antonakakis2014gratings}
\begin{subequations}\label{vsh-X-reciprocal-space}
\begin{equation}
\sum_{\mathbf{R}}\mathbf{M}^{(3)}_{jm}\big(k(\mathbf{r} - \mathbf{R}) \big)e^{{\rm i}\mathbf{k}_{\parallel}\cdot\mathbf{R}}=\frac{2\pi {\rm i}^{-j}}{Ak}\sum_{\mathbf{G}} \frac{ \mathbf{X}_{jm}(\mathbf{k}^{\pm}_{\rm G})}{|k^{\pm}_{{\mathbf{G}},z}|} e^{\mathrm{i}\mathbf{k}^{\pm}_{\mathbf{G}}\cdot\mathbf{r}},
\end{equation}
\begin{equation}
\sum_{\mathbf{R}}\mathbf{N}^{(3)}_{jm}\big(k(\mathbf{r} - \mathbf{R}) \big)e^{\mathrm{i}\mathbf{k}_{\parallel}\cdot\mathbf{R}}=\frac{2\pi {\rm i}^{1-j}}{Ak}\sum_{\mathbf{G}} \frac{ {\hat{\mathbf{r}}}\times \mathbf{X}_{jm}(\mathbf{k}_\mathbf{G}^{\pm})}{|k^{\pm}_{\mathbf{G},z}|} e^{{\rm i}\mathbf{k}_{\mathbf{G}}^{\pm}\cdot\mathbf{r}},
\end{equation}
\text{with}
\begin{equation}
 \hat{\mathbf{r}}\times\mathbf{X}_{jm}(kr,\theta,\phi) =   \gamma_{jm}\,  [\tau_{jm}({\rm cos}\theta)\hat{\bm{\theta}} +{\rm i}\pi_{jm}({\rm cos}\theta)\hat{\bm{\phi}}]\,e^{{\rm i}m\phi} =  \mathbf{Z}_{jm}(kr,\theta,\phi),
\end{equation}
\text{\rm and}
\begin{equation}
\mathbf{k}^{\pm}_{\mathbf{G}} = \mathbf{k}_{\parallel} + \mathbf{G} \pm \hat{\mathbf{z}} \sqrt{k^2 - (\mathbf{k}_{\parallel} + \mathbf{G})^2 } = \mathbf{k}_{\parallel} + \mathbf{G} \pm  k_{\mathbf{G},z}\,\hat{\mathbf{z}},
\end{equation}
\end{subequations}
where $\mathbf{G} = n_1 \mathbf{u}'_1 + n_2 \mathbf{u}'_2$, with $\mathbf{u}'_1 = \frac{2\pi}{A} (\mathbf{u}_2 \times \hat{\mathbf{z}})$ and $\mathbf{u}'_2 = \frac{2\pi}{A} (\hat{\mathbf{z}} \times \mathbf{u}_1)$, is the reciprocal lattice vector, and $A=(\mathbf{u}_1 \times \mathbf{u}_2)\cdot \hat{\mathbf{z}}$. After substituting  of \eqref{vsh-X} in \eqref{scat-field-total-2D}, one arrives to
\begin{equation}\label{scat-field-total-2D-vsh-x-z}
\mathbf{E}^{\rm sca}_s(\mathbf{r}) = \sum_{j=1}^{\infty} \sum_{m=-j}^{j}\frac{2\pi {\rm i}^{-j}}{Ak}\sum_{\mathbf{G}}
\left[
b^{\rm e}_{jm} {\rm i} \mathbf{Z}_{jm}(\mathbf{k}_\mathbf{G}^{\pm}) + b^{\rm m}_{jm}  \mathbf{X}_{jm}(\mathbf{k}_\mathbf{G}^{\pm}) \right]
\frac{e^{{\rm i}\mathbf{k}_{\mathbf{G}}^{\pm}\cdot\mathbf{r}}}{|k^{\pm}_{\mathbf{G},z}|}.
\end{equation}
The formula \eqref{scat-field-total-2D-vsh-x-z} provides the total scattered field from 2D particle lattice.
\subsection{Propagating diffraction orders scattered from a 2D array}
Some simplifications will be gradually inserted into the general model to identify the propagating diffraction orders from the array. 
Considering the case of oblique incidence or $\mathbf{k}_{\parallel} = k_{\mathrm{inc},x}\,\mathbf{\hat{x}} + k_{\mathrm{inc},y}\,\mathbf{\hat{y}} \neq 0$, for a propagating mode, it must hold that, $k^2>|\mathbf{k}_{\parallel} + \mathbf{G}|^2$. Under the assumption of a {\bf{rectangular lattice}}, with  $|\mathbf{u}_1|= \Lambda_1$ and $|\mathbf{u}_2| = \Lambda_2$, then, the reciprocal lattice vector is calculated as $\mathbf{G} = \frac{2\pi n_1}{\Lambda_1} \hat{\mathbf{x}}  + \frac{2\pi n_2}{\Lambda_2}\hat{\mathbf{y}}$, while $A=\Lambda_1 \Lambda_2$. Hence, the condition for propagating modes generated from a rectangular lattice turns to 
\begin{equation}\label{propagated-modes-rect}
k^2> \left(k^{\mathrm{inc}}_x+ \frac{2\pi n_1}{\Lambda_1}\right)^2 + \left(k^{\mathrm{inc}}_y+ \frac{2\pi n_2}{\Lambda_2}\right)^2.
\end{equation}
It can be deduced from \eqref{propagated-modes-rect}, that $(n_1,n_2) = (0,0)$, or the zeroth diffraction order, is the only propagating mode under the condition of $\{\Lambda_1,\Lambda_2\} < \lambda/2$.
Thus, for each propagating mode and after further manipulation of \eqref{scat-field-total-2D-vsh-x-z}, one arrives to
\begin{equation}\label{scat-field-total-2D-gen}
\begin{split}
\mathbf{E}^{\rm sca}_{\mathbf{G},{\rm s}}(\mathbf{r}) = \frac{2\pi }{A k}\frac{e^{{\rm i}\mathbf{k}_{\mathbf{G}}^{\pm}\cdot\mathbf{r}}}{|k^{\pm}_{\mathbf{G},z}|}&\sum_{j=1}^{\infty} \sum_{m=-j}^{j}{\rm i}^{-j}\sqrt{\frac{2j+1}{4\pi j (j+1)}}\sqrt{\frac{(j-m)!}{(j+m)!}}e^{{\rm i} m \phi} \cdot\\
&\cdot\left[{\rm i}\left( 
 \mathbf{\tau}_{jm}({\rm cos}\theta)b^{\rm e}_{jm}  +  \mathbf{\pi}_{jm}({\rm cos}\theta) b^{\rm m}_{jm}   \right) \bm{\hat{\theta}}  
 - \left( 
 \mathbf{\pi}_{jm}({\rm cos}\theta)b^{\rm e}_{jm}  +  \mathbf{\tau}_{jm}({\rm cos}\theta) b^{\rm m}_{jm}   \right) \bm{\hat{\phi}}\right],
\end{split}
\end{equation}
where 
\begin{subequations}\label{modes-wavevectors-rectangular}
\begin{equation}
\mathbf{k}^{\pm}_{\mathbf{G}} =  k_{\mathbf{G},x}\mathbf{\hat{x}} + k_{\mathbf{G},y}\mathbf{\hat{y}} + k^{\pm}_{\mathbf{G},z}\, \hat{\mathbf{z}},
\end{equation}
\begin{equation}
k_{\mathbf{G},x} = k^{\mathrm{inc}}_x+ \frac{2\pi n_1}{\Lambda_1},\,\, k_{\mathbf{G},y} = k^{\mathrm{inc}}_y+ \frac{2\pi n_2}{\Lambda_2}, \,\, k^{\pm}_{\mathbf{G},z} =  \pm \sqrt{k^2 - \left(k^{\mathrm{inc}}_x+ \frac{2\pi n_1}{\Lambda_1}\right)^2 -  \left(k^{\mathrm{inc}}_y+ \frac{2\pi n_2}{\Lambda_2}\right)^2}=k\cos{\theta},
\end{equation}
\begin{equation}
{\rm cos}\theta = \frac{k^{\pm}_{{\mathbf{G}},z}}{|\mathbf{k}^{\pm}_{\mathbf{G}}|} \quad \text{and} \quad \phi = {\rm arctan}\left(\frac{k_{\mathbf{G},y}}{k_{\mathbf{G},x}}\right).
\end{equation}
\end{subequations}

For the common sub-case of the {\bf{square lattice}}, or $\Lambda_1 = \Lambda_2 = \Lambda$ and $A=\Lambda^2$, we can also write $\mathbf{k}_{\bf G}$ from \eqref{modes-wavevectors-rectangular} in the following form 
\begin{subequations}\label{modes-wavevectors-square}
\begin{equation}
\mathbf{k}^{\pm}_{\mathbf{G}} =  k_{\mathbf{G},x}\mathbf{\hat{x}} + k_{\mathbf{G},y}\mathbf{\hat{y}} +k^{\pm}_{\mathbf{G},z} \hat{\mathbf{z}},
\end{equation}
\text{with}
\begin{equation}
k_{\mathbf{G},x} = \frac{2\pi}{\lambda} \left( \sin{\theta_\mathrm{inc}}\cos{\phi_\mathrm{inc}} +\frac{\lambda}{\Lambda} n_1\right),\quad k_{\mathbf{G},y} =\frac{2\pi}{\lambda}  \left( \sin{\theta_\mathrm{inc}}\sin{\phi_\mathrm{inc}} +\frac{\lambda}{\Lambda} n_2\right),\\ 
\end{equation}
\begin{equation}
\centering
\qquad\qquad\,\, k^{\pm}_{\mathbf{G},z} =\pm\frac{2\pi}{\lambda} \sqrt{1 - \left( \sin{\theta_\mathrm{inc}}\cos{\phi_\mathrm{inc}} +\frac{\lambda}{\Lambda} n_1 \right)^2 -  \left(\sin{\theta_\mathrm{inc}}\sin{\phi_\mathrm{inc}} +\frac{\lambda}{\Lambda} n_2\right)^2} = k \cos{\theta}.
\end{equation}
\end{subequations}
Due to the symmetries of the square lattice, an incident wave with only $k_{\mathrm{inc},x}$ can be considered in the analysis, or $\theta \neq0$ and $n_2=\phi_{\rm inc}=0$. Hence, the diffraction angles for directions orders in transmission (also called here forward) or reflection (also called here backward) can be calculated via \eqref{modes-wavevectors-rectangular} using
\begin{align}\label{Diffraction_Angles-square-1}
   \theta^{\,\rm forward}_{\rm diffraction}= \arcsin{\left(\,|\sin{\theta_{\rm inc}}+\frac{\lambda}{\Lambda}n_1|\,\right)}, \qquad   \theta^{\rm backward}_{\rm diffraction}=\pi- \arcsin{\left(\,|\sin{\theta_{\rm inc}}+\frac{\lambda}{\Lambda}n_1|\,\right)},
\end{align}
while in general form, the polar angle of the wavevector of each diffraction order can be calculated as
\begin{subequations}\label{Diffraction_Angles-square-2}
\begin{equation}
   \theta_{\rm diffraction} = \arccos{\left[\pm\sqrt{1-(\sin{\theta_{\rm inc}}\cos{\phi_{\rm inc}}+\frac{\lambda}{\Lambda}n_1)^2-(\sin{\theta_{\rm inc}}\sin{\phi_{\rm inc}}+\frac{\lambda}{\Lambda}n_2)^2}\,\right]}, 
\end{equation}
\begin{equation}
   \phi_{\rm diffraction} =\arctan{\left(\frac{\sin{\theta_{\rm inc}}\sin{\phi_{\rm inc}}+\frac{\lambda}{\Lambda}n_2}{\sin{\theta_{\rm inc}}\cos{\phi_{\rm inc}}+\frac{\lambda}{\Lambda}n_1}\right)}.
\end{equation}
\end{subequations}

If now the {\bf{normal incidence on a square lattice}} is considered, or $\theta_{\rm inc}=0$, \eqref{Diffraction_Angles-square-1} and \eqref{Diffraction_Angles-square-2} simplify to
\begin{subequations}\label{Diffraction_Angles-square-normal}
\begin{align}
   \theta_{\rm diffraction} =\arccos{\left[\pm\sqrt{1-\left(\frac{\lambda}{\Lambda}\right)^2\left(n_1^2+n_2^2\right)}\,\right]},& \qquad \theta^{\,\rm forward}_{\rm diffraction} = \arcsin\left({\frac{\lambda}{\Lambda}\sqrt{n_1^2+n_2^2}}\right), \\
\theta^{\,\rm backward}_{\rm diffraction} = \pi-\arcsin\left({\frac{\lambda}{\Lambda}\sqrt{n_1^2+n_2^2}}\right), & \qquad
   \phi_{\rm diffraction} = \arctan{\left(\frac{n_2}{n_1}\right)}.
\end{align}
\end{subequations}
Note that in $\pm$, the plus sign is for the forward direction and the minus sign is for the backward direction. 
\subsection{Scattering multipolar contribution}
\paragraph{Dipole:}
Let us begin our calculations from \eqref{scat-field-total-2D-gen} starting with the {\bf{dipole case}}, or for $j=1$. Hence, for each propagating diffraction order, it holds that
\begin{equation}\label{scat-field-total-2D-dipole-0}
\begin{split}
\mathbf{E}_{\mathbf{G},{\rm s}}^{\rm sca, d}(\mathbf{r}) = \frac{{\rm i}^{-1}\pi }{Ak}&\sqrt{\frac{3}{2\pi}}\frac{e^{{\rm i}\mathbf{k}^{\pm}_{\mathbf{G}}\cdot\mathbf{r}}}{|k^{\pm}_{\mathbf{G},z}|} \sum_{m=-1}^{+1}\sqrt{\frac{(1-m)!}{(1+m)!}}e^{{\rm i} m\phi} \cdot\\
&\cdot\left[{\rm i}\left( 
 \mathbf{\tau}_{1m}({\rm cos}\theta)b^{\rm e}_{1m}  +  \mathbf{\pi}_{1m}({\rm cos}\theta) b^{\rm m}_{1m}   \right) \bm{\hat{\theta}}  
 - \left( 
 \mathbf{\pi}_{1m}({\rm cos}\theta)b^{\rm e}_{1m}  +  \mathbf{\tau}_{1m}({\rm cos}\theta) b^{\rm m}_{1m}   \right) \bm{\hat{\phi}}\right],
\end{split}
\end{equation}
Hence, by calculating the summation in \eqref{scat-field-total-2D-dipole-0}, the scattered field from a dipole array at oblique incidence reads as
\begin{subequations}\label{scat-field-total-2D-dipole}
\begin{equation}
\mathbf{E}^{\rm sca,d}_{\mathbf{G},{\rm s}}(\mathbf{r}) = \frac{{\rm i}^{-1}}{Ak}\frac{\sqrt{3\pi}}{2} \frac{e^{{\rm i}\mathbf{k}^{\pm}_{\mathbf{G}}\cdot\mathbf{r}}}{|k^{\pm}_{\mathbf{G},z}|} \Big [
{\rm i}\left(\mathbf{W}_{1} \mathbf{b}^{\rm e}_1 + \mathbf{W}^{\prime}_{1}\mathbf{b}^{\rm m}_1\right) \hat{\bm{\theta}} - \left(\mathbf{W}^{\prime}_{1}\mathbf{b}^{\rm e}_1 + \mathbf{W}_{1}\mathbf{b}^{\rm m}_1\right) \hat{\bm{\phi}}\Big ],
\end{equation}
\text{\rm where}
\begin{equation}
\mathbf{W}_{1} = \left[e^{-{\rm i}\phi}{\rm cos}\theta \quad -\sqrt{2}{\rm sin}\theta \quad -e^{{\rm i}\phi}{\rm cos}\theta  
\right],
\qquad \mathbf{W}^{\prime}_{1} = \left[-e^{-{\rm i}\phi}\quad 0 \quad -e^{{\rm i}\phi}  
\right].
\end{equation}
\end{subequations}
From \eqref{scat-field-total-2D-dipole} one gets the scattered field from a dipole array expressed in the spherical coordinate system. However, for many applications and a better understanding of physical mechanisms, sometimes it is desirable to express lattice scattering in Cartesian coordinates. Let us, first, express \eqref{scat-field-total-2D-dipole} with Cartesian dipoles. 
If the modified transformations between spherical and Cartesian coordinates for multipoles \cite{mun2020describing} are applied  ($\rightarrow$ \textit{Appendix E - Main Text}),
they lead to
\begin{equation}\label{scat-field-total-2D-dipole-cartesian}
\begin{split}
\mathbf{E}^{\rm sca,d}_{\mathbf{G},{\rm s}}(\mathbf{r}) = \frac{{\rm i} k^2}{A} \frac{e^{{\rm i}\mathbf{k}^{\pm}_{\mathbf{G}}\cdot\mathbf{r}}}{|k^{\pm}_{\mathbf{G},z}|} \frac{\sqrt{3\pi}}{2\zeta_1} \Big[
{\rm i}^{1-1}\big(\mathbf{W}_{1}\bar{\bar{F}}_1^{\,-1}\mathbf{p}/\varepsilon &+ \mathbf{W}^{\prime}_{1}\bar{\bar{F}}_1^{\,-1}\,{\rm i}\eta\mathbf{m}\big)\, \hat{\bm{\theta}} \\ 
&- {\rm i}^{-1} \big(\mathbf{W}^{\prime}_{1}\bar{\bar{F}}_1^{\,-1}\mathbf{p}/\varepsilon + \mathbf{W}_{1}\bar{\bar{F}}_1^{\,-1}\,{\rm i}\eta\mathbf{m}\big)\, \hat{\bm{\phi}}\Big].
\end{split}
\end{equation}
Moreover, if a transformation to Cartesian coordinates is performed and the resulting  $\mathbf{E}^{\rm sca,d}_{\mathbf{G},{\rm c}}$ is broken down to its vector components for convenience after some algebra one arrives at the expression of
\begin{equation}\label{scat-field-total-2D-dipole-cartesian-matrix-v2}
\left[\begin{array}{c}
     {E}^{\rm \, sca, d}_{\mathbf{G}\,x}  \\
     {E}^{\rm \, sca, d}_{\mathbf{G}\,y} \\
     {E}^{\rm \, sca, d}_{\mathbf{G}\,z} 
\end{array}\right] = \frac{{\rm i} k^2\sqrt{\pi}}{2A} \frac{e^{{\rm i}\mathbf{k}^{\pm}_{\mathbf{G}}\cdot\mathbf{r}}}{|k^{\pm}_{\mathbf{G},z}|} \,\bar{\bar{S}}_1\,\left[\begin{array}{r}
      (\varepsilon\zeta^{-1})\,\mathbf{p}  \\
      {\rm i}\eta\zeta^{-1}\,\mathbf{m} 
\end{array}\right]=\frac{{\rm i} k\sqrt{\pi}}{2A} \frac{e^{\mathrm{i}\mathbf{k}^{\pm}_{\mathbf{G}}\cdot\mathbf{r}}}{|\cos{\theta}|}
\left[\begin{array}{cc}
     \mathbf{S}_{1}^{\rm xe} & \mathbf{S}_{1}^{\rm xm} \\
     \mathbf{S}_{1}^{\rm ye} & \mathbf{S}_{1}^{\rm ym} \\
     \mathbf{S}_{1}^{\rm ze} & \mathbf{S}_{1}^{\rm zm}
\end{array}\right]\,\left[\begin{array}{r}
      (\varepsilon\zeta^{-1})\,\mathbf{p}  \\
      {\rm i}\eta\zeta^{-1}\,\mathbf{m} 
\end{array}\right],
\end{equation}
with
\begin{subequations}\label{scat-field-total-2D-dipole-cartesian-5}
\begin{equation}
\mathbf{S}_{1}^{\rm xe} = \sqrt{3}\left(\mathbf{W}_{1}{\rm cos}\theta{\rm cos}\phi - {\rm i}\mathbf{W}^{\prime}_{1}{\rm sin}\phi\right)\bar{\bar{F}}_1^{\,-1}, 
\quad
\mathbf{S}_{1}^{\rm xm} =\sqrt{3} \left(\mathbf{W}^{\prime}_{1}{\rm cos}\theta{\rm cos}\phi  - {\rm i} \mathbf{W}_{1}{\rm sin}\phi\right)\bar{\bar{F}}_1^{\,-1}, 
\end{equation}
\begin{equation}
\mathbf{S}_{1}^{\rm ye} = \sqrt{3}\left(\mathbf{W}_{1}{\rm cos}\theta{\rm sin}\phi + {\rm i}\mathbf{W}^{\prime}_{1}{\rm cos}\phi\right)\bar{\bar{F}}_1^{\,-1}, 
\quad
\mathbf{S}_{1}^{\rm ym} = \sqrt{3}\left(\mathbf{W}^{\prime}_{1}{\rm cos}\theta{\rm sin}\phi + {\rm i}\mathbf{W}_{1}{\rm cos}\phi\right)\bar{\bar{F}}_1^{\,-1}, 
\end{equation}
\begin{equation}
\mathbf{S}_{1}^{\rm ze}= -\sqrt{3}\,\mathbf{W}_{1}{\rm sin}\theta\,\bar{\bar{F}}_1^{\,-1},
\quad
\mathbf{S}_{1}^{\rm zm} = -\sqrt{3}\,\mathbf{W}^{\prime}_{1}{\rm sin}\theta\,\bar{\bar{F}}_1^{\,-1},
\end{equation}
\end{subequations}
or, 
\begin{subequations}\label{scat-field-total-2D-dipole-cartesian-5-b}
\begin{equation}
\mathbf{S}_{1}^{\rm xe} = \sqrt{3}\,\left[     
{\rm sin}^2\phi + {\rm cos}^2\theta{\rm cos}^2\phi
\quad
-\frac{1}{2}{\rm sin}^2\theta{\rm sin}(2\phi)
\quad
-\frac{1}{2}{\rm sin}(2\theta){\rm cos}\phi   \right],
\end{equation}
\begin{equation}
\mathbf{S}_{1}^{\rm xm} = \sqrt{3}\,\Big[ 
0
\quad
-{\rm i}\,{\rm cos}\theta
\quad
{\rm i}\, {\rm sin}\theta{\rm sin}\phi   \Big],
\end{equation}
\begin{equation}
\mathbf{S}_{1}^{\rm ye} = \sqrt{3}\,\left[     
-\frac{1}{2}{\rm sin}^2\theta{\rm sin}(2\phi)
\quad 
{\rm cos}^2\phi+{\rm cos}^2\theta{\rm sin}^2\phi 
\quad 
-\frac{1}{2}{\rm sin}(2\theta){\rm sin}\phi   \right],
\end{equation}
\begin{equation}
\mathbf{S}_{1}^{\rm ym} = \sqrt{3}\,\Big[     
{\rm i}\,{\rm cos}\theta 
\quad 
0
\quad 
-{\rm i}\,{\rm sin}\theta{\rm cos}\phi   \Big],
\end{equation}
\begin{equation}
\mathbf{S}_1^{\rm ze} =  \sqrt{3}\,\left[     
-\frac{1}{2}{\rm sin}(2\theta){\rm cos}\phi
\quad 
-\frac{1}{2}{\rm sin}(2\theta){\rm sin}\phi 
\quad 
{\rm sin}^2\theta   \right],
\end{equation}
\begin{equation}
\mathbf{S}_{1}^{\rm zm} =\sqrt{3}\, \Big[     
-{\rm i}\,{\rm sin}\theta{\rm sin}\phi
\quad 
{\rm i}\,{\rm sin}\theta{\rm cos}\phi  
\quad 
0  \Big].
\end{equation}
\end{subequations}
\paragraph{Quadrupole:}

Let us continue the calculations with the {\bf{quadrupole case}}, or for $j=2$. Hence, \eqref{scat-field-total-2D-gen} turns to
\begin{equation}\label{scat-field-total-2D-quadrupole-0}
\begin{split}
\mathbf{E}_{\mathbf{G}, {\rm s}}^{\rm sca,q}(\mathbf{r}) = \frac{{\rm i}^{-2}\pi }{Ak}&\sqrt{\frac{5}{6\pi}}\frac{e^{{\rm i}\mathbf{k}^{\pm}_{\mathbf{G}}\cdot\mathbf{r}}}{|k^{\pm}_{\mathbf{G},z}|} \sum_{m=-2}^{2}\sqrt{\frac{(2-m)!}{(2+m)!}}e^{{\rm i}m\phi} \cdot\\
&\cdot\left[{\rm i}\left( 
 \mathbf{\tau}_{2m}({\rm cos}\theta)b^{\rm e}_{2m}  +  \mathbf{\pi}_{2m}({\rm cos}\theta) b^{\rm m}_{2m}   \right) \bm{\hat{\theta}}  
 - \left( 
 \mathbf{\pi}_{2m}({\rm cos}\theta)b^{\rm e}_{2m}  +  \mathbf{\tau}_{2m}({\rm cos}\theta) b^{\rm m}_{2m}   \right) \bm{\hat{\phi}}\right].
\end{split}
\end{equation}
Hence, by calculating the summation above, the scattered field from a quadrupole array at oblique incidence reads as
\begin{subequations}\label{scat-field-total-2D-quadrupole}
\begin{equation}
\mathbf{E}^{\rm sca, q}_{\mathbf{G},{\rm s}}(\mathbf{r}) = \frac{{\rm i}^{-2}}{Ak}\frac{\sqrt{5\pi}}{2} \frac{e^{{\rm i}\mathbf{k}^{\pm}_{\mathbf{G}}\cdot\mathbf{r}}}{|k^{\pm}_{\mathbf{G},z}|} \, \Big [
{\rm i}\left(\mathbf{W}_{2} \mathbf{b}^{\rm e}_2 + \mathbf{W}^{\prime}_{2} \mathbf{b}^{\rm m}_2\right) \hat{\bm{\theta}} - \left(\mathbf{W}^{\prime}_{2}\mathbf{b}^{\rm e}_2 + \mathbf{W}_{2}\mathbf{b}^{\rm m}_2\right) \hat{\bm{\phi}}\Big ],
\end{equation}
\text{\rm where}
\begin{equation}
\mathbf{W}_{2} = \left[\frac{1}{2}
e^{-{\rm i}2\phi}{\rm sin}2\theta \quad 
e^{-{\rm i}\phi}{\rm cos}2\theta \quad - \sqrt{\frac{3}{2}}{\rm sin}2\theta  \quad 
-e^{{\rm i}\phi}{\rm cos}2\theta   \quad \frac{1}{2}
e^{{\rm i}2\phi}{\rm sin}2\theta
\right],
\end{equation}
\begin{equation}
\mathbf{W}^{\prime}_{2} = \left[-
e^{-{\rm i}2\phi}{\rm sin}\theta \quad -
e^{-{\rm i}\phi}{\rm cos}\theta \quad 0 \quad -
e^{{\rm i}\phi}{\rm cos}\theta  \quad 
e^{{\rm i}2\phi}{\rm sin}\theta 
\right].
\end{equation}
\end{subequations}
If the modified transformations between spherical and Cartesian coordinates for multipoles \cite{mun2020describing} are applied  ($\rightarrow$ \textit{Appendix E - Main Text}), they lead to 
\begin{equation}\label{scat-field-total-2D-quadrupole-cartesian}
\begin{split}
\mathbf{E}^{\rm sca, q}_{\mathbf{G},{\rm s}}(\mathbf{r}) = \frac{{\rm i}k^2}{A}\frac{\sqrt{5\pi}}{2\zeta_2} \frac{e^{{\rm i}\mathbf{k}^{\pm}_{\mathbf{G}}\cdot\mathbf{r}}}{|k^{\pm}_{\mathbf{G},z}|} \Big [
{\rm i}^{1-2}\big(\mathbf{W}_{2}\bar{\bar{F}}_2^{\,-1}k\mathbf{Q}^{\rm e}/\varepsilon &+ \mathbf{W}^{\prime}_{2}\bar{\bar{F}}_2^{\,-1}\,{\rm i}\eta k\mathbf{Q}^{\rm m}\big) \hat{\bm{\theta}} \\
&- {\rm i}^{-2}\big(\mathbf{W}^{\prime}_{2}\bar{\bar{F}}_2^{\,-1}k\mathbf{Q}^{\rm e}/\varepsilon + \mathbf{W}_{2}\bar{\bar{F}}_2^{\,-1} \,{\rm i}\eta k \mathbf{Q}^{\rm m}\big) \hat{\bm{\phi}}\Big ].
\end{split}
\end{equation}
After utilizing spherical to Cartesian coordinates conversion and breaking down ${E}^{\rm \, sca, q}_{\mathbf{G}, c}$ to its vector components, we arrive at the expression of \eqref{scat-field-total-2D-quadrupole} for Cartesian coordinates
\begin{equation}\label{scat-field-total-2D-quadrupole-cartesian-matrix}
\left[\begin{array}{c}
     {E}^{\rm \, sca, q}_{\mathbf{G},\,x}  \\
     {E}^{\rm \, sca, q}_{\mathbf{G},\,y} \\
     {E}^{\rm \, sca, q}_{\mathbf{G},\,z} 
\end{array}\right] = \frac{\mathrm{i}k^2\sqrt{\pi}}{2A} \frac{e^{{\rm i}\mathbf{k}^{\pm}_{\mathbf{G}}\cdot\mathbf{r}}}{|k^{\pm}_{\mathbf{G},z}|} \,\bar{\bar{S}}_2\,\left[\begin{array}{r}
      k(\varepsilon\zeta_2)^{-1}\mathbf{Q}^{\rm e}  \\
      {\rm i}\eta k \zeta_2^{-1} \mathbf{Q}^{\rm m} 
\end{array}\right]=\frac{\mathrm{i}k\sqrt{\pi}}{2A} \frac{e^{{\rm i}\mathbf{k}^{\pm}_{\mathbf{G}}\cdot\mathbf{r}}}{|\cos{\theta}|}
\left[\begin{array}{cc}
     \mathbf{S}_{2}^{\rm xe} & \mathbf{S}_{2}^{\rm xm} \\
     \mathbf{S}_{2}^{\rm ye} & \mathbf{S}_{2}^{\rm ym} \\
     \mathbf{S}_{2}^{\rm ze} & \mathbf{S}_{2}^{\rm zm}
\end{array}\right]\,\left[\begin{array}{r}
      k(\varepsilon\zeta_2)^{-1}\mathbf{Q}^{\rm e}  \\
      {\rm i}\eta k \zeta_2^{-1} \mathbf{Q}^{\rm m} 
\end{array}\right],
\end{equation}
with
\begin{subequations}\label{scat-field-total-2D-quadrupole-cartesian-5-b}
\begin{equation}
\mathbf{S}_{2}^{\rm xe} = \sqrt{5}\left({\rm i}\mathbf{W}_{2}{\rm cos}\theta{\rm cos}\phi + \mathbf{W}^{\prime}_{2}{\rm sin}\phi\right)\bar{\bar{F}}_2^{\,-1}, 
\quad
\mathbf{S}_{2}^{\rm xm} = \sqrt{5}\left({\rm i}\mathbf{W}^{\prime}_{2}{\rm cos}\theta{\rm cos}\phi + \mathbf{W}_{2}{\rm sin}\phi\right)\bar{\bar{F}}_2^{\,-1}, 
\end{equation}
\begin{equation}
\mathbf{S}_{2}^{\rm ye} = \sqrt{5}\left({\rm i}\mathbf{W}_{2}{\rm cos}\theta{\rm sin}\phi - \mathbf{W}^{\prime}_{2}{\rm cos}\phi\right)\bar{\bar{F}}_2^{\,-1}, 
\quad
\mathbf{S}_{2}^{\rm ym} = \sqrt{5}\left({\rm i}\mathbf{W}^{\prime}_{2}{\rm cos}\theta{\rm sin}\phi - \mathbf{W}_{2}{\rm cos}\phi\right)\bar{\bar{F}}_2^{\,-1}, 
\end{equation}
\begin{equation}
\mathbf{S}_{2}^{\rm ze}= -{\rm i}\sqrt{5}\,\mathbf{W}_{2}{\rm sin}\theta\,\bar{\bar{F}}_2^{\,-1},
\quad
\mathbf{S}_{2}^{\rm zm} = -{\rm i}\sqrt{5}\,\mathbf{W}^{\prime}_{2}{\rm sin}\theta\,\bar{\bar{F}}_2^{\,-1},
\end{equation}
\end{subequations}
or,
\begin{subequations}
\begin{equation} \small
\mathbf{S}_{2}^{\rm xe} = \sqrt{5}\left[\begin{array}{c}
     -{\rm i}\,{\rm sin}\theta\,{\rm sin}\phi\left({\rm cos}(2\phi) - 2{\rm cos}^2\theta\,{\rm cos}^2\phi  \right)  \\
     -{\rm i}\,{\rm sin}^2\theta\,{\rm cos}\theta\,{\rm sin}(2\phi)  \\
     -{\rm i}\sqrt{3}\,{\rm cos}^2\theta\,{\rm sin}\theta\,{\rm cos}\phi \\
     {\rm i}\,{\rm cos}\theta\,\left({\rm cos}(2\theta)\,{\rm cos}^2\phi + {\rm sin}^2\phi\right) \\
     {\rm i}\,{\rm sin}\theta\,{\rm cos}\phi\left({\rm cos}^2\theta\,{\rm cos}(2\phi) + 2 {\rm sin}^2\phi\right),
\end{array}\right]^T,\quad \mathbf{S}_{2}^{\rm xm} = \sqrt{5}\left[\begin{array}{c}
     \frac{1}{2}{\rm sin}(2\theta)\,{\rm cos}\phi  \\
     {\rm cos}(2\theta)\,{\rm sin}^2\phi + {\rm cos}^2\theta\,{\rm cos}^2\phi\\
    -\frac{\sqrt{3}}{2}{\rm sin}(2\theta)\,{\rm sin}\phi \\
     -\frac{1}{2}{\rm sin}^2\theta\,{\rm sin}(2\phi) \\
     -\frac{1}{2}{\rm sin}(2\theta)\,{\rm sin}\phi
\end{array}\right]^T, 
\end{equation}
\begin{equation} \small
\mathbf{S}_{2}^{\rm ye} = \sqrt{5}\left[\begin{array}{c}
     {\rm i}\,{\rm sin}\theta\left({\rm cos}(2\theta)\,{\rm sin}^2\phi\,{\rm cos}\phi +  {\rm cos}^3\phi \right)\\
         {\rm i}\,{\rm cos}\theta\,\left({\rm sin}^2\theta\,{\rm cos}(2\phi) + {\rm cos}^2\theta\right)  \\
     -{\rm i}\sqrt{3}\,{\rm cos}^2\theta\,{\rm sin}\theta\,{\rm sin}\phi  \\
     -{\rm i}\,{\rm sin}^2\theta\,{\rm cos}\theta\,{\rm sin}(2\phi) \\
     -{\rm i}\,{\rm sin}\theta\,{\rm sin}\phi\left(2{\rm cos}^2\phi - {\rm cos}^2\theta\,{\rm cos}(2\phi) \right)
\end{array}\right]^T,\,\, \mathbf{S}_{2}^{\rm ym} = \sqrt{5}\left[\begin{array}{c}
     -\frac{1}{2} {\rm sin}(2\theta)\,{\rm sin}\phi  \\
     \frac{1}{2} {\rm sin}^2\theta\,{\rm sin}(2\phi)  \\
     \frac{\sqrt{3}}{2}{\rm sin}(2\theta)\,{\rm cos}\phi \\
     -\left({\rm cos}(2\theta)\,{\rm cos}^2\phi + {\rm cos}^2\theta\,{\rm sin}^2\phi\right) \\
     -\frac{1}{2}{\rm sin}(2\theta)\,{\rm cos}\phi
\end{array}\right]^T, 
\end{equation}
\begin{equation} \small
\mathbf{S}_{2}^{\rm ze} = \sqrt{5}\left[\begin{array}{c}
     -{\rm i}\,{\rm sin}^2\theta\,{\rm cos}\theta\,{\rm sin}(2\phi) \\
     \frac{{\rm i}}{2} {\rm sin}\phi\,\left({\rm sin}\theta-{\rm sin}(3\theta)\right) \\
    {\rm i}\sqrt{3}\,{\rm sin}^2\theta\,{\rm cos}\theta  \\
    \frac{{\rm i}}{2}{\rm cos}\phi\left({\rm sin}\theta -  {\rm sin}(3\theta) \right) \\
     -{\rm i}\,{\rm sin}^2\theta\,{\rm cos}\theta\,{\rm cos}(2\phi)
\end{array}\right]^T, 
\quad
\mathbf{S}_{2}^{\rm zm} = \sqrt{5}\left[\begin{array}{c}
     - {\rm sin}^2\theta\,{\rm cos}(2\phi) \\
     - \frac{1}{2} {\rm sin}(2\theta)\,{\rm cos}\phi \\
    0 \\
     \frac{1}{2}{\rm sin}(2\theta)\,{\rm sin}\phi\\
     {\rm sin}^2\theta\,{\rm sin}(2\phi)
\end{array}\right]^T. 
\end{equation}
\end{subequations}

\paragraph{Octupole:}
Finally,  the calculations with the {\bf{octupole case}}, or for $j=3$, are performed. Hence, \eqref{scat-field-total-2D-gen} turns to
\begin{equation}\label{scat-field-total-2D-octupole-0}
\begin{split}
\mathbf{E}_{\mathbf{G},{\rm s}}^{\rm sca,o}(\mathbf{r}) = \frac{{\rm i}\pi }{2Ak}\sqrt{\frac{7}{3\pi}}&\frac{e^{{\rm i}\mathbf{k}^{\pm}_{\mathbf{G}}\cdot\mathbf{r}}}{|k^{\pm}_{{\mathbf{G}},z}|} \sum_{m=-3}^{3}\sqrt{\frac{(3-m)!}{(3+m)!}}e^{{\rm i}m\phi} \cdot\\
&\cdot\left[{\rm i}\left( 
 \mathbf{\tau}_{3m}({\rm cos}\theta)b^{\rm e}_{3m}  +  \mathbf{\pi}_{3m}({\rm cos}\theta) b^{\rm m}_{3m}   \right) \bm{\hat{\theta}}  
 - \left( 
 \mathbf{\pi}_{3m}({\rm cos}\theta)a_{3m}  +  \mathbf{\tau}_{3m}({\rm cos}\theta) b_{3m}   \right) \bm{\hat{\phi}}\right].
\end{split}
\end{equation}
Hence, by calculating the summation above, the scattered field from an octupole array at oblique incidence reads as
\begin{subequations}\label{scat-field-total-2D-octupole}
\begin{equation}
\mathbf{E}_{\mathbf{G},{\rm s}}^{\rm sca,o}(\mathbf{r}) = \frac{{\rm i}^{-3}}{Ak}\frac{\sqrt{7\pi}}{2} \frac{e^{{\rm i}\mathbf{k}^{\pm}_{\mathbf{G}}\cdot\mathbf{r}}}{|k^{\pm}_{{\mathbf{G}},z}|} \, \Big [
{\rm i}\left(\mathbf{W}_{3} \mathbf{b}^{\rm e}_3 + \mathbf{W}^{\prime}_{32}\mathbf{b}^{\rm m}_3\right) \hat{\bm{\theta}} - \left(\mathbf{W}^{\prime}_{32}\mathbf{b}^{\rm e}_3 + \mathbf{W}_{3}\mathbf{b}^{\rm m}_3\right) \hat{\bm{\phi}}\Big ],
\end{equation}
\text{\rm where}
\begin{equation}
\mathbf{W}_{3} = 
\left[\begin{array}{c}
      \frac{\sqrt{15}}{4} e^{-{\rm i}3\phi}{\rm sin}^2\theta\,{\rm cos}\theta \\
      \frac{1}{4}\sqrt{\frac{5}{2}} e^{-{\rm i}2\phi}{\rm sin}\theta\,\left(3{\rm cos}2\theta + 1\right) \\
      \frac{1}{16} e^{-{\rm i}\phi}\left({\rm cos}\theta + 15{\rm cos}3\theta\right) \\
      -\frac{\sqrt{3}}{8} \left({\rm sin}\theta + 5{\rm sin}3\theta\right)  \\
      -\frac{1}{16} e^{-{\rm i}\phi}\left({\rm cos}\theta + 15{\rm cos}3\theta\right) \\
      \frac{1}{4}\sqrt{\frac{5}{2}} e^{-{\rm i}2\phi}{\rm sin}\theta\,\left(3{\rm cos}2\theta + 1\right)   \\
      -\frac{\sqrt{15}}{4} e^{-{\rm i}3\phi}{\rm sin}^2\theta\,{\rm cos}\theta 
\end{array}\right], \qquad
\mathbf{W}^{\prime}_{3} = 
\left[\begin{array}{c}
      -\frac{\sqrt{15}}{4} e^{-{\rm i}3\phi}{\rm sin}^2\theta \\
      -\frac{1}{2}\sqrt{\frac{5}{2}} e^{-{\rm i}2\phi}{\rm sin}2\theta \\
      -\frac{1}{4} e^{-{\rm i}\phi}\left(5{\rm cos}^2\theta - 1\right) \\
      0  \\
      -\frac{1}{4} e^{{\rm i}\phi}\left(5{\rm cos}^2\theta - 1\right) \\
      \frac{1}{2}\sqrt{\frac{5}{2}} e^{{\rm i}2\phi}{\rm sin}2\theta   \\
      -\frac{\sqrt{15}}{4} e^{{\rm i}3\phi}{\rm sin}^2\theta 
\end{array}\right].
\end{equation}
\end{subequations}
If the modified transformations between spherical and Cartesian coordinates for multipoles \cite{mun2020describing} are applied  ($\rightarrow$ \textit{Appendix E - Main Text}), they lead to
\begin{equation}\label{scat-field-total-2D-octupole-cartesian}
\begin{split}
\mathbf{E}_{\mathbf{G},{\rm s}}^{\rm sca,o}(\mathbf{r})  = \frac{{ \rm i} k^2 }{A}\, \frac{\sqrt{7\pi}}{2\zeta_3} \frac{e^{{\rm i}\mathbf{k}^{\pm}_{\mathbf{G}}\cdot\mathbf{r}}}{|k^{\pm}_{{\mathbf{G}},z}|} \Big [{ \rm i}^{1-3}
\big(\mathbf{W}_{3}\bar{\bar{F}}_3^{\,-1}\mathbf{O}^{\rm e} /\varepsilon &+ \mathbf{W}^{\prime}_{3}\bar{\bar{F}}_3^{\,-1}\,{ \rm i}\eta\mathbf{O}^{\rm m}\big) \hat{\bm{\theta}} \\
&- { \rm i}^{-3}\big(\mathbf{W}^{\prime}_{3}\bar{\bar{F}}_3^{\,-1}\mathbf{O}^{\rm e}/\varepsilon + \mathbf{W}_{3}\bar{\bar{F}}_3^{\,-1}\,{ \rm i}\eta\mathbf{O}^{\rm m}\big) \hat{\bm{\phi}}\Big ].
\end{split}
\end{equation}
Let us now for convenience break down $\mathbf{E}_{\mathbf{G},{\rm s}}^{\rm sca,o}$ to its vector components. Thus,
\begin{equation}\label{scat-field-total-2D-octupole-cartesian-matrix}
\left[\begin{array}{c}
     {E}^{\rm \, sca, o}_{\mathbf{G},\,x}  \\
     {E}^{\rm \, sca, o}_{\mathbf{G},\,y} \\
     {E}^{\rm \, sca, o}_{\mathbf{G},\,z} 
\end{array}\right] = \frac{\mathrm{i} k^2 \sqrt{\pi} }{2A}\, \frac{e^{{\rm i}\mathbf{k}^{\pm}_{\mathbf{G}}\cdot\mathbf{r}}}{|k^{\pm}_{{\mathbf{G}},z}|}\, \bar{\bar{S}}_3\,\left[\begin{array}{r}
      k^2(\varepsilon \zeta_3^{-1})\mathbf{O}^{\rm e} \\
      { \rm i}\eta k^2 \zeta_3^{-1}\mathbf{O}^{\rm m} 
\end{array}\right] = \frac{\mathrm{i} k \sqrt{\pi} }{2A}\, \frac{e^{i\mathbf{k}^{\pm}_{\mathbf{G}}\cdot\mathbf{r}}}{|\cos{\theta}|}
\left[\begin{array}{cc}
     \mathbf{S}_{3}^{\rm xe} & \mathbf{S}_{3}^{\rm xm} \\
     \mathbf{S}_{3}^{\rm ye} & \mathbf{S}_{3}^{\rm ym} \\
     \mathbf{S}_{3}^{\rm ze} & \mathbf{S}_{3}^{\rm zm}
\end{array}\right]\,\left[\begin{array}{r}
      k^2(\varepsilon \zeta_3^{-1})\mathbf{O}^{\rm e} \\
      { \rm i}\eta k^2 \zeta_3^{-1}\mathbf{O}^{\rm m} 
\end{array}\right],
\end{equation}
with
\begin{subequations}\label{scat-field-total-2D-octupole-cartesian-5}
\begin{equation}
\mathbf{S}_{3}^{\rm xe} = -\sqrt{7}\left(\mathbf{W}_{3}{\rm cos}\theta{\rm cos}\phi - {\rm i} \mathbf{W}^{\prime}_{3}{\rm sin}\phi\right)\bar{\bar{F}}_3^{\,-1}, \quad
\mathbf{S}_{3}^{\rm xm} = -\sqrt{7}\left(\mathbf{W}^{\prime}_{3}{\rm cos}\theta{\rm cos}\phi - {\rm i} \mathbf{W}_{3}{\rm sin}\phi\right)\bar{\bar{F}}_3^{\,-1}, 
\end{equation}
\begin{equation}
\mathbf{S}_{3}^{\rm ye} = -\sqrt{7}\left(\mathbf{W}_{3}{\rm cos}\theta{\rm sin}\phi + {\rm i} \mathbf{W}^{\prime}_{3}{\rm cos}\phi\right)\bar{\bar{F}}_3^{\,-1}, \quad
\mathbf{S}_{3}^{\rm ym} = -\sqrt{7}\left(\mathbf{W}^{\prime}_{3}{\rm cos}\theta{\rm sin}\phi + {\rm i} \mathbf{W}_{3}{\rm cos}\phi\right)\bar{\bar{F}}_3^{\,-1}, 
\end{equation}
\begin{equation}
\mathbf{S}_{3}^{\rm ze}= \sqrt{7}\, \mathbf{W}_{3}{\rm sin}\theta\,\bar{\bar{F}}_3^{\,-1},
\quad
\mathbf{S}_{3}^{\rm zm} = \sqrt{7} \, \mathbf{W}^{\prime}_{3}{\rm sin}\theta\,\bar{\bar{F}}_3^{\,-1}.
\end{equation}
\end{subequations}

Hence, we can now derive the general equations for the fields scattered from the surface, both for spherical and Cartesian coordinates. For the spherical case from \eqref{scat-field-total-2D-dipole}, \eqref{scat-field-total-2D-quadrupole}, \eqref{scat-field-total-2D-octupole}, the general equation for each scattered mode for the metasurface is derived as
\begin{subequations}\label{spherical-general}
\begin{equation}
\mathbf{E}_{\mathbf{G}, {\rm s}}^{\rm sca}(\mathbf{r}) = \left[\begin{array}{c}
     {E}^{\rm \, sca}_{\mathbf{G}, \theta}  \\
     {E}^{\rm \, sca}_{\mathbf{G}, \phi} 
\end{array}\right] = \sum_{j=1}^{3}\frac{{\rm i}^{-j}}{Ak}\frac{\sqrt{(2j+1)\pi}}{2} \frac{e^{{\rm i}\mathbf{k}^{\pm}_{\mathbf{G}}\cdot\mathbf{r}}}{|k^{\pm}_{\mathbf{G},z}|} \, \Big [
\mathrm{i}\left(\mathbf{W}_{j} \mathbf{b}^{\rm e}_j + \mathbf{W}^{\prime}_{j} \mathbf{b}^{\rm m}_j\right) \hat{\bm{\theta}} - \left(\mathbf{W}^{\prime}_{j}\mathbf{b}^{\rm e}_j + \mathbf{W}_{j}\mathbf{b}^{\rm m}_j\right) \hat{\bm{\phi}}\Big ],
\end{equation}
\begin{equation}
\mathbf{W}_{1} = \left[e^{-{\rm i}\phi}{\rm cos}\theta \quad -\sqrt{2}{\rm sin}\theta \quad -e^{{\rm i}\phi}{\rm cos}\theta  
\right],
\qquad \mathbf{W}^{\prime}_{1} = \left[-e^{-{\rm i}\phi}\quad 0 \quad -e^{{\rm i}\phi}  
\right],
\end{equation}
\begin{equation}
\mathbf{W}_{2} = \left[\frac{1}{2}
e^{-{\rm i}2\phi}{\rm sin}2\theta \quad 
e^{-{\rm i}\phi}{\rm cos}2\theta \quad - \sqrt{\frac{3}{2}}{\rm sin}2\theta  \quad 
-e^{{\rm i}\phi}{\rm cos}2\theta   \quad \frac{1}{2}
e^{{\rm i}2\phi}{\rm sin}2\theta
\right],
\end{equation}
\begin{equation}
\mathbf{W}^{\prime}_{2} = \left[-
e^{-{\rm i}2\phi}{\rm sin}\theta \quad -
e^{-{\rm i}\phi}{\rm cos}\theta \quad 0 \quad -
e^{{\rm i}\phi}{\rm cos}\theta  \quad 
e^{{\rm i}2\phi}{\rm sin}\theta 
\right],
\end{equation}
\begin{equation}
\mathbf{W}_{3} = 
\left[\begin{array}{c}
      \frac{\sqrt{15}}{4} e^{-{\rm i}3\phi}{\rm sin}^2\theta\,{\rm cos}\theta \\
      \frac{1}{4}\sqrt{\frac{5}{2}} e^{-{\rm i}2\phi}{\rm sin}\theta\,\left(3{\rm cos}2\theta + 1\right) \\
      \frac{1}{16} e^{-{\rm i}\phi}\left({\rm cos}\theta + 15{\rm cos}3\theta\right) \\
      -\frac{\sqrt{3}}{8} \left({\rm sin}\theta + 5{\rm sin}3\theta\right)  \\
      -\frac{1}{16} e^{-{\rm i}\phi}\left({\rm cos}\theta + 15{\rm cos}3\theta\right) \\
      \frac{1}{4}\sqrt{\frac{5}{2}} e^{-{\rm i}2\phi}{\rm sin}\theta\,\left(3{\rm cos}2\theta + 1\right)   \\
      -\frac{\sqrt{15}}{4} e^{-{\rm i}3\phi}{\rm sin}^2\theta\,{\rm cos}\theta 
\end{array}\right]^T,
\quad
\mathbf{W}^{\prime}_{3} = 
\left[\begin{array}{c}
      -\frac{\sqrt{15}}{4} e^{-{\rm i}3\phi}{\rm sin}^2\theta \\
      -\frac{1}{2}\sqrt{\frac{5}{2}} e^{-{\rm i}2\phi}{\rm sin}2\theta \\
      -\frac{1}{4} e^{-{\rm i}\phi}\left(5{\rm cos}^2\theta - 1\right) \\
      0  \\
      -\frac{1}{4} e^{{\rm i}\phi}\left(5{\rm cos}^2\theta - 1\right) \\
      \frac{1}{2}\sqrt{\frac{5}{2}} e^{{\rm i}2\phi}{\rm sin}2\theta   \\
      -\frac{\sqrt{15}}{4} e^{{\rm i}3\phi}{\rm sin}^2\theta 
\end{array}\right]^T.
\end{equation}
\end{subequations}
By applying a spherical-to-Cartesian transformation or by combining \eqref{scat-field-total-2D-dipole-cartesian-matrix-v2}, \eqref{scat-field-total-2D-quadrupole-cartesian-matrix}, and \eqref{scat-field-total-2D-octupole-cartesian-matrix}, one arrives at
\begin{equation}\label{scat-field-total-2D-general-cartesian-matrix}
\mathbf{E}^{\rm sca}_{\mathbf{G}, {\rm c}} (\mathbf{r}) = 
\left[\begin{array}{c}
     {E}^{\rm \, sca}_{\mathbf{G},\,x}  \\
     {E}^{\rm \, sca}_{\mathbf{G},\,y} \\
     {E}^{\rm \, sca}_{\mathbf{G},\,z} 
\end{array}\right] = \frac{\mathrm{i} k \sqrt{\pi}}{2A}\, \frac{e^{i\mathbf{k}^{\pm}_{\mathbf{G}}\cdot\mathbf{r}}}{|\cos{\theta}|}\,
\bar{\bar{S}} \,\left[\begin{array}{c}
      (\varepsilon\zeta_1)^{-1}\,\mathbf{p}  \\
        k(\varepsilon\zeta_2)^{-1}\mathbf{Q}^{\rm e}  \\
       k^2(\varepsilon\zeta_3)^{-1}\mathbf{O}^{\rm e}  \\{\rm i}\eta\zeta_1^{-1}\,\mathbf{m} \\ {\rm i}\eta k\zeta_2^{-1} \mathbf{Q}^{\rm m} \\
      {\rm i}\eta k^2\zeta_3^{-1} \mathbf{O}^{\rm m} 
\end{array}\right] = \frac{\mathrm{i} \sqrt{\pi}}{k^2 2A}\, \frac{e^{i\mathbf{k}^{\pm}_{\mathbf{G}}\cdot\mathbf{r}}}{|\cos{\theta}|}\,
\bar{\bar{S}} \,{\bar{\bar{\widetilde{\alpha}}}_{\rm eff}} 
\left[\begin{array}{c}
\zeta_1\mathbf{E}_1 \\[0.05cm]
k^{-1}\zeta_2\mathbf{E}_2\\[0.05cm]
k^{-2}\zeta_3\mathbf{E}_3\\[0.05cm]
{\rm i}\eta\zeta_1\,\mathbf{H}_1\\[0.05cm]
{\rm i}k^{-1}\eta\zeta_2\,\mathbf{H}_2\\[0.05cm]
{\rm i}k^{-2}\eta\,\zeta_3\mathbf{H}_3
\end{array}\right],
\end{equation}
\begin{subequations}\label{General}
\begin{equation}
\mathbf{S}_{j}^{\rm xe} = \mathrm{i}^{\,j-1}\sqrt{(2j+1)} \left(\mathbf{W}_{j}\,{\rm cos}\theta{\rm cos}\phi - \mathrm{i}  \mathbf{W}^{\prime}_{j}\,{\rm sin}\phi\right)\bar{\bar{F}}_j^{\,-1}, 
\end{equation}
\begin{equation}
\mathbf{S}_{j}^{\rm xm} =\mathrm{i}^{\,j-1}\sqrt{(2j+1)} \left(\mathbf{W}^{\prime}_{j}\,{\rm cos}\theta{\rm cos}\phi - \mathrm{i}  \mathbf{W}_{j}\,{\rm sin}\phi\right)\bar{\bar{F}}_j^{\,-1},
\end{equation}
\begin{equation}
\mathbf{S}_{j}^{\rm ye} =\mathrm{i}^{\,j-1}\sqrt{(2j+1)} \left(\mathbf{W}_{j}\,{\rm cos}\theta{\rm sin}\phi + \mathrm{i}   \mathbf{W}^{\prime}_{j}\,{\rm cos}\phi\right)\bar{\bar{F}}_j^{\,-1}, 
\end{equation}
\begin{equation}
\mathbf{S}_{j}^{\rm ym} =\mathrm{i}^{\,j-1}\sqrt{(2j+1)} \left(\mathbf{W}^{\prime}_{j}\,{\rm cos}\theta{\rm sin}\phi +\mathrm{i}   \mathbf{W}_{j}\,{\rm cos}\phi\right)\bar{\bar{F}}_j^{\,-1}, 
\end{equation}
\begin{equation}
\mathbf{S}_{j}^{\rm ze}= -\mathrm{i}^{\,j-1}\sqrt{(2j+1)}\, \mathbf{W}_{j}\,{\rm sin}\theta\,\bar{\bar{F}}_j^{\,-1},
\quad
\mathbf{S}_{j}^{\rm zm} = -\mathrm{i}^{\,j-1}\sqrt{(2j+1)}\, \mathbf{W}^{\prime}_{j}\,{\rm sin}\theta\,\bar{\bar{F}}_j^{\,-1},
\end{equation}

\begin{equation}
\bar{\bar{S}}  = \left[\begin{array}{cccccc}
\mathbf{S}_{1}^{\rm xe} & \mathbf{S}_{2}^{\rm xe} & \mathbf{S}_{3}^{\rm xe} & \mathbf{S}_{1}^{\rm xm} & \mathbf{S}_{2}^{\rm xm} & \mathbf{S}_{3}^{\rm xm} \\[0.05cm]
\mathbf{S}_{1}^{\rm ye} & \mathbf{S}_{2}^{\rm ye} & \mathbf{S}_{3}^{\rm xe} & \mathbf{S}_{1}^{\rm ym} & \mathbf{S}_{2}^{\rm ym} & \mathbf{S}_{3}^{\rm ym} \\[0.05cm]
\mathbf{S}_{1}^{\rm ze} & \mathbf{S}_{2}^{\rm ze} & \mathbf{S}_{3}^{\rm ze} & \mathbf{S}_{1}^{\rm zm} & \mathbf{S}_{2}^{\rm zm} & \mathbf{S}_{3}^{\rm zm}  
\end{array}\right].
\end{equation}
\end{subequations}
Concisely, the total scattered field from all multipoles up to octupoles as a function of the incident field, the lattice interaction, and the consisting polarizabilities can be calculated as,
\begin{equation}\label{scat-field-total-2D-octupole-cartesian-matrix-2-sup}
   \mathbf{E}^{\rm \, sca}_{\mathbf{G}, {\rm c}} (\mathbf{r}) = \left[\begin{array}{r}
 \mathbf{E}^{\rm \, sca}_{\mathbf{G},\,x} \\[0.05cm]
 \mathbf{E}^{\rm \, sca}_{\mathbf{G},\,y}\\[0.05cm]
 \mathbf{E}^{\rm \, sca}_{\mathbf{G},\,z}
\end{array}\right] = \frac{\mathrm{i}\sqrt{\pi}}{2A k^2} \frac{e^{\mathrm{i}\mathbf{k}^{\pm}_{\mathbf{G}}\cdot\mathbf{r}}}{|\cos{\theta}|}\,
\bar{\bar{S}} \left(\bar{\bar{I}} - \bar{\bar{\widetilde{\alpha}}}_0\bar{\bar{C}}\right)^{-1} \bar{\bar{\widetilde{\alpha}}}_0
\left[\begin{array}{r}
 \zeta_1\mathbf{E}_1 \\[0.05cm]
k^{-1}\zeta_2\mathbf{E}_2\\[0.05cm]
k^{-2}\zeta_3\mathbf{E}_3\\[0.05cm]
\mathrm{i}\eta\zeta_1\mathbf{H}_1\\[0.05cm]
\mathrm{i}\eta k^{-1}\zeta_2 \mathbf{H}_2\\[0.05cm]
\mathrm{i}\eta  k^{-2} \zeta_3\mathbf{H}_3
\end{array}\right].
\end{equation}
%
%
\subsection{Computation of the lattice coupling matrix $\bar{\bar{C}}_s$ for vector spherical harmonics}
The lattice coupling matrix $\bar{\bar{C}}_s$ is a function depending on the lattice geometry and the wave vector $\mathbf{k}^\text{inc}$ illuminating plane wave direction. The lattice is described by the vectors $\mathbf{R}=n_1\mathbf{u}_1+n_2\mathbf{u}_2$, with $n_1, n_2 \in \mathbb{Z}$. In the two-dimensional lattice, we consider the particle at the origin. Then, the matrix $\bar{\bar{C}}_s$ is the summation of the translation coefficients from each point of the lattice to the origin.

To translate the regular to the outgoing vector spherical harmonics, we use \cite{tsang1985}
	\begin{eqnarray} 
		\mathbf{M}_{jm}^{(3)}(k{\mathbf{r}}-k{\mathbf{R}})= \sum_{\iota=1}^{\infty}\sum_{\mu=-\iota}^{\iota} 
		A_{\iota\mu jm}(-k{\mathbf{R}}) \mathbf{M}_{\iota\mu}^{(1)}(k{\mathbf{r}})
		+ B_{\iota\mu jm}(-k{\mathbf{R}}) \mathbf{N}_{\iota\mu}^{(1)}(k{\mathbf{r}}), \\ 
		\mathbf{N}_{jm}^{(3)}(k{\mathbf{r}}-k{\mathbf{R}}) = \sum_{\iota=1}^{\infty}\sum_{\mu=-\iota}^{\iota} 
		B_{\iota\mu jm}(-k{\mathbf{R}}) \mathbf{M}_{\iota\mu}^{(1)}(k{\mathbf{r}})
		+ A_{\iota\mu jm}(-k{\mathbf{R}}) \mathbf{N}_{\iota\mu}^{(1)}(k{\mathbf{r}}),
	\end{eqnarray}
\noindent where
\begin{subequations} \label{eq:transltsang}
	\begin{gather}
		\begin{split}
			A_{\iota\mu jm}(kr, \theta, \phi) =
			\frac{\gamma_{jm}}{\gamma_{\iota\mu}}(-1)^m\frac{2\iota+1}{\iota(\iota+1)}\mathrm{i}^{\iota-j}
			\sqrt{\pi\frac{(j+m)!(\iota-\mu)!}{(j-m)!(\iota+\mu)!}}
			\sum_{p} \mathrm{i}^p\sqrt{2p+1} h_p^{(1)}(kr) 
			 Y_{p,m-\mu}(\theta, \phi) \times \\
			\begin{pmatrix}
				j & \iota & p \\
				m & -\mu & -m+\mu
			\end{pmatrix}
			\begin{pmatrix}
				j & \iota & p \\
				0 & 0 & 0
			\end{pmatrix}
			\left[j(j+1)+\iota(\iota+1) - p(p+1)\right],
		\end{split} \\
		\begin{split}
			B_{\iota\mu jm}(kr, \theta, \phi) =
			\frac{\gamma_{jm}}{\gamma_{\iota\mu}}
			(-1)^m\frac{2\iota +1}{\iota(\iota+1)}\mathrm{i}^{\iota-j}
			\sqrt{\pi\frac{(j+m)!(\iota-\mu)!}{(j-m)!(\iota+\mu)!}} 
			\sum_{p} \mathrm{i}^p\sqrt{2p+1} h_p^{(1)}(kr) Y_{p,m-\mu}(\theta, \phi)\times \\
			\begin{pmatrix}
				j & \iota & p \\
				m & -\mu & -m+\mu
			\end{pmatrix} 
			\begin{pmatrix}
				j & \iota & p-1 \\
				0 & 0 & 0
			\end{pmatrix}
			\sqrt{\left[(j+\iota+1)^2-p^2\right]\left[p^2-(j-\iota)^2\right]}.
		\end{split}
	\end{gather}
\end{subequations}

The brackets $\begin{pmatrix}
				j_1 & j_2 & j_3 \\
				m_1 & m_2 & m_3
			\end{pmatrix}$ are the Wigner 3j-symbols \cite{wigner1993matrices,messiah1962clebsch}. The summation over $p$ runs through all possible values determined by the selection rule in  the Wigner 3j-symbols. $\gamma$ is defined in \eqref{vsh-X}. $Y$ is the spherical harmonics, and $h_p^{(1)}$ is the spherical Hankel function of the first kind. For a complete discussion on the addition theorem, refer to Ref.~\cite{xu1998efficient}

With a suitable order of the modes, the matrix $\bar{\bar{C}}_s$ can be expressed as
\begin{equation} \label{eq:cs:translation}
	\bar{\bar{C}}_s = \sideset{}{'}\sum_{\mathbf{R}}
	\begin{pmatrix}
		\bar{\bar{A}}(-k\mathbf{R}) & \bar{\bar{B}}(-k\mathbf{R}) \\
		\bar{\bar{B}}(-k\mathbf{R}) & \bar{\bar{A}}(-k\mathbf{R})
	\end{pmatrix}
e^{\rm i \mathbf{k}_\parallel \cdot\mathbf{R}},
\end{equation}
where the rows and columns of the matrices $\bar{\bar{A}}$ and $\bar{\bar{B}}$ can be indexed with $j \in \mathbb{N}$ and $m \in \{-j, -j+1, \dots, j\}$. The exponential function, where $\mathbf{k}_\parallel$ is the part of $\mathbf{k}^\text{inc}$ tangential to the lattice, accounts to the different phase of the illuminating plane wave on each particle. As a reminder, that we omit the origin in the summation over all lattice points, we use a prime next to the summation symbol.

Now, we single out the part of each element in Eq.~(\ref{eq:cs:translation}) that depends on $\mathbf{R}$, which is
\begin{equation} \label{eq:dlm}
	D_{jm}(\mathbf{k}) = \sideset{}{'}\sum_{\mathbf{R}}h_j^{(1)}(kR) Y_{jm}\left(\theta_{-\mathbf{R}}, \phi_{-\mathbf{R}}\right) e^{\rm i \mathbf{k}_\parallel \cdot\mathbf{R}}\,.
\end{equation}
The computation of this quantity is the main challenge when computing the lattice coupling matrix. The infinite series converges very slowly when summed directly. However, using the Ewald summation method \cite{ewald1921berechnung}, it is possible to separate long and short-range parts of this series that can be summed separately. The short-range contribution is evaluated in real space, whereas the long-range contribution is summed in Fourier space. These transformed series are quickly converging, making a fast and efficient computation possible. Conventionally, these parts $D_{jm} = D_{jm}^{(1)} + D_{jm}^{(2)} + D_{jm}^{(3)}$ are written as \cite{kambe1967}
\begin{subequations}\label{eq:dlms}
	\begin{align}
		D_{jm}^{(1)} &=
		\frac{ \sqrt{(2j+1)(j-m)!(j+m)!}}
		{A k \,\mathrm{i}^{-m}}
		\sum_{\mathbf{G}}
		\left(\frac{|\mathbf{k}_\parallel + \mathbf{G}|}{2k}\right)^j
		\frac{e^{\mathrm{i} m \varphi_{\mathbf{k}_\parallel + \mathbf{G}}}}{k_{\mathbf{G},z}^{+}}
		\sum_{\lambda=0}^{\frac{j-|m|}{2}}
		\frac{\left(\frac{k_{\mathbf{G},z}^{+}}{|\mathbf{k}_\parallel +\mathbf{G}|}\right)^{2\lambda} \Gamma\left(\frac{1}{2}-\lambda, -\frac{(k_{\mathbf{G},z}^{+})^2}{4T^2}\right)}
		{\lambda! \left(\frac{j+m}{2} - \lambda\right)!\left(\frac{j-m}{2} - \lambda\right)!},
		\\
		D_{jm}^{(2)} &=
		\frac{-\mathrm{i} (-1)^\frac{j+m}{2} \sqrt{(2j+1)(j-m)!(j+m)!}}{2^{j+1}\pi\frac{j-m}{2}!\frac{j+m}{2}!}
		\sideset{}{'}\sum_{\mathbf{R}} e^{\mathrm{i} \mathbf{k}_\parallel\cdot \mathbf{R}+\mathrm{i} m \varphi_{-\mathbf{R}}} \frac{1}{k} \left(\frac{2R}{k}\right)^j \int\limits_{T^2}^\infty \, u^{j-\frac{1}{2}} e^{-R^2u+\frac{k^2}{4u}} \mathrm{d} u,\\
		D_{jm}^{(3)} &= \frac{\delta_{j0}}{4\pi}  \Gamma\left(-\frac{1}{2}\,,\, -\frac{k^2}{4T^2}\right),
	\end{align}
\end{subequations}

\noindent where the first and the second term are the Fourier and real space summations. The third term is a correction term that appears to account for the missing lattice point at the origin when applying the transformation from real to Fourier space.

In $D_{jm}^{(1)}$ the summation runs over the reciprocal lattice points $\mathbf{G}$. The series contains the upper incomplete Gamma function $\Gamma$ where $k_{\mathbf{G},z}^{+}=\sqrt{k^2-(\mathbf{k}_\parallel +\mathbf{G})^2}$. The parameter $T$ defines the splitting between the real and Fourier space summation. $T=\sqrt{\pi/A}$ is considered as a proper splitting factor \cite{beutel2021efficient}. The integral in $D_{jm}^{(2)}$ can be computed by a recursion relation. $D_{jm}^{(3)}$ contains the Kronecker delta $\delta_{ij}$.
Having solved the lattice summation with Eq.~(\ref{eq:dlms}), it is possible to insert the result in Eq.~(\ref{eq:cs:translation}) and, hence, to obtain the lattice coupling matrix finally.
\newpage 
\subsection{Verification}
\begin{figure*}[ht]
    \centering
    \subfigure{\includegraphics[scale=0.5]{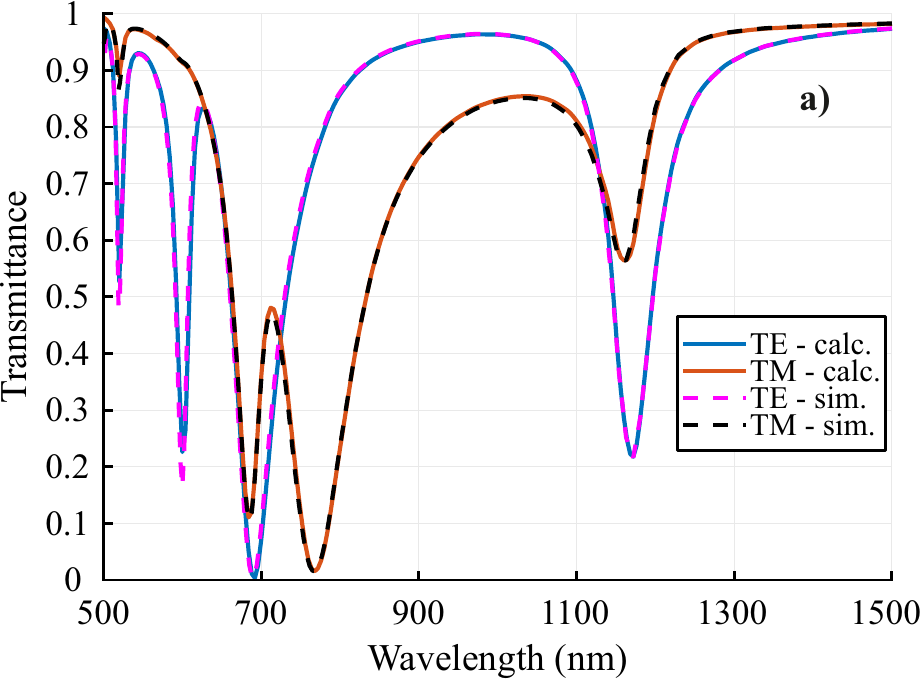}}
    \subfigure{\includegraphics[scale=0.5]{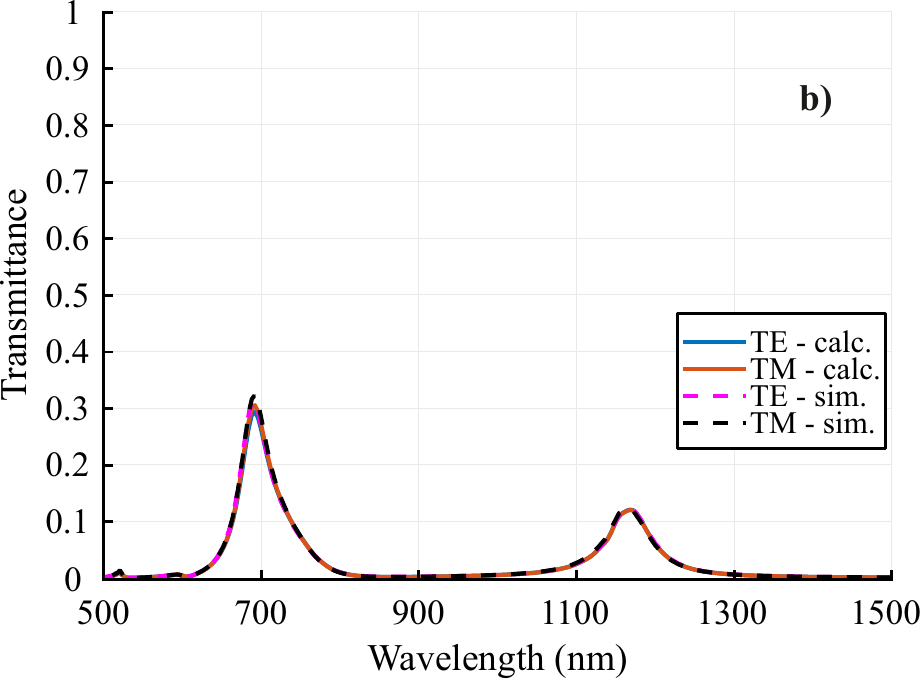}}
    \subfigure{\includegraphics[scale=0.5]{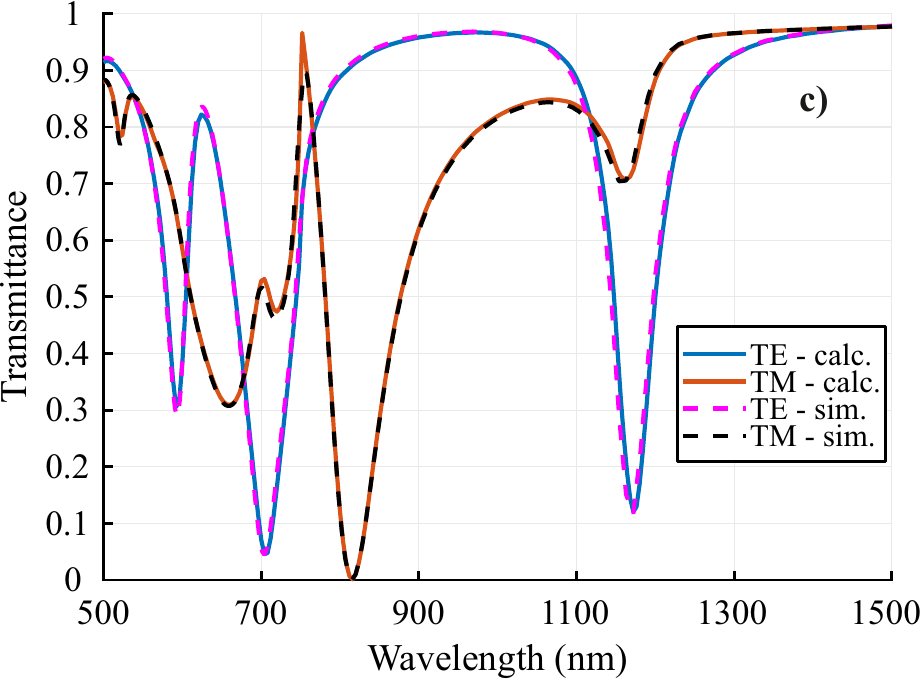}}
    \subfigure{\includegraphics[scale=0.5]{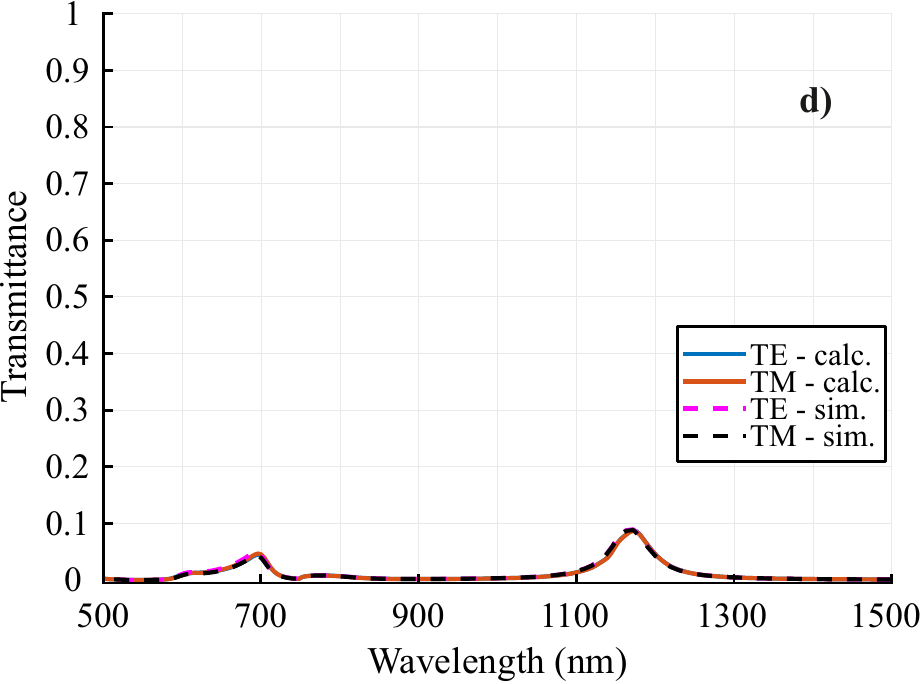}}
    \caption{\textbf{Verification with a helical particle:} Calculated and simulated transmittance versus wavelength from an infinite 2D square array of Ag Helices in vacuum under plane wave illumination, for the primary mode; a) co-polarized and b) cross-polarized scattered wave from a TE or TM illumination and $(\theta_{\rm inc},\phi_{\rm inc}) = (0,0)$, and c) co-polarized and d) cross-polarized scattered wave from a TE or TM illumination and $(\theta_{\rm inc},\phi_{\rm inc}) = (\pi/6,0)$.}
\label{FIG:ver-helix}
\end{figure*}

In this section, the metasurface model presented above is evaluated, herein, via simulations of two characteristic examples, using COMSOL$^{{\footnotesize {\rm TM}}}$ Multiphysics. As a first example, a symmetry-broken case of a metallic helix under oblique incidence is studied. The dimensions of the helix are selected as, large radius, $R_l = 80$ nm, small radius, $R_s = 20$ nm, helix pitch, $P_h = 150$ nm, and number of turns, $N=2$. The lattice is set as square with a periodicity $\Lambda = 500$ nm. Silver (Ag) is selected as helix material \cite{mcpeak2015plasmonic}, while the surrounding material is the vacuum. Comparative results are depicted in Figs.~\ref{FIG:ver-helix}
for the zeroth diffraction order or $(n_1,n_2) = (0,0)$, and show an excellent agreement with negligible small discrepancies. 
 \begin{figure*}[ht]
    \centering
    \subfigure{\includegraphics[scale=0.53]{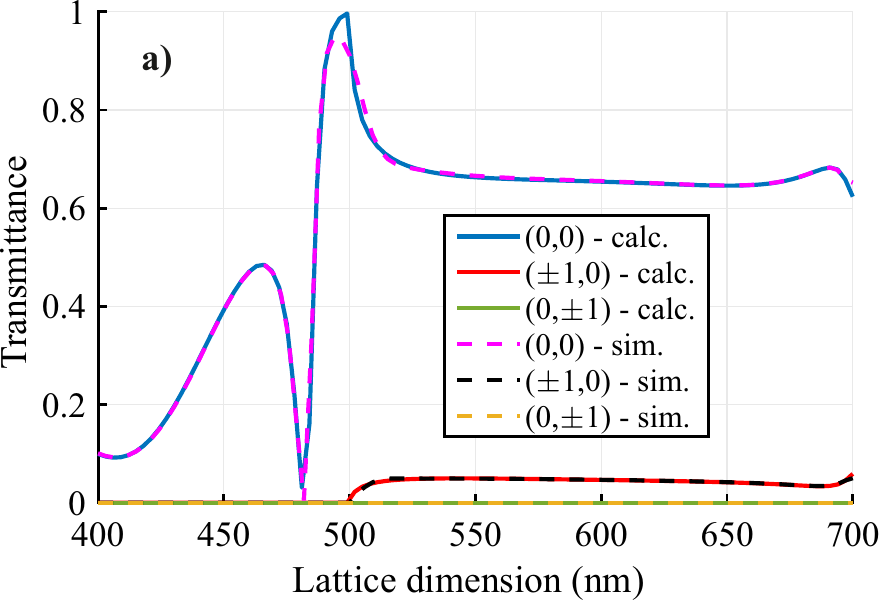}}
    \subfigure{\includegraphics[scale=0.53]{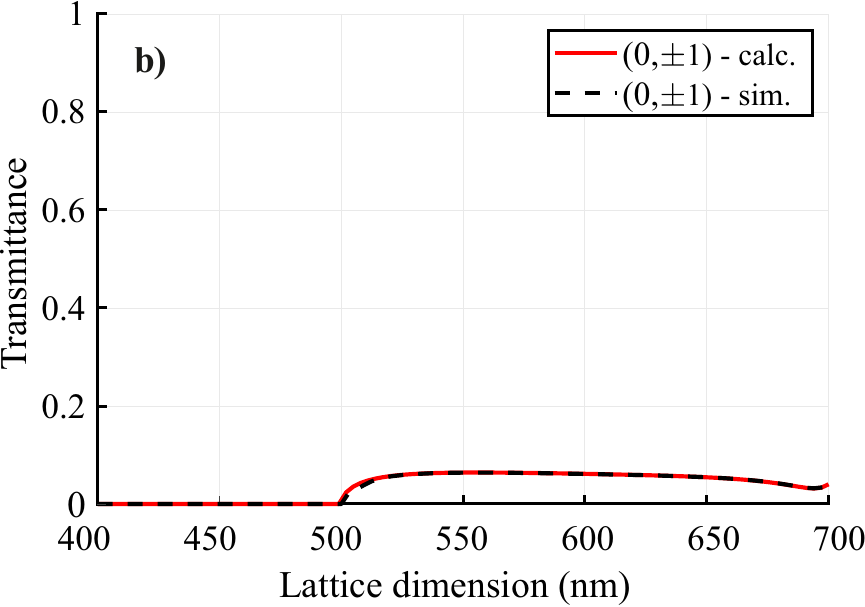}}
    \subfigure{\includegraphics[scale=0.53]{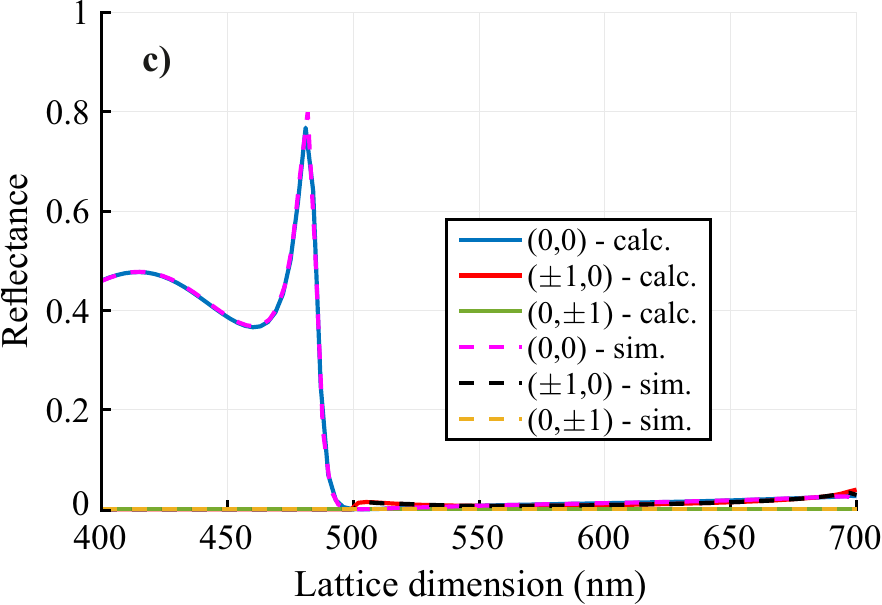}}
    \subfigure{\includegraphics[scale=0.53]{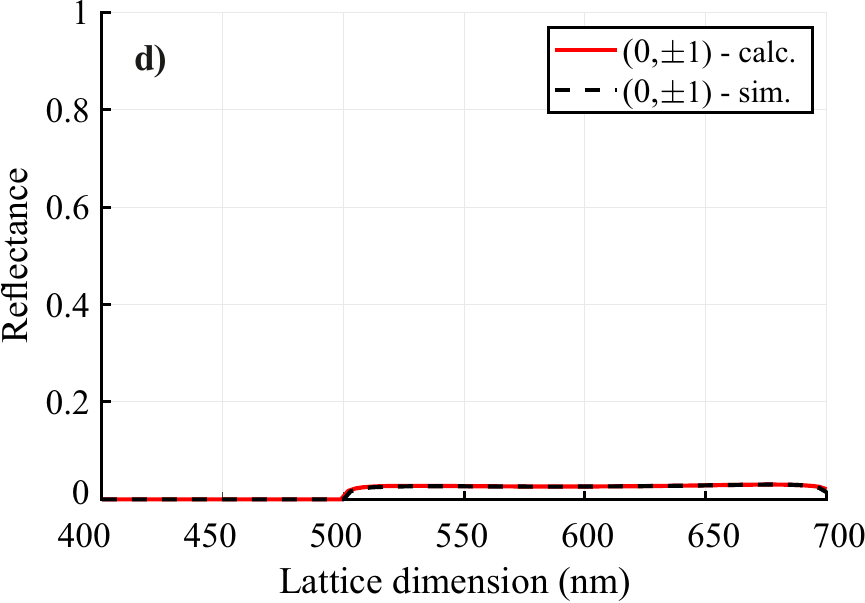}}
    \caption{\textbf{Verification with an isotropic particle:} Calculated and simulated response versus lattice dimension from an infinite 2D square array of dielectric spheres in a vacuum under linearly-polarized, normally-incident plane wave illumination; a) transmittance from co-polarized and b) cross-polarized scattered modes, and c) reflectance from co-polarized and d) cross-polarized scattered modes.}
\label{FIG:ver-sphere}
\end{figure*}

\vspace{-0mm}
Although the zeroth diffraction mode is most of the time sufficient for the analysis of metasurfaces, the proposed method can calculate the amplitudes from possible diffraction orders of the array, something handy for metagrating applications. For this purpose, a square lattice of dielectric spheres is simulated under normally plane wave incidence at  $\lambda = 500$ nm, and with a varying lattice dimension, $\Lambda$. The sphere material is set as, $\varepsilon_r = 9 + 0.2i$ and $\mu_r = 1$, while the surrounding material is vacuum. Results are depicted in Fig.~\ref{FIG:ver-sphere} for a propagation higher diffraction order with $\{n_1,n_2\} = \{-1,0,1\}$. The results show excellent agreement for transmittance and reflectance for the performed lattice sweep, with only some discrepancies around the unit-cell dimension where the onset of a new propagation order is encountered, i.e., around $\Lambda = \lambda$. This behavior is attributed to the tendency to infinity of interaction coefficients when they are calculated via the semi-analytic fast-converging series techniques of \cite{beutel2021efficient}, but do not affect the general validity of the proposed approach.

\vspace{10mm}
\section{\quad Propagating diffraction orders from a 2D lattice excited by a plane wave}
\setcounter{equation}{41} 
In this section, the lattice and incident wavelength relation for a propagating diffraction order from a 2D array illuminated with a plane wave is derived for two interesting cases: a square and a hexagonal lattice. The propagating diffraction orders for the 2D lattice can be obtained from \eqref{vsh-X-reciprocal-space} and the propagation constraint, $k^2 > \left|\mathbf{k}_\parallel + \mathbf{G}\right|$. Thus, after some substitutions, for a propagating diffraction order, it must hold that
\begin{equation}\label{propagated-modes-lattice}
\begin{split}
k^2> \left(k^{\mathrm{inc}}_x+ \frac{2\pi n_1}{A}{\rm sin}\psi\right)^2 +& \left(k^{\mathrm{inc}}_y+ \frac{2\pi n_2}{A}(1 - {\rm cos}\psi)\right)^2 \Rightarrow \\
k^2> \left(k^{\mathrm{inc}}_x+ \frac{2\pi n_1}{\Lambda}\right)^2 +& \left(k^{\mathrm{inc}}_y+ \frac{2\pi n_2}{\Lambda}\frac{1 - {\rm cos}\psi}{{\rm sin}\psi}\right)^2,
\end{split}
\end{equation}
where $\psi$ is the angle between the vectors of the 2D lattice of the metasurface. Herein, it is also assumed that $\mathbf{u_1}=\Lambda_1\,\hat{\mathbf{x}}$ , $\mathbf{u_2}=\Lambda_2({\rm cos}\psi\, \hat{\mathbf{x}}+{\rm sin}\psi  \,\hat{\mathbf{y}})$, with $|u_1|=|u_2| = \Lambda$. Additionally,  $\mathbf{G} = n_1\mathbf{u'_1} + n_2\mathbf{u'_2} =\frac{2\pi n_1}{A}\,(\mathbf{u}_2 \times  \hat{\mathbf{z}}) + \frac{2\pi n_2}{A}\,(  \hat{\mathbf{z}} \times \mathbf{u}_1  )  =\frac{2\pi n_1}{A}\Lambda\,{\rm sin}\psi\, \hat{\mathbf{x}}  + \frac{2\pi n_2}{A}\Lambda (1 - {\rm cos}\psi) \,\hat{\mathbf{y}}$  and $A=(\mathbf{u}_1 \times \mathbf{u}_2)\cdot \hat{\mathbf{z}} = \Lambda^2\, {\rm sin}\psi$. 

At this point it becomes convenient for the analysis to split the incident field to a $k_{\mathrm{inc},y}=0$ and a $k_{\mathrm{inc},x}=0$ part. With $k_{\mathrm{inc},y}=0 $ and the dimension of the lattice is a fraction of the wavelength, or $\Lambda=t\lambda$ with $t \in (0,1)$, \eqref{propagated-modes-lattice} turns to
\begin{equation}\label{propagated-modes-kx}
\begin{multlined}
 \left(\frac{2\pi}{\lambda}\right)^2> \left(\frac{2\pi}{\lambda} {\rm sin}\theta + \frac{2\pi n_1}{t\lambda}\right)^2 + \left( \frac{2\pi n_2}{t\lambda}\frac{1 - {\rm cos}\psi}{{\rm sin}\psi}\right)^2 \Rightarrow
1 > \left({\rm sin}\theta + \frac{n_1}{t}\right)^2 + \left( \frac{n_2}{t}\frac{1 - {\rm cos}\psi}{{\rm sin}\psi}\right)^2.
\end{multlined}
\end{equation}
For the case of a {\bf{square lattice}}, or $\psi = \pi/2$, \eqref{propagated-modes-kx} turns to,
\begin{equation}\label{propagated-modes-90-kx}
1 > \left({\rm sin}\theta + \frac{n_1}{t}\right)^2 + \left( \frac{n_2}{t}\right)^2.
\end{equation}
For $n_2=0$, only the $n_1=0$ propagates for $t<1/2$, because $-1<{\rm sin}\theta<1$. Moreover, for $n_1=0$, only the $n_2=0$ propagates for $t<1$. Thus, for the case of the square lattice the $(n_1,n_2) = (0,0)$, or the zeroth diffraction order,  is the only propagating diffraction order, if $\Lambda<\lambda/2$.
For the case of a  {\bf{hexagonal lattice}}, or  $\psi = \pi/3$, \eqref{propagated-modes-kx} turns to,
\begin{equation}\label{propagated-modes-60-kx}
1 > \left({\rm sin}\theta + \frac{n_1}{t}\right)^2 + \left( \frac{n_2}{t}(2-\sqrt{3})\right)^2.
\end{equation}
In this case, if only the  $(n_1, n_2)=(0,0)$ diffraction order should be propagating, it should hold that
\begin{equation}\label{propagated-modes-60-kx-m1}
\frac{n_2}{t}\frac{1 - {\rm cos}\psi}{{\rm sin}\psi}<1 \Rightarrow \frac{n_2}{t}(2-\sqrt{3}) <1.
\end{equation}
Hence, for a normal hexagonal lattice the zeroth diffraction order is the only propagating one if $\Lambda < (2-\sqrt{3})\,\lambda$ or, approximately, $\Lambda < 0.2679 \lambda$.

For the case of $k_{\mathrm{inc},x}=0$ and the dimension of the lattice is a fraction of the wavelength, or $\Lambda=t\lambda$ with $t \in (0,1)$, \eqref{propagated-modes-lattice} turns to
\begin{equation}\label{propagated-modes-ky}
\begin{multlined}
 \left(\frac{2\pi}{\lambda}\right)^2> \left( \frac{2\pi n_1}{t\lambda}\right)^2 + \left(\frac{2\pi}{\lambda} {\rm sin}\theta + \frac{2\pi n_2}{t\lambda}\frac{1 - {\rm cos}\psi}{{\rm sin}\psi}\right)^2 \Rightarrow
1 > \left( \frac{n_1}{t}\right)^2 + \left( {\rm sin}\theta + \frac{n_2}{t}\frac{1 - {\rm cos}\psi}{{\rm sin}\psi}\right)^2.
\end{multlined}
\end{equation}
For a {\bf{square lattice}}, or $\psi = \pi/2$, the propagation mode constraint is,
\begin{equation}\label{propagated-modes-90-ky}
1 > \left(\frac{n_1}{t}\right)^2 + \left( {\rm sin}\theta + \frac{n_2}{t}\right)^2. 
\end{equation}
Due to symmetry this case arrives to the same conclusions as in the previous $k_{\mathrm{inc},y}=0$ one. Hence, the $(n_1,n_2) = (0,0)$ is the only propagating mode for $\Lambda<\lambda/2$.
For a {\bf{hexagonal lattice}}, or  $\psi = \pi/3$, \eqref{propagated-modes-ky} turns to,
\begin{equation}\label{propagated-modes-60-ky}
1 > \left(\frac{n_1}{t}\right)^2 + \left({\rm sin}\theta + \frac{n_2}{t}(2-\sqrt{3})\right)^2.
\end{equation}
Following the same analysis as in the previous section, for a hexagonal lattice the $(n_1,n_2) = (0,0)$ is the only propagating diffraction order for $\Lambda < (2-\sqrt{3})/2\,\lambda$ or, approximately, $\Lambda < 0.13397 \lambda$.
%
%
\section{\quad Calculations of the $\zeta_j$ prefactors via the radiation fields}
\setcounter{equation}{49} 
The prefactors will be derived by comparing a scatterer's far-field expressions in Cartesian and spherical coordinates.
The scattering far-field in Cartesian is defined as
\begin{subequations}\label{radiation-fields-cart-def-S}
\begin{equation}
\begin{split}
\mathbf{E}^c = \frac{k^2}{4\pi}\frac{e^{ikr}}{r}\,\Bigg[\hat{\mathbf{r}}\times\left(\frac{1}{{\varepsilon}}\mathbf{p}\times\hat{\mathbf{r}}\right) &+ \left(\eta\,\mathbf{m}\times\hat{\mathbf{r}}\right) - \frac{ik}{6} \mathbf{r}\times\left(\frac{1}{\varepsilon}\boldsymbol{\mathcal{Q}}^{\rm e}\times\hat{\mathbf{r}}\right) - \frac{ik}{6} \left(\eta\,\boldsymbol{\mathcal{Q}}^{\rm m}\times\hat{\mathbf{r}}\right) \\
&- \frac{k^2}{16} \hat{\mathbf{r}}\times\left(\frac{1}{\varepsilon}\boldsymbol{\mathcal{O}}^{\rm e}\times\hat{\mathbf{r}}\right) - \frac{k^2}{16} \left(\eta\,\boldsymbol{\mathcal{O}}^{\rm m}\times\hat{\mathbf{r}}\right)\Bigg],
\end{split}
\end{equation}
where
\begin{equation}
\hat{\mathbf{r}} = 
{\rm sin}\theta{\rm cos}\phi\,\hat{\mathbf{x}} + {\rm sin}\theta{\rm sin}\phi\,\hat{\mathbf{y}} + {\rm cos}\theta\,\hat{\mathbf{z}},
\end{equation}
\begin{equation}
\mathcal{Q}^v_\alpha = \sum_\beta Q_{\alpha\beta}\,\hat{r}_\beta,\quad \mathcal{{O}}^v_\alpha = \sum_{\beta\gamma} O_{\alpha\beta\gamma}\,\hat{r}_\beta\,\hat{r}_\gamma,
\end{equation}
\end{subequations}
with  $\{\alpha,\beta,\gamma\}=\{x,y,z\}$ and $v = \{ {\rm {e, m}} \}$. Moreover the far-field is expressed in spherical coordinates as
\begin{subequations}\label{radiation-fields-sph-def}
\begin{equation}
\begin{split}
\mathbf{E}^s = \frac{e^{\mathrm{i}kr}}{r}\,\sum_{j=1}^{\infty}\sum_{m=-j}^{j}a_{jm}\,(-i)^j\,\mathbf{Z}_{jm}(\theta,\phi) + (-i)^{j+1}\,b_{jm}\,\mathbf{X}_{jm}(\theta,\phi),
\end{split}
\end{equation}
where
\begin{equation}
\mathbf{Z}_{jm}(\theta,\phi) = \gamma_{jm}\left[\tau_{jm}({\rm cos}\theta)\, \bm{\hat{\theta}} + \mathrm{i}\,\pi_{jm}({\rm cos}\theta)\, \bm{\hat{\phi}} \right],
\end{equation}
\begin{equation}
\mathbf{X}_{jm}(\theta,\phi) = \gamma_{jm}\left[\mathrm{i}\,\pi_{jm}({\rm cos}\theta)\, \bm{\hat{\theta}} - \tau_{jm}({\rm cos}\theta)\, \bm{\hat{\phi}} \right].
\end{equation}
\end{subequations}
Let's check the {\bf{dipole}} case. The prefactor $\zeta_1$ can be obtained by comparing \eqref{radiation-fields-cart-def-S} and \eqref{radiation-fields-sph-def}. For $j=1$, \eqref{radiation-fields-sph-def} turns to
\begin{equation}\label{radiation-fields-sph-dip}
\mathbf{E}^s_1 = -\frac{e^{\mathrm{i}kr}}{r}\frac{\mathrm{i}}{2}\sqrt{\frac{3}{2\pi}}\,\left[\left(\mathbf{W}_{11}\,\mathbf{a}_1 + \mathbf{W}^{\prime}_{1}\,\mathbf{b}_1\right)\bm{\hat{\theta}} + \mathrm{i}\,\left(\mathbf{W}^{\prime}_{1}\,\mathbf{a}_1 + \mathbf{W}_{11}\,\mathbf{b}_1\right)\bm{\hat{\phi}} \right],
\end{equation}
with the related vectors defined in \eqref{scat-field-total-2D-dipole}. Let us now assume only the excitation of the electric dipole $p_x$. If also observations are set to $\theta = \phi = 0$, then from \eqref{radiation-fields-cart-def-S}, \eqref{radiation-fields-sph-def}, and comparing only the $x$-component, one gets
\begin{equation}\label{c1-derive}
\mathbf{E}^s_1 = \mathbf{E}^c_1 \Rightarrow 
\frac{-\mathrm{i}}{2}\sqrt{\frac{3}{2\pi}}\frac{\sqrt{2}}{2}\frac{e^{\mathrm{i}kr}}{kr}\left(a_{1-1} - a_{11}\right) \mathbf{\hat{x}} = \frac{k^3}{4\pi}\frac{e^{\mathrm{i}kr}}{kr}\frac{p_x}{\varepsilon_0}\mathbf{\hat{x}} \Rightarrow \frac{p_x}{\varepsilon_0} = \frac{\sqrt{3\pi}}{\mathrm{i}k^3}\left(a_{1-1} - a_{11}\right), 
\end{equation}
which in turn gives $\zeta_1 = \sqrt{6\pi}$, if the formulation of \eqref{scat-field-total-2D-dipole-cartesian} and the $\bar{\bar{F}}_1$ tensor are used ($\rightarrow$ \textit{Appendix E - Main Text}). The same result could be obtained via magnetic dipoles following a similar procedure.

Moving on to the {\textbf{quadrupole}} case, 
For $j=2$, \eqref{radiation-fields-sph-def} turns to
\begin{equation}\label{radiation-fields-sph-quad}
\mathbf{E}^s_{2} = -\frac{e^{\mathrm{i}kr}}{r}\frac{1}{2}\sqrt{\frac{5}{6\pi}}\left[\left(\mathbf{W}_{21}\,\mathbf{a}_2 + \mathbf{W}^{\prime}_{2}\,\mathbf{b}_2\right)\,\bm{\hat{\theta}} + \mathrm{i}\,\left(\mathbf{W}^{\prime}_{2}\,\mathbf{a}_1 + \mathbf{W}_{21}\,\mathbf{b}_1\right)\,\bm{\hat{\phi}} \right],
\end{equation}
with the related vectors are defined in \eqref{scat-field-total-2D-quadrupole}. Let us now assume only the excitation of the electric quadrupole $Q_{xz}$. If also observations are set to $\theta = \phi = 0$, then from \eqref{radiation-fields-cart-def-S}, \eqref{radiation-fields-sph-def}, and comparing only the $x$-component, one gets
\begin{equation}\label{c2-derive}
\begin{split}
\mathbf{E}^s_{2} = \mathbf{E}^c_{2} \quad \Rightarrow \quad 
\frac{-1}{2}\sqrt{\frac{5}{6\pi}}\sqrt{\frac{3}{2}}\frac{e^{\mathrm{i}kr}}{kr}\left(a_{2-1} - a_{21}\right) \mathbf{\hat{x}} &= -\frac{k^3}{4\pi}\frac{e^{ikr}}{kr}\frac{ikQ_{xz}}{6\varepsilon_0}\mathbf{\hat{x}}\\
\Rightarrow \quad  \frac{kQ_{xz}}{\varepsilon_0\sqrt{3}} &= \frac{\sqrt{120\pi}}{ik^3}\left(a_{2-1} - a_{21}\right), 
\end{split}
\end{equation}
which in turn gives $\zeta_2 = \sqrt{120\pi}$, if the formulation of \eqref{scat-field-total-2D-quadrupole-cartesian} and the $\bar{\bar{F}}_2$ tensor are used ($\rightarrow$ \textit{Appendix E - Main Text}). The same result could be obtained via magnetic quadrupoles following a similar procedure.

Moving on to the {\textbf{octupole}} case, 
For $j=3$, \eqref{radiation-fields-sph-def} turns to
\begin{equation}\label{radiation-fields-sph-oct}
\mathbf{E}^s_{1} = \frac{e^{\mathrm{i}kr}}{r}\frac{\mathrm{i}}{4}\sqrt{\frac{7}{3\pi}}\,\left[\left(\mathbf{W}_{3}\,\mathbf{a}_3 + \mathbf{W}^{\prime}_{3}\,\mathbf{b}_3\right)\,\bm{\hat{\theta}} + \mathrm{i}\,[\left(\mathbf{W}^{\prime}_{3}\,\mathbf{a}_3 + \mathbf{W}_{3}\,\mathbf{b}_3\right)\,\bm{\hat{\phi}} \right],
\end{equation}
with the related vectors defined in \eqref{scat-field-total-2D-quadrupole}. Let us now assume only the excitation of the electric quadrupole $O_{xzz}$. If also observations are set to $\theta = \phi = 0$, then from \eqref{radiation-fields-cart-def-S}, \eqref{radiation-fields-sph-def}, and for the $x$-component, one gets
\begin{equation}\label{c3-derive-1}
\begin{split}
\mathbf{E}^s_{\rm oct} \hspace{-0.5mm}=\hspace{-0.5mm} \mathbf{E}^c_{\rm oct} \hspace{-0.5mm}\Rightarrow\hspace{-0.5mm} 
\frac{i}{4}\sqrt{\frac{7}{3\pi}}\sqrt{3}\frac{e^{ikr}}{kr}\hspace{-0.3mm}\left(a_{3-1}\hspace{-0.5mm} -\hspace{-0.5mm} a_{31}\hspace{-0.3mm}\right)\hspace{-0.3mm} \mathbf{\hat{x}} \hspace{-0.5mm}=\hspace{-0.5mm} -\frac{k^3}{4\pi}\frac{e^{ikr}}{kr}\frac{ik^2\,O_{xzz}}{16\varepsilon_0}\mathbf{\hat{x}} \hspace{-0.5mm}\Rightarrow\hspace{-0.5mm} \frac{k^2\,O_{xzz}}{\varepsilon_0} \hspace{-0.5mm}=\hspace{-0.5mm} \frac{16\sqrt{7\pi}}{ik^3}\left(a_{3-1}\hspace{-0.5mm} - \hspace{-0.5mm}a_{31}\right),
\end{split}
\end{equation}
where $\zeta_3 = 12\sqrt{35\pi}$ is set,  if the formulation of \eqref{scat-field-total-2D-octupole-cartesian} and the $\bar{\bar{F}}_3$ tensor are used ($\rightarrow$ \textit{Appendix E - Main Text}). The same result could be obtained via magnetic octupoles using a similar procedure. 

By combining the results above the general formula for the $\zeta$ prefactors is finally composed as, $\zeta_j = \sqrt{\left(2j+1\right)!\,\pi}$, where $j=\{1,2,3\}$.
%
\section{\quad Transformations of polarizability matrices}
\setcounter{equation}{57} 
The quadrupole matrix for an arbitrary particle is defined as \cite{Jackson1999}
\begin{equation}\label{Q-matrix-def}
\bar{\bar{Q}}^{v} = 
\left[\begin{array}{ccc}
Q^{ v}_{xx} & Q^{ v}_{xy} & Q^{ v}_{xz} \\
Q^{ v}_{yx} & Q^{ v}_{yy} & Q^{ v}_{yz} \\
Q^{ v}_{zx} & Q^{ v}_{zy} & Q^{ v}_{zz}
\end{array} \right], 
\end{equation}
where $v = \{ \rm e, m\}$. However, due to the symmetries that are implied in the definition of these moments ($\rightarrow$ \textit{Appendix E - Main Text}), $Q^{v}_{ij} =Q^{v}_{ji}$ and because the $\bar{\bar{Q}}^{v}$ is traceless, thus, 
$Q^{v}_{xx} + Q^{v}_{yy} + Q^{v}_{zz} = 0$, only five terms from \eqref{Q-matrix-def} are required to sufficiently describe the quadrupolar response of a particle. Hence, the properly selected five quadrupole moments are expressed in the main text as a $\mathbf{Q}^{v}$ vector.

With the modified transformations between spherical and Cartesian coordinates for multipoles \cite{mun2020describing} ($\rightarrow$ \textit{Appendix E - Main Text}), one can deduce the polarizability matrix of an arbitrary particle from its T matrix and vice versa. With the specific selection of $\bar{\bar{F}}_j,\,\, i=\{1,2,3\}$, for dipoles, quadrupoles, and octupoles, respectively, as presented in \textit{Appendix D} of the \textit{Main Text}, any symmetry is retained after transformation. However, the resulting polarizability matrix, as defined in this work, provides connections between fields and combinations of multipole moments that are not always easy to interpret physically, e.g., $Q_{xx} - Q_{yy}$. This can be alleviated by using simple matrix transformations, that the quadrupole matrix is traceless, and that fields are free of divergence. Let us demonstrate the procedure while beginning with the electric quadrupole. Multipole moments and field irreducible basis definitions in \textit{Appendix E} of the \textit{Main Text} can be changed as
\begin{subequations}\label{transformation-quadrupole-M}
\begin{equation}\label{transformation-quadrupole-M-a}
\breve{\mathbf{Q}}^{{\rm e}} = \bar{\bar{M}}_{2\rm{m}} \mathbf{Q}^{\rm e} \Rightarrow \left[\begin{array}{c}
Q^{\rm e}_{xy}\\
Q^{\rm e}_{yz}\\
Q^{\rm e}_{xz}\\
Q^{\rm e}_{xx}\\
Q^{\rm e}_{yy} \end{array} \right]=
\bar{\bar{M}}_{2\rm{m}}\,\,\frac{1}{\sqrt{3}}\left[\begin{array}{c}
Q^{\rm e}_{xy}\\
Q^{\rm e}_{yz}\\
\sqrt{3}/2\,Q^{\rm e}_{zz}\\
Q^{\rm e}_{xz}\\
\left(Q^{\rm e}_{xx} - Q^{\rm e}_{yy}\right)/2 \end{array} \right], 
\end{equation}
\begin{equation}\label{transformation-quadrupole-M-b}
\breve{\mathbf{E}}_2 = \bar{\bar{M}}_{2\rm{f}}\, \mathbf{E}_2 \Rightarrow 
\frac{1}{2}\left[\begin{array}{c}
\partial_y E_x + \partial_x E_y \\ 
\partial_y E_z + \partial_z E_y \\
\partial_x E_z + \partial_z E_x \\
2\,\partial_x E_x \\
2\, \partial_y E_y\end{array}\right]=
\bar{\bar{M}}_{2\rm{f}}\,\,\frac{1}{2\sqrt{3}}
\left[\begin{array}{c}
\partial_y E_x + \partial_x E_y \\ 
\partial_y E_z + \partial_z E_y \\
\sqrt{3}\, \partial_z E_z  \\
\partial_x E_z + \partial_z E_x \\
\partial_x E_x - \partial_y E_y\end{array}\right],
\end{equation}
\text{where}
\begin{equation}\label{transformation-quadrupole-M-c}
\bar{\bar{M}}_{2\rm{m}} = \bar{\bar{M}}_{2\rm{f}} = \left[\begin{array}{cccccc}
\sqrt{3} & 0 & 0 & 0 & 0 \\
0 & \sqrt{3} & 0 & 0 & 0 \\
0 & 0 & 0 & \sqrt{3} & 0 \\
0 & 0 & -1 & 0 & \sqrt{3} \\
\sqrt{3} & 0 & -1 & 0 & -\sqrt{3} 
\end{array} \right].
\end{equation}
\end{subequations}
Under this procedure, the quadrupole moments are written individually on the left side of \eqref{transformation-quadrupole-M-a}, and the fields are written on the left side of \eqref{transformation-quadrupole-M-b} in the form presented in \cite{babicheva2019analytical, feshbach2019methods}.  While transformations are performed separately, the same $\bar{\bar{M}}_{2\rm{m}}$ and $\bar{\bar{M}}_{2\rm{f}}$ matrices appear in both quadrupole moments and field transformations. This observation can be utilized for the octupole case. Thus, the quadrupole polarizabilities for the individual quadrupoles and fields of \eqref{transformation-quadrupole-M} are acquired as
\begin{equation}\label{transformation-quadrupole-pol}
\begin{split}
\mathbf{Q}^{\rm e} = \bar{\bar{\alpha}}^{vv'}_{22}\mathbf{E}_2 \Rightarrow 
\bar{\bar{M}}_{2\rm{m}}^{-1}\breve{\mathbf{Q}}^{\rm e} &= \bar{\bar{\alpha}}^{vv'}_{22}\bar{\bar{M}}_{2\rm{f}}^{-1}\breve{\mathbf{E}}_2 \Rightarrow \\
\breve{\mathbf{Q}}^{{\rm e}} = \bar{\bar{M}}_{2\rm{m}}\,\bar{\bar{\alpha}}^{vv'}_{22}\,\bar{\bar{M}}_{2\rm{f}}^{-1}\,\,\breve{\mathbf{E}}_2
\Rightarrow \breve{\bar{\bar{\alpha}}}^{vv'}_{22} &= \bar{\bar{M}}_{2\rm{m}}\,\bar{\bar{\alpha}}^{vv'}_{22}\,\bar{\bar{M}}_{2\rm{f}}^{-1},\quad \{v,v'\} = \{{\rm e},\,{ \rm m}\}.
\end{split}
\end{equation}
The same procedure can be applied for the magnetic quadrupole moments and fields.

Moving to the octupole case, the same procedure as above for quadrupoles is repeated. The octupole matrix $\bar{\bar{O}}^v$, analogously to \eqref{Q-matrix-def}, is a 3D matrix. Again, symmetries can be applied, i.e. $O^{v}_{ijk} = O^{v}_{jik} = O^{v}_{kji} = O^{v}_{kij} = O^{v}_{ikj}$, and the traceless identity, or $O^{v}_{ixx} + O^{v}_{iyy} + O^{v}_{izz} = 0$, with $\{i,j,k\} = \{x,y,z\}$. Thus, only seven appropriately selected octupole moments suffice to describe the octupolar response of the particle understudy, which in turn are written in vector form as $\mathbf{O}^v$.
As in the quadrupole case, the octupole moments combinations for the utilized $\mathbf{O}^v$ vector and the resulting polarizability matrix as defined in this work ($\rightarrow$ \textit{Appendix E} - \textit{Main Text} ) are not always easy to interpret physically, e.g. $3\,{O}^{\rm e}_{yxx} - O^{\rm e}_{yyy}$. This can be alleviated by using simple matrix transformations, as performed above. Let us demonstrate the procedure, beginning with the electric octupole moment. Multipole moments and field basis definitions proposed herein, can be changed as \cite{mun2020describing},
\begin{subequations}\label{transformation-octupole-M}
\begin{equation}\label{transformation-octupole-M-a}
\breve{\mathbf{O}}^{\rm e} = \bar{\bar{M}}_{3\rm{m}} \mathbf{O}^{\rm e} \Rightarrow
\left[\begin{array}{c}
O^{\rm e}_{xyy}\\
O^{\rm e}_{xzz}\\
O^{\rm e}_{yzz}\\
O^{\rm e}_{yxx}\\
O^{\rm e}_{zxx}\\
O^{\rm e}_{zyy}\\
O^{\rm e}_{xyz} 
\end{array} \right]   = 
\bar{\bar{M}}_{3\rm{m}}\,\frac{\sqrt{6}}{4}\left[\begin{array}{c}
     {1}/{2}\,\left(3\,{O}^{\rm e}_{yxx} - O^{\rm e}_{yyy} \right)\\
     {\sqrt{6}}/{4}\,\,{O}^{\rm e}_{xyz} \\
     {\sqrt{15}}/{2}\,\,{O}^{\rm e}_{yzz} \\
     \sqrt{{5}/{2}}\,\,{O}^{\rm e}_{zzz} \\
     {\sqrt{15}}/{2}\,\,{O}^{\rm e}_{xzz} \\
     {\sqrt{6}}/{2}\,\,\left({O}^{\rm e}_{xxz} - {O}^{\rm e}_{zyy}\right) \\
      {1}/{2}\,\left({O}^{\rm e}_{xxx} - 3\,O^{\rm e}_{xyy}\right)   
\end{array}\right],
\end{equation}
\text{where}
\begin{equation}\label{transformation-octupole-M-c}
\bar{\bar{M}}_{3\rm{m}} = \frac{4}{\sqrt{6}}\left[\begin{array}{ccccccccccc}
0 & 0 & 0 & 0 & -1/(2\sqrt{15}) & 0 & -1/2  \\
0 & 0 & 0 & 0 & 2/\sqrt{15} & 0 & 0 \\
0 & 0 & 2/\sqrt{15} & 0 & 0  & 0 & 0  \\
1/2 & 0 & -1/(2\sqrt{15}) & 0 & 0 & 0 & 0  \\
0 & 0 & 0 & -1/\sqrt{10} &  0 & 1/\sqrt{6}  & 0  \\
0 & 0 & 2/\sqrt{15} & 0 &  0 & 0  & 0  \\
0 & 4/\sqrt{6} & 0 & 0  & 0 & 0 & 0
\end{array} \right],
\end{equation}
\end{subequations}
with the transformation above applying, also, for magnetic octupoles. The same procedure will be followed for the octupolar fields using the E-field arrangements as in \cite{mun2020describing}. Thus, the E-field vector can be transformed as,
\begin{subequations}\label{transformation-octupole-M-2}
\begin{equation}\label{transformation-octupole-M-2-a}
\begin{split}
\breve{\mathbf{E}}_3 &= \bar{\bar{M}}_{3\rm{f}} \mathbf{E}_3 \Rightarrow \\
\left[\hspace{-1mm}
\begin{array}{c}
\partial^2_{y}E_x - \partial^2_{x}E_x + 2\partial_{xy}E_y \\
\partial^2_{z}E_x - \partial^2_{x}E_x + 2\partial_{xz}E_z \\
\partial^2_{z}E_x - \partial^2_{y}E_y + 2\partial_{yz}E_z \\
\partial^2_{x}E_y - \partial^2_{y}E_y + 2\partial_{xy}E_x \\
\partial^2_{z}E_z - \partial^2_{z}E_z + 2\partial_{xz}E_x \\
\partial^2_{y}E_z - \partial^2_{z}E_z + 2\partial_{yz}E_y \\
2\left(\partial_{yz}E_x + \partial^2_{x}E_x + 2\partial_{xy}E_y\right) 
\end{array} \hspace{-1mm}\right] \hspace{-0.8mm}  &= \hspace{-0.8mm}
\bar{\bar{M}}_{32}\frac{\sqrt{6}}{24}\left[\hspace{-1mm}\begin{array}{c}
\frac{1}{2}\left( \partial^2_{x} E_y  + 2\partial_{xy} E_x - \partial^2_{y} E_y \right) \\ [0mm]
\frac{2}{\sqrt{6}}\left(\partial_{yz} E_x  + \partial_{xz} E_y + \partial_{xy} E_z \right) \\[0mm]
\frac{2}{\sqrt{15}}\left( \partial^2_{z} E_y  + 2\partial_{yz} E_z - \frac{1}{4}\partial^2_{x} E_y - \frac{1}{2} \partial_{xy} E_x - \frac{3}{4} \partial^2_{y} E_y \right)  \\[0mm]
\frac{1}{\sqrt{10}}\left(2\partial^2_{z} E_z  - \partial^2_{x} E_z  -  2\partial_{xz} E_x - \partial^2_{y} E_z - 2\partial_{yz} E_y \right)  \\[0mm]
\frac{2}{\sqrt{15}}\left(\partial^2_{z} E_x  + 2\partial_{xz} E_z - \frac{1}{4}\partial^2_{y} E_x - \frac{1}{2}\partial_{xy} E_y - \frac{3}{4}\partial^2_{x} E_x \right)  \\[0mm]
\frac{1}{\sqrt{6}}\left(\partial^2_{x} E_z + 2\partial_{xz} E_x - \partial^2_{y} E_z - 2\partial_{yz} E_y  \right) \\[0mm]
\frac{1}{2}\left(\partial^2_{x} E_x  - 2\partial_{xy} E_y - \partial^2_{y} E_x \right)
\end{array}\hspace{-1mm}\right],
\end{split}
\end{equation}
\text{where}
\begin{equation}\label{transformation-octupole-M-2-c}
\bar{\bar{M}}_{3\rm{f}} = \frac{24}{\sqrt{6}}\left[\begin{array}{ccccccccccc}
0 & 0 & 0 & 0 & 0 & 0 & -2  \\
0 & 0 & 0 & 0 & \sqrt{15}/2 & 0 & -1/2 \\
1/2 & 0 & \sqrt{15}/2 & 0 & 0  & 0 & 0  \\
2 & 0 & 0 & 0 & 0 & 0 & 0  \\
0 & 0 & 0 & \sqrt{10}/2 &  0 & \sqrt{6}/2  & 0  \\
0 & 0 & 0 & \sqrt{10}/2 &  0 & -\sqrt{6}/2  & 0  \\
0 & \sqrt{6} & 0 & 0  & 0 & 0 & 0
\end{array} \right].
\end{equation}
\end{subequations}
Therefore, the octupole polarizabilities for the individual octupoles and fields of \eqref{transformation-octupole-M} and \eqref{transformation-octupole-M-2} are acquired as,
\begin{equation}\label{transformation-octupole-pol}
\begin{split}
\mathbf{O}^{\rm e} = \bar{\bar{\alpha}}^{vv'}_{33}\mathbf{E}_3 \Rightarrow 
\bar{\bar{M}}_{3\rm{m}}^{-1}\breve{\mathbf{O}}^{\rm e} &= \bar{\bar{\alpha}}^{vv'}_{33}\bar{\bar{M}}_{32}^{-1}\breve{\mathbf{E}}_3 \Rightarrow \\
\breve{\mathbf{O}}^{\rm e} = \bar{\bar{M}}_{3\rm{m}}\,\bar{\bar{\alpha}}^{vv'}_{33}\,\bar{\bar{M}}_{3\rm{f}}^{-1}\,\,\breve{\mathbf{E}}_3
\Rightarrow \breve{\bar{\bar{\alpha}}}^{vv'}_{33} &= \bar{\bar{M}}_{3\rm{m}}\,\bar{\bar{\alpha}}^{vv'}_{33}\,\bar{\bar{M}}_{3\rm{f}}^{-1},\quad \{v,v'\} = \{{\rm e},\,{ \rm m}\}.
\end{split}
\end{equation}
The transformations for the magnetic multipoles and fields are performed equivalently. 

Finally, if the same procedure is followed with dipoles and octupoles as with quadrupoles, with $\bar{\bar{M}}_{1\rm{m}}$ and $\bar{\bar{M}}_{1\rm{f}}$ being identity matrices, then the final transformation formula for the full polarizability matrix is
\begin{equation}\label{transformation-pol}
\breve{\alpha}^{vv'}_{ij} = \bar{\bar{M}}_{i\rm{m}}\,\bar{\bar{\alpha}}^{vv'}_{ij}\,\bar{\bar{M}}_{j\rm{f}}^{-1},
\quad \{v,v'\} = \{{\rm e},\,{\rm m}\}\,\,\text{and}\,\,\{i,\,j\} = \{1,2,3\}.
\end{equation}
This methodology can also construct other transformation matrices for different multipole basis selection.
The matrices $\bar{\bar{M}}_{i\rm{m}}$ and $\bar{\bar{M}}_{i\rm{f}}$ derived above in this section for the specific combinations of multipole moments and field derivatives can be used on the formulas of Ref.~\cite{mun2020describing}, in order to derive the transformation matrices $\bar{\bar{F}}_{i}$, $i=\{2,3\}$  utilized in this work ($\rightarrow$ \textit{Appendix E - Main Text}). 
%
%
%
\section{\quad Transformations between spherical and Cartesian coordinates for optical cross-sections}
\setcounter{equation}{64} 
The scattering cross-section is defined as \cite{mishchenko2002scattering}
\begin{equation}\label{scattering-cross-section-def}
\sigma_{\rm sca}^{\rm s} = \frac{1}{|E_0|^2k^2}\sum_{j=1}^{\infty}\left(\,|\mathbf{b}^{\rm e}_j|^2 +|\mathbf{b}^{\rm m}_j|^2\,\right),
\end{equation}
where $|\cdot|^2$ is the 2-norm of a vector or a matrix. In this work, investigations are limited to the octupole or $j=3$ order. Thus, \eqref{scattering-cross-section-def} is simplified to
\begin{equation}\label{scattering-cross-section-def-oct-S}
\sigma^{\rm s}_{\rm sca} = \frac{1}{|E_0|^2k^2}\sum_{j=1}^{3}\left(\,|\mathbf{b}^{\rm e}_j|^2 +|\mathbf{b}^{\rm m}_j|^2\,\right).
\end{equation}
To express $\sigma_{\rm sca}$ as a function of the multipole moments in Cartesian coordinates, the transformations of \textit{Appendix E} - \textit{Main Text} ) are employed. Therefore,
\eqref{scattering-cross-section-def-oct-S} is re-written as
%
\begin{equation}\label{scattering-cross-section-def-cart}
\begin{split}
\sigma^{\rm c}_{\rm sca} =  \,\frac{k^4\varepsilon^{-2}}{|E_0|^2\zeta_1^2}\,|\bar{\bar{F}}_1^{-1}|^2\,|\mathbf{p}|^2 +& \frac{k^4\eta^2}{|E_0|^2\zeta_1^2}\,|\bar{\bar{F}}_1^{-1}|^2\,|\mathbf{m}|^2  
 + \frac{k^6\varepsilon^{-2}}{|E_0|^2\zeta_2^2}\,|\bar{\bar{F}}_2^{-1}|^2\,|\mathbf{Q}^{\rm e}|^2  + \\ &\frac{k^6\eta^2}{|E_0|^2\zeta_2^2}\,|\bar{\bar{F}}_2^{-1}|^2\,|\mathbf{Q}^{\rm m}|^2 
 +  \frac{k^{8}\varepsilon^{-2}}{|E_0|^2\zeta_3^2}\,|\bar{\bar{F}}_3^{-1}|^2\,|\mathbf{O}^{\rm e}|^2 + \frac{k^{8}\eta^2}{|E_0|^2\zeta_3^2}\,|\bar{\bar{F}}_3^{-1}|^2\,|\mathbf{O}^{\rm m}|^2,
\end{split}
\end{equation}
%
with the multipole moments expressed in Cartesian coordinates, defined in Appendix A. However, for the specific transformation matrices $|[F_j]^{-1}|^2 = 1$, and \eqref{scattering-cross-section-def-cart} can be manipulated further if $\zeta_j$ are, also, calculated. Therefore,
%
\begin{equation}\label{scattering-cross-section-def-cart-2-S}
\sigma^{\rm c}_{\rm sca} =  \,\frac{1}{|E_0|^2}\Big(\frac{k^4\varepsilon^{-2}}{6\pi}\,|\mathbf{p}|^2 + \frac{k^4\eta^2}{6\pi}\,|\mathbf{m}|^2  
 + \frac{k^6\varepsilon^{-2}}{120\pi}\,|\mathbf{Q}^{\rm e}|^2  + \frac{k^6\eta^2}{120\pi}\,|\mathbf{Q}^{\rm m}|^2 
 +  \frac{k^{8}\varepsilon^{-2}}{5040\pi}\,|\mathbf{O}^{\rm e}|^2 + \frac{k^{8}\eta^2}{5040\pi}\,|\mathbf{O}^{\rm m}|^2\Big).
\end{equation}
%
Similarly, the extinction cross-section can be calculated as a function of the multipole moments and the fields represented in Cartesian coordinates. The extinction cross-section is defined as \cite{mishchenko2002scattering}
\begin{equation}\label{extinction-cross-section-def-1}
\sigma_{\rm ext}^{\rm s} = -\frac{1}{|E_0|^2k^2} \Re \left\{\sum_{j=1}^{\infty}\left[\mathbf{q}^{\rm e}_j\cdot\mathbf{b}^{\rm e,*}_j + \mathbf{q}^{\rm m}_j\cdot\mathbf{b}^{\rm m,*}_j \right] \right\},
\end{equation}
where the superscript $^*$ denotes the conjugate operation. For up to $j=3$ order,
\begin{equation}\label{extinction-cross-section-def-2-S}
\sigma_{\rm ext}^{\rm s} = -\frac{1}{|E_0|^2k^2}{\Re} \Bigg\{\sum_{j=1}^{3}\left[\mathbf{q}^{\rm e}_j\cdot\mathbf{b}^{\rm e,*}_j + \mathbf{q}^{\rm m}_j\cdot\mathbf{b}^{\rm m,*}_j \right]
\Bigg\}
= -\frac{1}{|E_0|^2k^2}{\Re} \Bigg\{\sum_{j=1}^{3}\left[\,\mathbf{q}^{\rm e,T}_j\,\mathbf{b}^{\rm e,*}_j + \mathbf{q}^{\rm m,T}_j\,\mathbf{b}^{\rm m,*}_j \right]
\Bigg\},
\end{equation}
where the superscript $^{\rm T}$ denotes the transpose operation. If the transformations of fields and multipole moments 
between Cartesian and spherical coordinates 
are employed ($\rightarrow$ \textit{Appendix E} - \textit{Main Text} ), \eqref{extinction-cross-section-def-2-S} turns to
%
\begin{equation}\label{extinction-cross-section-cart-1}
\begin{split}
\sigma_{\rm ext}^{\rm c} &=-\frac{1}{|E_0|^2k^2}{\rm Re} \Bigg\{
-\frac{{\rm i}k^3}{\varepsilon}\mathbf{E}^T\bar{\bar{F}}_1^{-1,{\rm T}}\bar{\bar{F}}_1^{-1,{*}}\mathbf{p}^*_1
- {\rm i}k^3 \eta^2\mathbf{H}^T\bar{\bar{F}}_1^{-1,{\rm T}}\bar{\bar{F}}_1^{-1,{*}}\mathbf{m}^*_1 
- \frac{{\rm i}k^3}{\varepsilon}\mathbf{E}^{T}\bar{\bar{F}}_2^{-1,{\rm T}}\bar{\bar{F}}_2^{-1,{*}}\mathbf{Q}^{\mathrm{e},*}_2 \\
&- {\rm i}k^3 \eta^2\mathbf{H}^{T}\bar{\bar{F}}_2^{-1,{\rm T}}\bar{\bar{F}}_2^{-1,{*}}\mathbf{Q}^{\mathrm{m},*}_2 
- \frac{{\rm i}k^3}{\varepsilon}\mathbf{E}^{T}\bar{\bar{F}}_3^{-1,{\rm T}}\bar{\bar{F}}_3^{-1,{*}}\mathbf{O}^{\mathrm{e},*}_3
- {\rm i}k^3 \eta^2\mathbf{H}^{T}\bar{\bar{F}}_3^{-1,{\rm T}}\bar{\bar{F}}_3^{-1,{*}}\mathbf{O}^{\mathrm{m},*}_3
\Bigg\}.
\end{split}
\end{equation}
%
After the identity $\bar{\bar{F}}_j^{-1,{\rm T}}\bar{\bar{F}}_j^{-1,{*}} = \bar{\bar{I}}$ is employed, \eqref{extinction-cross-section-cart-1} finally can be rewritten as
%
\begin{equation}\label{extinction-cross-section-cart-2-S}
\begin{split}
&\sigma_{\rm ext}^{\rm c}=\\ &\hspace{-0.8mm}=\hspace{-0.8mm}-\frac{1}{|E_0|^2k^2}{\rm Re}\hspace{-0.8mm} \left\{\hspace{-0.8mm}-{\rm i}k^3 \left(\hspace{-0.3mm}
\frac{1}{\varepsilon}\mathbf{E}_1\hspace{-0.5mm}\cdot\hspace{-0.5mm}\mathbf{p}^* \hspace{-0.5mm}+ \hspace{-0.5mm} {\eta^2}\mathbf{H}_1\hspace{-0.5mm}\cdot\hspace{-0.5mm}\mathbf{m}^* 
\hspace{-0.5mm}+\hspace{-0.5mm} \frac{1}{\varepsilon}\mathbf{E}_2\hspace{-0.5mm}\cdot\hspace{-0.5mm}\mathbf{Q}^{\mathrm{e},*}
\hspace{-0.5mm}+\hspace{-0.5mm} {\eta^2}\mathbf{H}_2\hspace{-0.5mm}\cdot\hspace{-0.5mm}\mathbf{Q}^{\mathrm{m},*} 
\hspace{-0.5mm}+\hspace{-0.5mm} \frac{1}{\varepsilon}\mathbf{E}_3\hspace{-0.5mm}\cdot\hspace{-0.5mm}\mathbf{O}^{\mathrm{e},*}
\hspace{-0.5mm}+ \hspace{-0.5mm}{\eta^2}\mathbf{H}_3\hspace{-0.5mm}\cdot\hspace{-0.5mm}\mathbf{O}^{\mathrm{m},*}\hspace{-0.5mm}\right)\hspace{-0.5mm}
\right\}  \\
&= -\frac{k}{|E_0|^2}\,{\rm Im} \left\{
\frac{1}{\varepsilon}\mathbf{E}_1\cdot\mathbf{p}^* +  \eta^2\mathbf{H}_1\cdot\mathbf{m}^* 
+ \frac{1}{\varepsilon}\mathbf{E}_2\cdot\mathbf{Q}^{\mathrm{e},*}
+ {\eta^2}\mathbf{H}_2\cdot\mathbf{Q}^{\mathrm{m},*} 
+ \frac{1}{\varepsilon}\mathbf{E}_3\cdot\mathbf{O}^{\mathrm{e},*}
+ {\eta^2}\mathbf{H}_3\cdot\mathbf{O}^{\mathrm{m},*}
\right\} .
\end{split}
\end{equation}
%
The absorption cross-section can be, afterwards, calculated from \eqref{scattering-cross-section-def-cart-2-S} and \eqref{extinction-cross-section-cart-2-S}, as $\sigma_{\rm abs} = \sigma_{\rm ext} - \sigma_{\rm sca}$.
%
%
%
\section{\quad Useful Legendre function formulas}
\setcounter{equation}{71} 
The Legendre functions, $P_j^{\rm m}(x)$, have certain useful recurrence properties, 
\begin{subequations}\label{recurrence-formulas-legendre-function}
\begin{equation}\label{recurrence-formulas-legendre-function-a}
\frac{1}{\sqrt{1-x^2}}P^{\rm m}_j(x) = \left(-\frac{1}{2m} \right)\bigg[P^{m+1}_{j-1}(x) + (j+m-1)(j+m)P^{m-1}_{j-1}(x)\bigg],
\end{equation}
\begin{equation}\label{recurrence-formulas-legendre-function-b}
\sqrt{1-x^2}\frac{d}{dx}P^{\rm m}_j(x) = \frac{1}{2}\bigg[ (j-m+1)(j+m)P^{m-1}_{j}(x) - P^{m+1}_{j}(x)  \bigg].
\end{equation}
\end{subequations}
For $x={\rm cos}\theta \Rightarrow dx = -({\rm sin}\theta\, d\theta)$, \eqref{recurrence-formulas-legendre-function} turn to
\begin{subequations}\label{pi-tau-legendre-function}
\begin{equation}\label{pi-tau-legendre-function-a}
\pi_{jm}({\rm cos}\theta) = \frac{m}{{\rm sin}\theta} P^{\rm m}_j({\rm cos}\theta) = \left(-\frac{1}{2} \right)\bigg[P^{m+1}_{j-1}({\rm cos}\theta) + (j+m-1)(j+m)P^{m-1}_{j-1}({\rm cos}\theta)\bigg],
\end{equation}
\begin{equation}\label{pi-tau-legendre-function-b}
\tau_{jm}({\rm cos}\theta) = \frac{d}{dx}P^{\rm m}_j({\rm cos}\theta) = \left(-\frac{1}{2} \right)\bigg[ (j-m+1)(j+m)P^{m-1}_{j}({\rm cos}\theta) - P^{m+1}_{j}({\rm cos}\theta)  \bigg].
\end{equation}
\end{subequations}
Thus, \eqref{vsh-X} are also calculated in terms of recurrence formulas.

Another useful Legendre function values are those for the arguments $x={\rm cos\theta}=\pm 1$, with $\theta=0$ and $\pi$ , respectively. Therefore,
%
\begin{equation}\label{plus-minus-one-legendre-function}
P_{j}^{m}(1) = 
  \begin{cases}
    \,\, 0       &,\quad m \neq 0 \\
    \,\, 1       &, \quad m =0
  \end{cases},\qquad
P_{j}^{m}(-1) = 
  \begin{cases}
    \,\, 0       &,\quad m \neq 0 \\
    \,\, (-1)^j       &, \quad m =0 
  \end{cases}.
\end{equation}
%
The first few associated Legendre functions up to $j=2$, or, equivalently, the quadrupole order, are
\begin{subequations}\label{associate-legendre-values}
\begin{equation}
P_{0}^{0}({\rm cos}\theta)  = 1,
\end{equation}
\begin{equation}
P_{1}^{1}({\rm cos}\theta)  = -{\rm sin}\theta, \, \, P_{1}^{0}({\rm cos}\theta) = {\rm cos}\theta,\, \,
P_{1}^{-1}({\rm cos}\theta)  = - \frac{1}{2} P_1^{1}({\rm cos}\theta) = \frac{1}{2}{\rm sin}\theta, 
\end{equation}
\begin{equation}
P_{2}^{2}({\rm cos}\theta)  = 3{\rm sin}^{2}\theta, \, \, P_{2}^{1}({\rm cos}\theta) = -3{\rm sin}\theta{\rm cos}\theta,\, \,P_{2}^{0}({\rm cos}\theta)  = \frac{3}{2}{\rm cos}^{2}\theta-\frac{1}{2},
\end{equation}
\begin{equation}
P_{2}^{-1}({\rm cos}\theta)  = \frac{1}{2}P_2^{1}(x) = \frac{1}{2}{\rm sin}\theta{\rm cos}\theta, \, \,  P_{2}^{-2}({\rm cos}\theta) = \frac{1}{24}P_2^{2}({\rm cos}\theta) = \frac{1}{8}{\rm sin}^{2}\theta.
\end{equation}
\end{subequations} 
If \eqref{associate-legendre-values} are substituted into \eqref{pi-tau-legendre-function}, one gets for the dipole ($j=1$) case
%
\begin{equation}\label{pi-tau-l1}
\pi_{1m}(x) = 
  \begin{cases}
    \,\, -\frac{1}{2}      &,\quad m = -1 \\
    \,\, 0       &, \quad m = 0   \\
    \,\, -1 &, \quad m = 1 
  \end{cases},\qquad
\tau_{1m}(x) = 
  \begin{cases}
    \,\, \frac{1}{2}{\rm cos}\theta      &,\quad m = -1 \\
    \,\, -{\rm sin}\theta      &, \quad m = 0   \\
    \,\, -{\rm cos}\theta &, \quad m = 1 
  \end{cases},
\end{equation}
%
and for the quadrupole one ($j=2$)
%
\begin{equation}\label{pi-tau-l2}
\pi_{2m}(x) = 
  \begin{cases}
  \,\, -\frac{1}{4}{\rm sin}\theta      &,\quad m = -2 \\
    \,\, -\frac{1}{2}{\rm cos}\theta      &,\quad m = -1 \\
    \,\, 0       &, \quad m = 0   \\
    \,\, -3{\rm cos}\theta      &,\quad m = 1 \\
    \,\, \quad 6{\rm sin}\theta &, \quad m = 2 
  \end{cases},\qquad
\tau_{2m}(x) = 
  \begin{cases}
  \,\, \frac{1}{4}{\rm sin}\theta{\rm cos}\theta      &,\quad m = -2 \\
    \,\, {\rm cos}^2\theta - \frac{1}{2}      &,\quad m = -1 \\
    \,\, -3{\rm sin}\theta{\rm cos}\theta       &, \quad m = 0   \\
    \,\, 3 - 6{\rm cos}^2\theta      &,\quad m = 1 \\
    \,\, 6{\rm sin}\theta{\rm cos}\theta &, \quad m = 2 
  \end{cases}.
\end{equation}
%
%
\section{\quad Deriving the imaginary part of the lattice coupling coefficients}
\setcounter{equation}{77} 
For a non-absorbing and non-diffracting metasurface, energy conservation dictates that $T+R=1$. If the metasurface consists of meta-atoms with only an electric dipole response, from \textbf{Table II A)} we can write:
 \begin{subequations}\label{CddIm-1}
 \begin{align}
r =\frac{-3}{4\pi\widetilde{\Lambda}^2}a_{1,\mathrm{eff}}=-A,\quad t=1-\frac{3}{4\pi\widetilde{\Lambda}^2}a_{1,\mathrm{eff}}=1-A.
 \end{align}
 \end{subequations}
Then, we can write
\begin{eqnarray} \small
|A|^{2}+|1-A|^{2}=2AA^{*}+1-A^{*}-A=1\Rightarrow AA^{*}=\Re\left(A^{*}\right) \Rightarrow 1=\Re\left(\frac{1}{A}\right).
\end{eqnarray}
Then, we can use the Mie angle model \cite{rahimzadegan2020minimalist} and write
\begin{eqnarray} \small
\Re\left(\frac{1}{\frac{3}{4\pi\widetilde{\Lambda}^2}a_{1,\mathrm{eff}}}\right)=\frac{4\pi\widetilde{\Lambda}^2}{3}\left[\Re\left(\frac{e^{-\mathrm{i}\theta_\mathrm{E1}}}{\cos\theta_\mathrm{E1}}\right)+\Im\left(C_{\rm dd}\right)\right] =\frac{4\pi\widetilde{\Lambda}^2}{3}\left[1+\Im\left(C_{\rm dd}\right)\right]=1.
\end{eqnarray}
Then, it can directly follow
\begin{eqnarray}
\Im\left(C_\mathrm{dd}\right)=\frac{3}{4\pi\widetilde{\Lambda}^2}-1.
\end{eqnarray}
\noindent where we have used the Mie model ($a_1=\cos\theta_\mathrm{E1} \exp{\mathrm{i}\theta_\mathrm{E1}}$) for non-absorbing spheres \cite{rahimzadegan2020minimalist}. Similarly, we can derive the imaginary parts of $C_\mathrm{QQ}$ and $C_\mathrm{dQ}$ as
\begin{subequations}\label{CQQIm}
\begin{align} 
\Im\left(C_\mathrm{QQ}\right)&=\frac{5}{4\pi\widetilde{\Lambda}^2}-1, \quad
\Im\left(C_\mathrm{dQ}\right)=\frac{\sqrt{15}}{4\pi\widetilde{\Lambda}^2}.
\end{align}
\end{subequations}

\section{\quad Effective T and polarizability matrices}
Here, we derive the equation for the effective polarizability matrix from the vector spherical counterpart relation. The effective T matrix is defined as
\begin{eqnarray} \label{T-eff-4}
\bar{\bar{T}}_\mathrm{eff}=	(\bar{\bar{I}}-\bar{\bar{T}}_0\bar{\bar{C}}_{\mathrm{s}})^{-1}\bar{\bar{T}}_0.
\end{eqnarray}
First, we define
\begin{equation}
\bar{\bar{F}}_{jj'} = \left[\begin{array}{c}
\bar{\bar{F}}_{j} \quad \bar{\bar{0}}\\
\bar{\bar{0}} \quad \bar{\bar{F}}_{j'}
\end{array} \right].
\end{equation}
From the above equations and the transformation formulas, we can derive the effective polarizability equations with the following step-by-step procedure
%
\begin{align}\label{T-eff-2}
 \mathrm{i}\,\bar{\bar{F}}_{jj'}^{-1}\,\bar{\bar{\widetilde{\alpha}}}_{_\mathrm{eff},jj'}^{pp'}\,\bar{\bar{F}}_{jj'}&=	(\bar{\bar{I}}_{jj'}-\mathrm{i}\bar{\bar{F}}_{jj'}^{-1}\,\bar{\bar{\widetilde{\alpha}}}_{_\mathrm{0},jj'}^{pp'}\,\bar{\bar{F}}_{jj'}(-\mathrm{i}\bar{\bar{F}}_{jj'}^{-1} \bar{\bar{C}}_{jj'}\bar{\bar{F}}_{jj'}))^{-1}\,\bar{\bar{F}}_{jj'}^{-1}\,\mathrm{i}\bar{\bar{\widetilde{\alpha}}}_{_\mathrm{0},jj'}^{pp'}\,\bar{\bar{F}}_{jj'} \Leftrightarrow \nonumber\\
\bar{\bar{F}}_{jj'}^{-1}\,\bar{\bar{\widetilde{\alpha}}}_{_\mathrm{eff},jj'}^{pp'}\,\bar{\bar{F}}_{jj'}&=	(\bar{\bar{I}}_{jj'}-\bar{\bar{F}}_{jj'}^{-1}\,\bar{\bar{\widetilde{\alpha}}}_{_\mathrm{0},jj'}^{pp'}\, \bar{\bar{C}}_{jj'}\bar{\bar{F}}_{jj'})^{-1}\,\bar{\bar{F}}_{jj'}^{-1}\,\bar{\bar{\widetilde{\alpha}}}_{_\mathrm{0},jj'}^{pp'}\,\bar{\bar{F}}_{jj'}\Leftrightarrow \nonumber\\
\bar{\bar{F}}_{jj'}^{-1}\,\bar{\bar{\widetilde{\alpha}}}_{_\mathrm{eff},jj'}^{pp'}\,\bar{\bar{F}}_{jj'}&=	[\bar{\bar{F}}_{jj'}^{-1}(\bar{\bar{I}}_{jj'}-\bar{\bar{\widetilde{\alpha}}}_{_\mathrm{0},jj'}^{pp'}\, \bar{\bar{C}}_{jj'})\bar{\bar{F}}_{jj'}]^{-1}\,\bar{\bar{F}}_{jj'}^{-1}\,\bar{\bar{\widetilde{\alpha}}}_{_\mathrm{0},jj'}^{pp'}\,\bar{\bar{F}}_{jj'}\Leftrightarrow \nonumber\\
\bar{\bar{F}}_{jj'}^{-1}\,\bar{\bar{\widetilde{\alpha}}}_{_\mathrm{eff},jj'}^{pp'}\,\bar{\bar{F}}_{jj'}&=	\bar{\bar{F}}_{jj'}^{-1}(\bar{\bar{I}}_{jj'}-\bar{\bar{\widetilde{\alpha}}}_{_\mathrm{0},jj'}^{pp'}\, \bar{\bar{C}}_{jj'})^{-1}\,\bar{\bar{\widetilde{\alpha}}}_{_\mathrm{0},jj'}^{pp'}\,\bar{\bar{F}}_{jj'} \Leftrightarrow\nonumber\\\nonumber\\
\bar{\bar{\widetilde{\alpha}}}_{_\mathrm{eff},jj'}^{pp'}&=(\bar{\bar{I}}_{jj'}-\bar{\bar{\widetilde{\alpha}}}_{_\mathrm{0},jj'}^{pp'}\, \bar{\bar{C}}_{jj'})^{-1}\,\bar{\bar{\widetilde{\alpha}}}_{_\mathrm{0},jj'}^{pp'},
\end{align}
%
\noindent where we have used the following property of square matrices $\mathbf{A}$, $\mathbf{B}$
\begin{eqnarray} \label{T-eff-3}
\left(\mathbf{A} \mathbf{B}\right)^{-1}= \mathbf{B}^{-1} \mathbf{A}^{-1}.
\end{eqnarray}
%
%
\section{\quad Analytical models for T-matrices and lattice coupling matrices}
\setcounter{equation}{85} 
\subsection{T matrix}
\paragraph{Spheres:} The T matrix of an isotropic meta-atom can be written as
\begin{eqnarray}
\bar{\bar{T}}_{ \mathrm{o}} =-\footnotesize\left[
\begin{array}{cccccccccccccccc}
 {a_1} & 0 & 0 & 0 & 0 & 0 & 0 & 0 & 0 & 0 & 0 & 0 & 0 & 0 & 0 & 0 \\
 0 & {a_1} & 0 & 0 & 0 & 0 & 0 & 0 & 0 & 0 & 0 & 0 & 0 & 0 & 0 & 0 \\
 0 & 0 & {a_1} & 0 & 0 & 0 & 0 & 0 & 0 & 0 & 0 & 0 & 0 & 0 & 0 & 0 \\
 0 & 0 & 0 & {a_2} & 0 & 0 & 0 & 0 & 0 & 0 & 0 & 0 & 0 & 0 & 0 & 0 \\
 0 & 0 & 0 & 0 & {a_2} & 0 & 0 & 0 & 0 & 0 & 0 & 0 & 0 & 0 & 0 & 0 \\
 0 & 0 & 0 & 0 & 0 & {a_2} & 0 & 0 & 0 & 0 & 0 & 0 & 0 & 0 & 0 & 0 \\
 0 & 0 & 0 & 0 & 0 & 0 & {a_2} & 0 & 0 & 0 & 0 & 0 & 0 & 0 & 0 & 0 \\
 0 & 0 & 0 & 0 & 0 & 0 & 0 & {a_2} & 0 & 0 & 0 & 0 & 0 & 0 & 0 & 0 \\
 0 & 0 & 0 & 0 & 0 & 0 & 0 & 0 & {b_1} & 0 & 0 & 0 & 0 & 0 & 0 & 0 \\
 0 & 0 & 0 & 0 & 0 & 0 & 0 & 0 & 0 & {b_1} & 0 & 0 & 0 & 0 & 0 & 0 \\
 0 & 0 & 0 & 0 & 0 & 0 & 0 & 0 & 0 & 0 & {b_1} & 0 & 0 & 0 & 0 & 0 \\
 0 & 0 & 0 & 0 & 0 & 0 & 0 & 0 & 0 & 0 & 0 & {b_2} & 0 & 0 & 0 & 0 \\
 0 & 0 & 0 & 0 & 0 & 0 & 0 & 0 & 0 & 0 & 0 & 0 & {b_2} & 0 & 0 & 0 \\
 0 & 0 & 0 & 0 & 0 & 0 & 0 & 0 & 0 & 0 & 0 & 0 & 0 & {b_2} & 0 & 0 \\
 0 & 0 & 0 & 0 & 0 & 0 & 0 & 0 & 0 & 0 & 0 & 0 & 0 & 0 & {b_2} & 0 \\
 0 & 0 & 0 & 0 & 0 & 0 & 0 & 0 & 0 & 0 & 0 & 0 & 0 & 0 & 0 & {b_2} \\
\end{array}
\right],
\end{eqnarray}
\noindent where $a_j$, and $b_j$ are the Mie coefficients. The Mie coefficients for a homogeneous sphere are calculated as ~\cite{Bohren2008,Jackson1999}
\begin{eqnarray}
a_{n}	&=&	\frac{\mu_{\rm em.}\eta^{2}j_{n}\left(\eta x\right)\left[xj_{n}\left(x\right)\right]^{\prime}-\mu_{\rm sph.}j_{n}\left(x\right)\left[\eta xj_{n}\left(\eta x\right)\right]^{\prime}}{\mu_{\rm em.}\eta^{2}j_{n}\left(\eta x\right)\left[xh_{n}^{\left(1\right)}\left(x\right)\right]^{\prime}-\mu_{\rm sph.}h_{n}^{\left(1\right)}\left(x\right)\left[\eta xj_{n}\left(\eta x\right)\right]^{\prime}},\nonumber\\
b_{n}	&=&	\frac{\mu_{\rm sph.}j_{n}\left(\eta x\right)\left[xj_{n}\left(x\right)\right]^{\prime}-\mu_{\rm em.}j_{n}\left(x\right)\left[\eta xj_{n}\left(\eta x\right)\right]^{\prime}}{\mu_{\rm sph.}j_{n}\left(\eta x\right)\left[xh_{n}^{\left(1\right)}\left(x\right)\right]^{\prime}-\mu_{\rm em.}h_{n}^{\left(1\right)}\left(x\right)\left[\eta xj_{n}\left(\eta x\right)\right]^{\prime}},
\label{eq:Miefull}
\end{eqnarray}
\noindent where subscript ${\rm em.}$ and ${\rm sph.}$ denote the properties of the embedding and the sphere, respectively. $j$ and $h$ denote the spherical Bessel and Hankel functions, respectively. We have used $n$ as the multipolar order so that it is not confused with the Bessel function. The primes are derivatives with respect to the argument. The parameters $x$ and $\eta$ are defined as 
\begin{equation}
x	=	\frac{\omega}{c} \sqrt{\epsilon_{\rm em.}\left(\omega\right)\mu_{\rm em.}\left(\omega\right)}R_{\rm sph.},\quad \text{and} \quad
\eta	=	\sqrt{\frac{\epsilon_{\rm sph.}\left(\omega\right)\mu_{\rm sph.}\left(\omega\right)}{\epsilon_{\rm em.}\left(\omega\right)\mu_{\rm em.}\left(\omega\right)}},
\end{equation}
\noindent where $R_{\rm sph.}$ is the radius of the sphere. 

\paragraph{Cylinders:} The T matrix of a cylinder can be written as the following model
\begin{eqnarray}
\bar{\bar{T}}_{ \mathrm{o}} =-\footnotesize\left[\hspace{-0.5mm}
\arraycolsep=1.2mm
\begin{array}{cccccccccccccccc}
 a_{11} & 0 & 0 & 0 & 0 & 0 & 0 & 0 & 0 & 0 & 0 & 0 & a_{\rm pM} & 0 & 0 & 0\\  0 & a_{12} & 0 & 0 & 0 & 0 & 0 & 0 & 0 & 0 & 0 & 0 & 0 & 0 & 0 & 0 \\  0 & 0 & a_{a11} & 0 & 0 & 0 & 0 & 0 & 0 & 0 & 0 & 0 & 0 & 0 & -a_{\rm pM} & 0    \\  0 & 0 & 0 & a_{21} & 0 & 0 & 0 & 0 & 0 & 0 & 0 & 0 & 0 & 0 & 0 & 0 \\  0 & 0 & 0 & 0 & a_{22} & 0 & 0 & 0 & a_{\rm Qm} & 0 & 0 & 0 & 0 & 0 & 0 & 0    \\  0 & 0 & 0 & 0 & 0 & a_{23} & 0 & 0 & 0 & 0 & 0 & 0 & 0 & 0 & 0 & 0 \\  0 & 0 & 0 & 0 & 0 & 0 & a_{22} & 0 & 0 & 0 & -a_{\rm Qm} & 0 & 0 & 0 & 0 & 0    \\  0 & 0 & 0 & 0 & 0 & 0 & 0 & a_{21} & 0 & 0 & 0 & 0 & 0 & 0 & 0 & 0 \\  0 & 0 & 0 & 0 & -a_{\rm Qm} & 0 & 0 & 0 & b_{11} & 0 & 0 & 0 & 0 & 0 & 0 & 0    \\  0 & 0 & 0 & 0 & 0 & 0 & 0 & 0 & 0 & b_{12} & 0 & 0 & 0 & 0 & 0 & 0 \\  0 & 0 & 0 & 0 & 0 & 0 & a_{\rm Qm} & 0 & 0 & 0 & b_{11} & 0 & 0 & 0 & 0 & 0
   \\  0 & 0 & 0 & 0 & 0 & 0 & 0 & 0 & 0 & 0 & 0 & b_{21} & 0 & 0 & 0 & 0 \\  -a_{\rm pM} & 0 & 0 & 0 & 0 & 0 & 0 & 0 & 0 & 0 & 0 & 0 & b_{22} & 0 & 0 & 0    \\  0 & 0 & 0 & 0 & 0 & 0 & 0 & 0 & 0 & 0 & 0 & 0 & 0 & b_{23} & 0 & 0 \\  0 & 0 & a_{\rm pM} & 0 & 0 & 0 & 0 & 0 & 0 & 0 & 0 & 0 & 0 & 0 & b_{22} & 0    \\  0 & 0 & 0 & 0 & 0 & 0 & 0 & 0 & 0 & 0 & 0 & 0 & 0 & 0 & 0 & b_{21}  \end{array}\hspace{-0.5mm}
\right],
\end{eqnarray}
\begin{figure*}
    \centering
    \includegraphics[width=\linewidth]{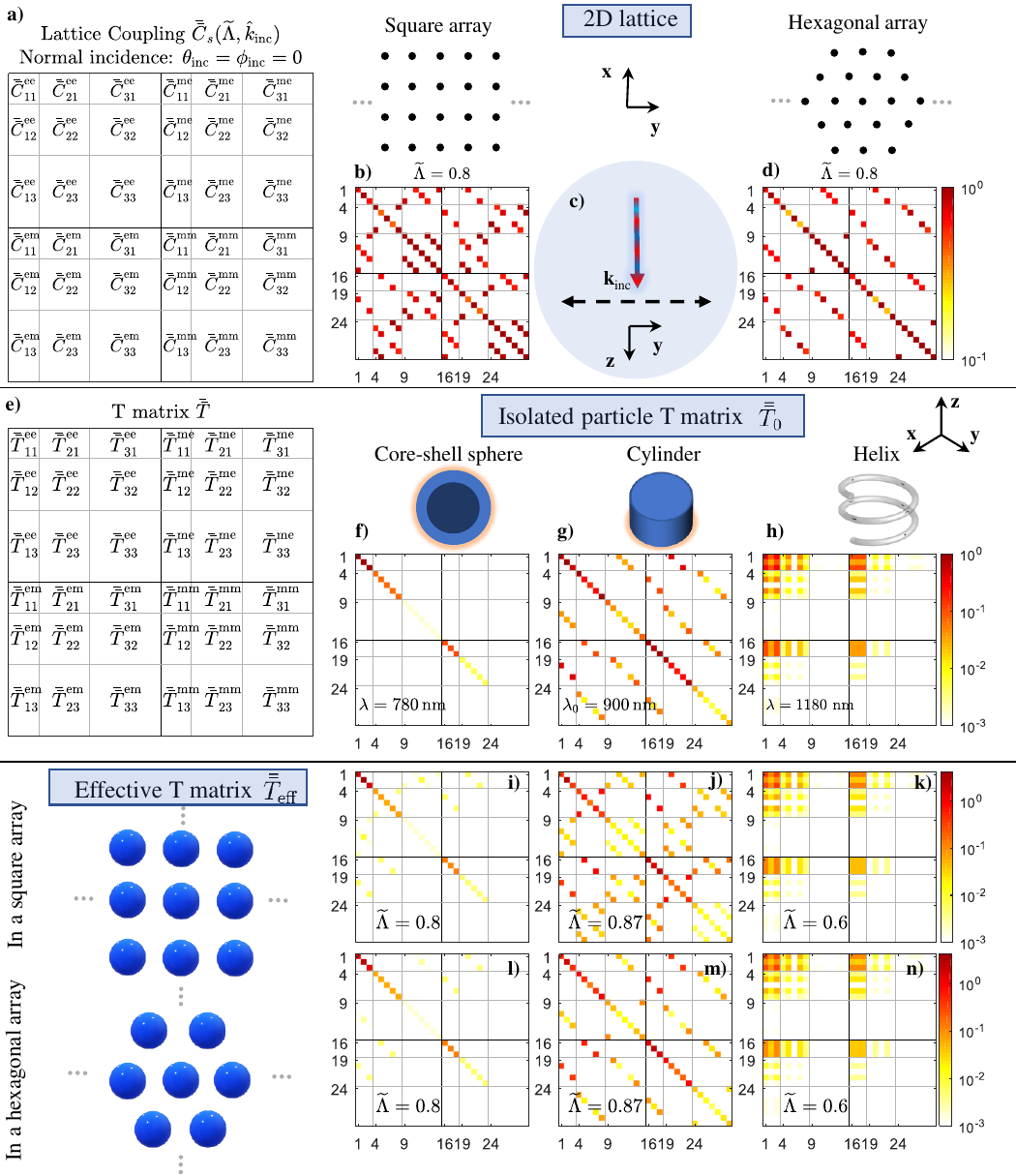}
    \caption{\textbf{Symmetries of scatterers and 2D lattices in VSH basis}: \textbf{Row I:} (a) The spherical lattice coupling matrix amplitude up to octupolar order for (b) a square array and d) Hexagonal array under normal incidence as shown in (c). The normalized periodicity for both arrays is 0.8. \textbf{Row II:} (e) The T-matrix amplitude for (f) an isolated Ag-core SiO2-shell particle ($r_\mathrm{core}=120,r_\mathrm{shell}=120+30$) in free space at $\lambda=780$~nm, and (g) an isolated cylinder ($r=291\,{\rm nm},h=211\,{\rm nm}$) embedded in silica ($n=1.44$) at $\lambda=900$~nm, and (h) an isolated helix ($R_{\rm axial}=80$~nm, $r_{\rm rod}=20$~nm, $P_{\rm pitch}=105$~nm and $N_{\rm turn}=2$) in free space at $\lambda=1180$~nm. \textbf{Row III:} The effective T-matrix amplitude of (i) \& (j) the core-shell sphere, (j) \& (m) the cylinder, and (k)\& (n) the helix inside an infinitely periodic (i)-(k) square array or (l)-(n) hexagonal array.}
\label{FIG:latticeFullT}
\end{figure*}
\subsection{C matrix}
\vspace{-5mm}
The Cartesian coupling matrix can be written as 
\begin{equation} 
\bar{\bar{C}}= \left[\begin{array}{cccccc}
\bar{\bar{C}}_{11}^{\rm ee} & \bar{\bar{C}}_{12}^{\rm ee} & \bar{\bar{C}}_{13}^{\rm ee} & \bar{\bar{C}}_{11}^{\rm em} & \bar{\bar{C}}_{12}^{\rm em} & \bar{\bar{C}}_{13}^{\rm em}  \\[0.05cm]
\bar{\bar{C}}_{21}^{\rm ee} & \bar{\bar{C}}_{22}^{\rm ee} & \bar{\bar{C}}_{23}^{\rm ee} & \bar{\bar{C}}_{21}^{\rm em} & \bar{\bar{C}}_{22}^{\rm em} & \bar{\bar{C}}_{23}^{\rm em}  \\[0.05cm]
\bar{\bar{C}}_{31}^{\rm ee} & \bar{\bar{C}}_{32}^{\rm ee} & \bar{\bar{C}}_{33}^{\rm ee} & \bar{\bar{C}}_{31}^{\rm em} & \bar{\bar{C}}_{32}^{\rm em} & \bar{\bar{C}}_{33}^{\rm em}  \\[0.05cm]
\bar{\bar{C}}_{11}^{\rm me} & \bar{\bar{C}}_{12}^{\rm me} & \bar{\bar{C}}_{13}^{\rm me} & \bar{\bar{C}}_{11}^{\rm mm} & \bar{\bar{C}}_{12}^{\rm mm} & \bar{\bar{C}}_{13}^{\rm mm} \\[0.05cm]
\bar{\bar{C}}_{21}^{\rm me} & \bar{\bar{C}}_{22}^{\rm me} &  \bar{\bar{C}}_{23}^{\rm me} & \bar{\bar{C}}_{21}^{\rm mm} & \bar{\bar{C}}_{22}^{\rm mm} & \bar{\bar{C}}_{23}^{\rm mm} \\[0.05cm]
\bar{\bar{C}}_{31}^{\rm me} & \bar{\bar{C}}_{32}^{\rm me} &  \bar{\bar{C}}_{33}^{\rm me} & \bar{\bar{C}}_{31}^{\rm mm} & \bar{\bar{C}}_{32}^{\rm mm} & \bar{\bar{C}}_{33}^{\rm mm}
\end{array}\right].
\end{equation}
Due to electromagnetic duality symmetry, $\bar{\bar{C}}_{jj'}^\mathrm{ee}=\bar{\bar{C}}_{jj'}^\mathrm{mm}$ and $\bar{\bar{C}}_{jj'}^\mathrm{me}=\bar{\bar{C}}_{jj'}^\mathrm{em}$. 
The dipolar and the dipolar-quadrupolar parts can be written as
\begin{equation} 
\bar{\bar{C}}_{\mathrm{D}} = \left[\begin{array}{cc}
\bar{\bar{C}}_{11}^{\rm ee} & \bar{\bar{C}}_{11}^{\rm em} \\[0.05cm]
\bar{\bar{C}}_{11}^{\rm me} & \bar{\bar{C}}_{11}^{\rm mm}
\end{array}\right]\qquad \text{and}\qquad
%
%
\bar{\bar{C}}_{\mathrm{D\&Q}}= \left[\begin{array}{cccccc}
\bar{\bar{C}}_{11}^{\rm ee} & \bar{\bar{C}}_{12}^{\rm ee} &  \bar{\bar{C}}_{11}^{\rm em} & \bar{\bar{C}}_{12}^{\rm em} \\[0.05cm]
\bar{\bar{C}}_{21}^{\rm ee} & \bar{\bar{C}}_{22}^{\rm ee} &  \bar{\bar{C}}_{21}^{\rm em} & \bar{\bar{C}}_{22}^{\rm em} \\[0.05cm]
\bar{\bar{C}}_{11}^{\rm me} & \bar{\bar{C}}_{12}^{\rm me} & \bar{\bar{C}}_{11}^{\rm me} & \bar{\bar{C}}_{12}^{\rm mm} \\[0.05cm]
\bar{\bar{C}}_{21}^{\rm me} & \bar{\bar{C}}_{22}^{\rm me} &  \bar{\bar{C}}_{21}^{\rm me} & \bar{\bar{C}}_{22}^{\rm mm} 
\end{array}\right].
\end{equation}

Different lattice symmetries result in different coupling matrix symmetries. Here, we show essential models of the C matrix based on the multipolar order, incidence orientation, and lattice arrangement. We will use simplified symbols for elements that would be later used in the equations. 
For a dipolar metasurface illuminated under normal incidence, the Cartesian C matrix for a square and a hexagonal arrangement can be written as 
\begin{eqnarray}
\footnotesize\bar{\bar{C}}_{\rm D}=\left[
\begin{array}{cccccc}
 C_{\rm dd}^{\rm xx} & 0 & 0 & 0 & 0 & 0 \\
 0 & C_{\rm dd}^{\rm yy}  & 0 & 0 & 0 & 0 \\
 0 & 0 &  C_{\mathrm{dd}}^{\rm zz} & 0 & 0 & 0 \\
 0 & 0 & 0 & C_{\rm dd}^{\rm xx} & 0 & 0 \\
 0 & 0 & 0 & 0 & C_{\rm dd}^{\rm yy} & 0 \\
 0 & 0 & 0 & 0 & 0 & C_{\mathrm{dd}}^{\rm zz} \\
\end{array}
\right] = \left[
\begin{array}{cccccc}
 C_{\rm dd} & 0 & 0 & 0 & 0 & 0 \\
 0 & C_{\rm dd}  & 0 & 0 & 0 & 0 \\
 0 & 0 &  C_{\mathrm{dd}}^{\rm zz} & 0 & 0 & 0 \\
 0 & 0 & 0 & C_{\rm dd} & 0 & 0 \\
 0 & 0 & 0 & 0 & C_{\rm dd} & 0 \\
 0 & 0 & 0 & 0 & 0 & C_{\mathrm{dd}}^{\rm zz} \\
\end{array}
\right],
\end{eqnarray}
\noindent where, for simplicity, and due to symmetry, we use $C_{\rm dd}^{\rm xx}= C_{\rm dd}^{\rm yy}= C_{\rm dd}$.
For a dipolar metasurface, illuminated at oblique incidence with a plane wave ($\phi_{\rm inc}=0$), the matrix for both the square and hexagonal arrangement is
\begin{eqnarray}
\footnotesize\bar{\bar{C}}_{\rm D}=\left[
\begin{array}{cccccc}
  C_{\rm dd}^{\rm xx} & 0 & 0 & 0 & 0 & 0 \\
 0 & C_{\rm dd}^{\rm yy} & 0 & 0 & 0 & C_{\rm dd}^{\rm yz} \\
 0 & 0 & C_{\rm dd}^{\rm zz} & 0 & C_{\rm dd}^{\rm zy} & 0 \\
 0 & 0 & 0 & C_{\rm dd}^{\rm xx} & 0 & 0 \\
 0 & 0 & C_{\rm dd}^{\rm yz} & 0 & C_{\rm dd}^{\rm yy} & 0 \\
 0 & C_{\rm dd}^{\rm zy} & 0 & 0 & 0 & C_{\rm dd}^{\rm zz} \\
\end{array}
\right]=\left[
\begin{array}{cccccc}
  C_{\rm xx} & 0 & 0 & 0 & 0 & 0 \\
 0 & C_{\rm yy} & 0 & 0 & 0 & C_{\rm yz} \\
 0 & 0 & C_{\rm zz} & 0 & -C_{\rm yz} & 0 \\
 0 & 0 & 0 & C_{\rm xx} & 0 & 0 \\
 0 & 0 & C_{\rm yz} & 0 & C_{\rm yy} & 0 \\
 0 & -C_{\rm yz} & 0 & 0 & 0 & C_{\rm zz} \\
\end{array}
\right],
\end{eqnarray}
\noindent where for simplicity we use  $ C_{\rm dd}^{\rm xx}= C_{\rm xx}$, $ C_{\rm dd}^{\rm yy}= C_{\rm yy}$,  $ C_{\rm dd}^{\rm zz}= C_{\rm zz}$, and $ C_{\rm dd}^{\rm yz}=- C_{\rm dd}^{\rm zy}= C_{\rm yz}$. It is anti-symmetric. 

For a dipolar-quadrupolar metasurface with a square lattice and illuminated at normal incidence ($\theta_{\rm inc}=\phi_{\rm inc}=0$), we can write the lattice coupling matrix as
\begin{flalign}
&\footnotesize\bar{\bar{C}}_{\mathrm{D\&Q}} \hspace{-0.5mm}=\hspace{-0.5mm}\footnotesize\left[\hspace{-0.5mm}
\arraycolsep=0.5mm \begin{array}{cccccccccccccccc}
 C_{\rm dd} & 0 & 0 & 0 & 0 & 0 & 0 & 0 & 0 & 0 & 0 & 0 & -C_{\rm dQ} & 0 & 0 & 0 \\
 0 &  C_{\rm dd} & 0 & 0 & 0 & 0 & 0 & 0 & 0 & 0 & 0 & 0 & 0 & 0 & C_{\rm dQ} & 0 \\
 0 & 0 &  C_{\mathrm{dd}}^{\rm zz}  & 0 & 0 & 0 & 0 & 0 & 0 & 0 & 0 & 0 & 0 & 0 & 0 & 0 \\
 0 & 0 & 0 & C_{\mathrm{QQ}}^{\rm xy} & 0 & 0 & 0 & 0 & 0 & 0 & 0 & 0 & 0 & 0 & 0 & 0 \\
 0 & 0 & 0 & 0 & C_{\rm QQ} & 0 & 0 & 0 & -C_{\rm dQ} & 0 & 0 & 0 & 0 & 0 & 0 & 0 \\
 0 & 0 & 0 & 0 & 0 & C_{\mathrm{QQ}}^{\rm zz} & 0 & 0 & 0 & 0 & 0 & 0 & 0 & 0 & 0 & 0 \\
 0 & 0 & 0 & 0 & 0 & 0 & C_{\rm QQ} & 0 & 0 & C_{\rm dQ} & 0 & 0 & 0 & 0 & 0 & 0 \\
 0 & 0 & 0 & 0 & 0 & 0 & 0 & C_{\mathrm{QQ}}^{\rm xxyy} & 0 & 0 & 0 & 0 & 0 & 0 & 0 & 0 \\
 0 & 0 & 0 & 0 & -C_{\rm dQ} & 0 & 0 & 0 &  C_{\rm dd} & 0 & 0 & 0 & 0 & 0 & 0 & 0 \\
 0 & 0 & 0 & 0 & 0 & 0 & C_{\rm dQ} & 0 & 0 &  C_{\rm dd} & 0 & 0 & 0 & 0 & 0 & 0 \\
 0 & 0 & 0 & 0 & 0 & 0 & 0 & 0 & 0 & 0 &  C_{\mathrm{dd}}^{\rm zz}  & 0 & 0 & 0 & 0 & 0 \\
 0 & 0 & 0 & 0 & 0 & 0 & 0 & 0 & 0 & 0 & 0 &  C_{\mathrm{QQ}}^{\rm xy}  & 0 & 0 & 0 & 0 \\
 -C_{\rm dQ} & 0 & 0 & 0 & 0 & 0 & 0 & 0 & 0 & 0 & 0 & 0 &  C_{\rm QQ} & 0 & 0 & 0 \\
 0 & 0 & 0 & 0 & 0 & 0 & 0 & 0 & 0 & 0 & 0 & 0 & 0 & C_{\mathrm{QQ}}^{\rm zz} & 0 & 0 \\
 0 & C_{\rm dQ}& 0 & 0 & 0 & 0 & 0 & 0 & 0 & 0 & 0 & 0 & 0 & 0 & C_{\rm QQ} & 0 \\
 0 & 0 & 0 & 0 & 0 & 0 & 0 & 0 & 0 & 0 & 0 & 0 & 0 & 0 & 0 & C_{\mathrm{QQ}}^{\rm xxyy} \\
\end{array}
\hspace{-0.5mm}\right]\hspace{-0.5mm}.
\end{flalign}
In Fig.~\ref{FIG:latticeFullT}, we have shown the T matrix, the C matrix, and the effective T matrix for a core-shell sphere, a cylinder, and a helix inside a square and a hexagonal lattice. The Cartesian counterpart is shown in Fig.~2 of the main manuscript.


\end{document}